\documentclass{aa}
\usepackage{graphicx}
\usepackage{txfonts}
\usepackage{xcolor}
\usepackage{amsmath}
\usepackage{multirow}
\usepackage{placeins}
\usepackage{float}

\begin{document}

  \title{Low-temperature optical constants of amorphous silicate dust analogues  \thanks{Data from this article are publicly-available through the STOPCODA (SpecTroscopy and Optical Properties of Cosmic Dust Analogues) database of the SSHADE infrastructure of solid spectroscopy (https://doi.org/10.26302/SSHADE/STOPCODA). The dataset are accessible via the following links: https://doi.org/10.26302/SSHADE/EXPERIMENT\_KD\_20220525\_002, https://doi.org/10.26302/SSHADE/EXPERIMENT\_KD\_20220525 and https://doi.org/10.26302/SSHADE/EXPERIMENT\_KD\_20220331.} }

   \subtitle{}

  \author{
           K. Demyk       \inst{1}
   \and  V. Gromov      \inst{2}
   \and  C. Meny        \inst{1}
   \and  N. Ysard       \inst{3}
   \and  D. Paradis     \inst{1}
   \and  A. P. Jones    \inst{3}
   \and D. Petitprez    \inst{4}
   \and P. Hubert       \inst{4, 5}
   \and  H. Leroux     \inst{6} 
   \and C. Nayral       \inst{7}
   \and F. Delpech     \inst{7} }
 \institute{Institut de Recherche en Astrophysique et Planétologie, Université de Toulouse, CNRS, UPS, IRAP, 9 Av. colonel Roche, BP 44346, 31028 Toulouse Cedex 4, France\\
              \email{karine.demyk@irap.omp.eu}
\and
Space Research Institute, RAS, 84/32 Profsoyuznaya, 117810 Moscow, Russia  
\and
Université Paris-Saclay, CNRS, Institut d'Astrophysique Spatiale, 91404 Orsay, France 
\and
 Physicochimie des Processus de Combustion et de l'Atmosphère, Université de Lille, CNRS, PC2A, F-59000 Lille, France
\and
 Laboratoire d'Optique Atmosphérique, Université de Lille, CNRS, LOA, F-59000 Lille, France
 \and
 Univ. Lille, CNRS, INRAE, Centrale Lille, UMR 8207 - UMET - Unité Matériaux et Transformations, F-59000 Lille      
 \and
Universit\'e de Toulouse, INSA, CNRS, LPCNO, F-31077 Toulouse, France  }

   \date{Received 19 April, 2022; accepted 05 September, 2022}

% \abstract{}{}{}{}{} 
% 5 {} token are mandatory
 
  \abstract
     {Cosmic dust models are key ingredients in advancing our understanding of astronomical environments as diverse as interstellar clouds in galaxies, circumstellar envelopes around evolved and young stars, and protoplanetary disks. Such models consist of several dust populations, each with different compositions and size distributions. They may also consider different grain shapes, although most models assume spherical grains.  All include a component of silicate dust. The absorption and emission properties of these dust components are calculated from the optical constants of each dust material which have various experimental, phenomenological, and theoretical origins depending on the model. }
   {We aim to provide the community with new sets of optical constants for amorphous silicate dust analogues at low temperatures. The analogues consist of four Mg-rich silicate samples of stoichiometry ranging from enstatite to olivine, and of eight samples of Mg- and Fe-rich silicates with a pyroxene stoichiometry and differing magnesium and iron content. }
   {We calculated the optical constants from transmission measurements using the Kramers-Kronig relations, assuming that the grains are small compared to the wavelength and prolate in shape with axis ratios of 1.5 and 2 for the Mg- and Fe-rich samples, respectively. }
   {New optical constants for silicate dust analogues of various compositions were calculated over the wavelength range from 5 to 800\,$\mu$m or $1000\, \mu$m, depending on the sample, and at temperatures of 10, 30, 100, 200, and 300 K. We determined the uncertainties on the derived optical constants based on the assumptions used to calculate them. To facilitate the use of these data in cosmic dust models, we provide optical constants extrapolated outside the measured spectral range into the ultraviolet(UV)/visual(VIS)/near-infrared(NIR) and millimetre and centimetre wavelength ranges, as well as formulae that can be used to interpolate the optical constants at any temperature in the range $10-300$\,K. }
   {}


   \maketitle
%
%-------------------------------------------------------------------

\section{Introduction}

Cosmic dust is almost ubiquitous in the Universe. It is produced in the envelopes and ejecta of dying stars and travels through the interstellar medium (ISM) where it is partly destroyed and re-condensed to eventually end its journey participating in the formation of new stars and planetary systems. The study of all astronomical environments requires calculation the absorption, scattering, and emission of dust grains. Cosmic dust models, providing astronomers with extinction and polarisation efficiencies for a range of dust components, have been in existence for many decades and are widely used to model the dust extinction and emission in external galaxies as well as in star-forming regions. These models are calibrated on observations from the X-ray to the millimetre (mm) domain {\citep[e.g.][]{zubko2004, draine2009, compiegne2011, jones2013, siebenmorgen2017,guillet2018} and the models principally differ from one another in the number of dust components that they consider as well as the nature of the dust (composition, structure, and size distribution). The building blocks of dust models are the optical constants of the  components that describe their interaction with light and which depend on their chemical composition and physical structure. Although the models adopt different carbonaceous grain populations, most of them \citep[e.g.][]{desert1990, draine2007, compiegne2011, guillet2018} use a silicate dust component based on the same optical constants, namely those of the {\it astrosil} from \cite{draine1984}. The {\it astrosil} optical constants are based on laboratory measurements from \cite{huffman1973} in the ultraviolet (UV) and on observations in the near-infrared (NIR) and mid-infrared (MIR) \citep[see][for details]{draine1984}, and are extrapolated beyond 20 $\mu$m using a power law with ${\lambda}^{-1.6}$ dependence for $\lambda \geqslant 250\, \mu$m \citep{li2001}. An alternative dust model, THEMIS \citep{jones2013,koehler2014,jones2017}, considers two silicate dust components based on the optical constants of amorphous MgSiO$_3$ and Mg$_2$SiO$_4$ from \cite{scottduley1996} in the UV, NIR, and MIR range and uses the same extrapolation as {\it astrosil} in the FIR; it also incorporates metallic Fe and FeS inclusions into the silicate grains in order to reproduce the NIR observations. The recent {\it Astrodust} model \citep{draine2021} considers a single grain component which is a mixture of silicate and carbonaceous matter with a dielectric constant  constrained by the observations presented by \citet{hensley2021}.

In parallel to the development of dust models, experimental studies measure the absorption properties of a wide range  of dust analog materials over the widest possible wavelength range and under conditions that are as close as possible to those of the cosmos. These studies provide astronomers with absorption properties and in some cases the optical constants required by astronomical dust models.\footnote{Various databases on silicate dust analogues for planetary and astronomical studies are gathered on the SSHADE (Solid Spectroscopy Hosting Architecture of Databases and Expertises) website https://www.sshade.eu/  \citep{schmitt2018}} Motivated by the {\it Herschel} and {\it Planck} space missions, which predominantly observed silicate grain emission, many experimental studies focused on silicate dust analogues in the FIR/mm range, from $\sim 100 \, \mu$m to $1-2$\,mm. Such studies showed that the absorption properties of amorphous silicates are different from those adopted in cosmic dust models in the FIR. First, the FIR spectral behaviour of the absorption cross section is different from a ${\lambda}^{-\beta}$ extrapolation, and second, it varies with the temperature of the grains \citep[e.g.][]{mennella1998, boudet2005, coupeaud2011, demyk2017a, demyk2017b}. However, these results are not yet taken into account in dust models, the main limitation being that often only the absorption coefficients are available and that the optical constants were not calculated. 

In this study, we calculated the optical constants of amorphous silicate dust analogues whose mass absorption coefficients were measured by \cite{demyk2017a} and \cite{demyk2017b} from the MIR to FIR range for various temperatures (10, 30, 100, 200, and 300\,K). The samples are presented in Section~\ref{sect_samples}. Section~\ref{sect_calculs} describes the method used and the assumptions adopted to calculate the optical constants. Sections ~\ref{sect_results} and ~\ref{sect_error} give the resulting optical constants and discuss the errors induced by the adopted assumptions. Sections~\ref{sect_extrapolation} and~\ref{sect_interpol_T} explain how the extrapolation of the optical constants was performed outside the experimental spectral range and how they may be interpolated at other temperatures. Section~\ref{sect_implications} discusses the implications of using these new optical constants.

\section{Studied amorphous silicate dust analogues}
\label{sect_samples}

The samples are the cosmic silicate dust analogues whose mass absorption coefficients (MACs) were measured from 5 to 800\,$mu$m or 1000\,$\mu$m, depending on the sample, and at low temperatures (10, 30, 100, 200, and 300\,K) using the ESPOIRS setup at IRAP \citep{demyk2017a} and on the AILES beam line at the SOLEIL synchrotron  \citep{brubach2010}. The main characteristics of the samples are briefly presented below; full details and the MAC for each sample are given in \cite{demyk2017a, demyk2017b}.

We studied 12 amorphous silicate samples of various compositions; four of them are magnesium-rich silicates and eight are magnesium- and iron-rich silicates (see Table ~\ref{table:extrapol}). The magnesium-rich silicates have a mean stoichiometry close to forsterite, the Mg-rich end-member of the olivine solid solution series (Mg$_{\rm{2}}$SiO$_{\rm{4}}$, sample X35), of enstatite, the Mg-rich end-member of the pyroxene solid solution series (MgSiO$_{\rm{3}}$, samples X50A and X50B), and an intermediate stoichiometry between the two (X40). These are glassy silicates characterised by a relatively compact and regular but amorphous structure at microscopic scales \citep[see][for details]{demyk2017a}. The Mg- and Fe-rich samples all have a mean stoichiometry on the pyroxene solid solutions series. They have four different iron contents, from 10\%\ to 40\% relative to magnesium. The oxidation state of the iron is Fe$^{\rm{3+}}$ for four samples (E10, E20, E30, E40). Iron is partly reduced in the four remaining samples with a mixture of Fe$^{\rm{3+}}$ and Fe$^{\rm{2+}}$ (E10R, E20R, E30R, E40R). These eight samples were produced with solgel methods and are characterised by microstructures that are more porous and irregular than the glassy samples \citep[see][for details of their structure]{demyk2017b, thompson2016}.

The samples are in the form of a powder of undefined shape and a large size distribution. Grain shapes and sizes were investigated using transmission electron microscopy (TEM). A small number of grains were dispersed on a substrate and images were  recorded at different resolutions. Figure~\ref{TEM_shape} shows that, even though the grains are agglomerated, most of them are submicron in size and that their shapes are irregular and characterised by axis ratios of the order of two or less. The grain size distribution was measured for the E30 sample. The particles were maintained in aerosol form  by mechanical agitation under a constant flow of pure nitrogen gas and their sizes measured using an Aerodynamic Particle Sizer (APS 3321 TSI): more details of the procedure can be found in the study by \cite{hubert2017}. Figure~\ref{TEM_shape} shows the measured size distribution of the sample, which can be fitted with a log-normal distribution. 

\begin{table*}[!t]
\caption {Sample composition and parameters for the computation of the optical constants and their extrapolation to short and long wavelengths.} 
\label{table:extrapol}
\begin{center}
\begin{tabular}{c c c c c c c c}
\hline 
\hline 
Sample   & Composition & Density                & n$_{\rm{vis}}$ & Shape &  Axis ratio & $ {\lambda}_{NIR} $ &   ${\lambda}_{FIR}$ \tablefootmark{(4)} \\ 
name      &                   &  (g.cm$^{-3}$)  &                         &              &  ($a/b$)         & ($\mu$m)                &   ($\mu$m)  \\ 
\hline 
X35  \tablefootmark{(1)}   &  (0.65)MgO - (0.35)SiO$_2$         & 2.7 & 1.61 & prolate & 1.5 & 5 & 1000 \\
X40   \tablefootmark{(2)}  &  (0.60)MgO - (0.40)SiO$_2$         & 2.7 & 1.52 & prolate & 1.5 & 5 & 1000/500/500 \\
X50A \tablefootmark{(3)}  &   (0.5)MgO - (0.5)SiO$_2$           & 2.7 & 1.55 & prolate & 1.5 & 5 & 750 \\
X50B  \tablefootmark{(3)} &   (0.5)MgO - (0.5)SiO$_2$           & 2.7 & 1.55 & prolate & 1.5 & 5 & 750/850/1000 \\
E10 & Mg$_{\rm{0.9}}$Fe$_{\rm{0.1}}$SiO$_{\rm{3}}$ ;   Fe$^{\rm{3+}}$  & 2.8  & 1.56 & prolate & 2 & 5 &  650 \\
E20 & Mg$_{\rm{0.8}}$Fe$_{\rm{0.2}}$SiO$_{\rm{3}}$ ;   Fe$^{\rm{3+}}$ &  2.9 & 1.58 & prolate & 2 & 5 &  1000 \\
E30 & Mg$_{\rm{0.7}}$Fe$_{\rm{0.3}}$SiO$_{\rm{3}}$ ;   Fe$^{\rm{3+}}$ &  3.0 & 1.62 & prolate & 2 & 5 &  650 \\
E40 & Mg$_{\rm{0.6}}$Fe$_{\rm{0.4}}$SiO$_{\rm{3}}$ ;   Fe$^{\rm{3+}}$ &  3.1 & 1.65 & prolate & 2 & 5.4  & 750 \\
E10R & Mg$_{\rm{0.9}}$Fe$_{\rm{0.1}}$SiO$_{\rm{3}}$ ;  Fe$^{\rm{2+}}$& 2.8  & 1.56 & prolate & 2 & 5 & 800  \\
E20R & Mg$_{\rm{0.8}}$Fe$_{\rm{0.2}}$SiO$_{\rm{3}}$ ;  Fe$^{\rm{2+}}$& 2.9  & 1.58 & prolate & 2 & 5 & 800  \\
E30R & Mg$_{\rm{0.7}}$Fe$_{\rm{0.3}}$SiO$_{\rm{3}}$ ;  Fe$^{\rm{2+}}$& 3.0  & 1.62 & prolate & 2 & 5 &  900\\
E40R & Mg$_{\rm{0.6}}$Fe$_{\rm{0.4}}$SiO$_{\rm{3}}$ ;  Fe$^{\rm{2+}}$& 3.1  & 1.65 & prolate & 2 & 5 & 850 \\
\hline
\hline 
\end{tabular} 
\end{center}
\tablefoot{
 \tablefoottext{(1) }{Mean composition close to  Mg$_{\rm{2}}$SiO$_{\rm{4}}$}.   \tablefoottext{(2) }{Intermediate composition between MgSiO$_{\rm{3}}$ and Mg$_{\rm{2}}$SiO$_{\rm{4}}$}. \tablefoottext{(3) }{Mean composition of MgSiO$_{\rm{3}}$; the two samples have different structures at the nanometer scale.} \tablefoottext{(4) }{If more than one value is indicated, they correspond to the spectra measured at 300 K, 100 K, and 30-10 K respectively, otherwise the same value for ${\lambda}_{FIR}$  is adopted for all the temperatures.}}
\end{table*}

\begin{figure*}[!h]
\centering
\includegraphics[width=0.3\textwidth]{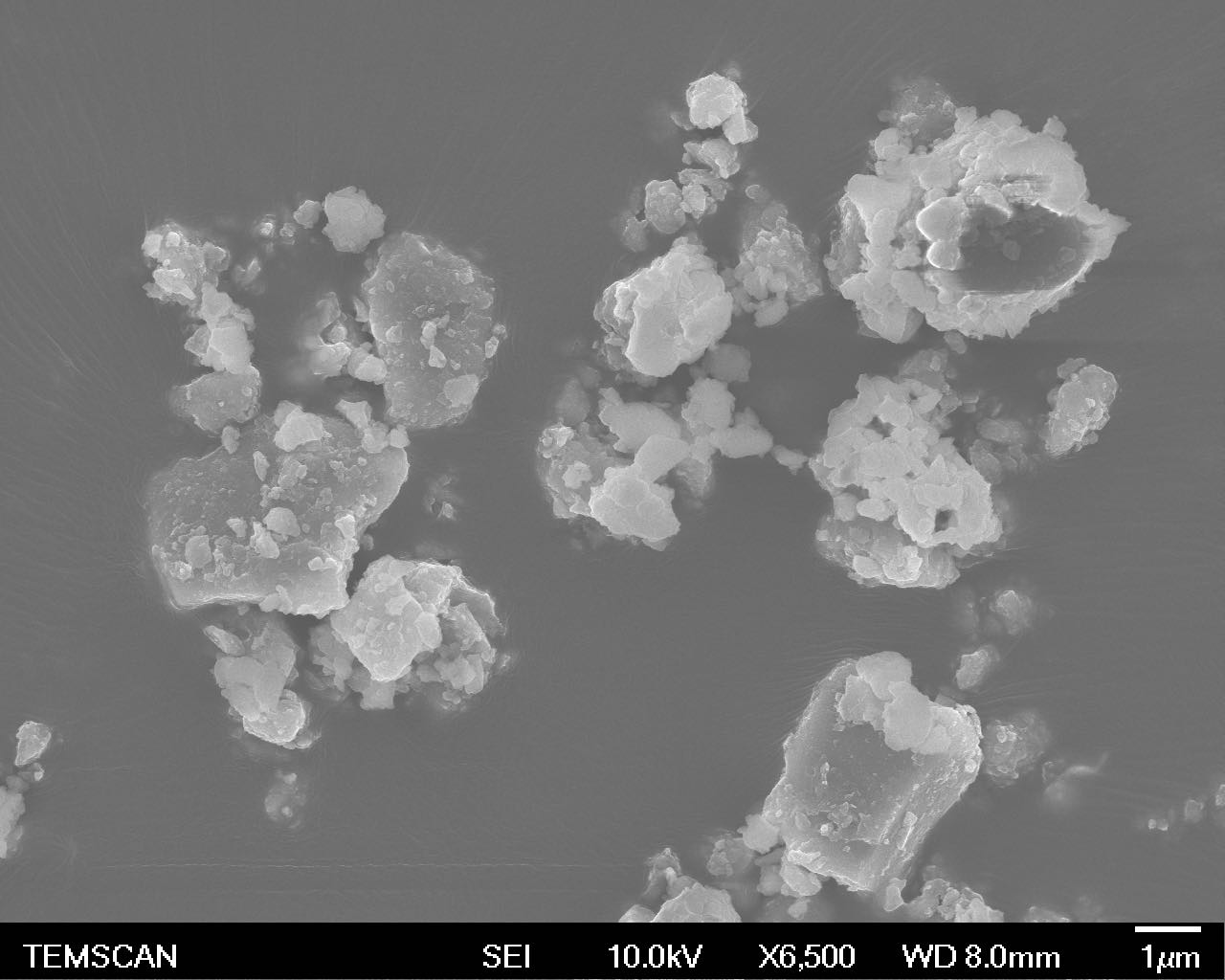}
\includegraphics[width=0.31\textwidth]{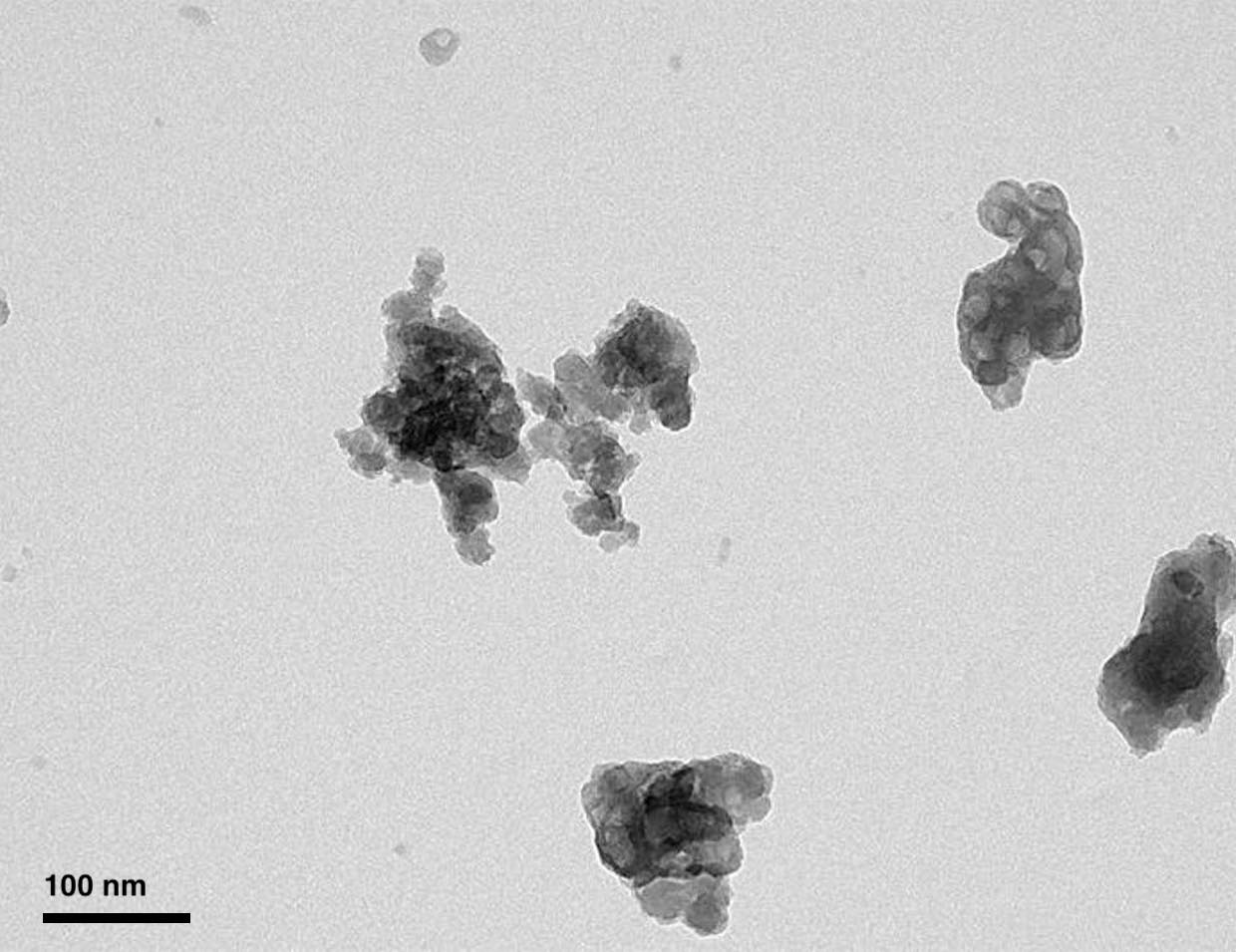}
\includegraphics[trim={1cm 0.5cm 0.5cm 1.5cm}, width=0.35\textwidth]{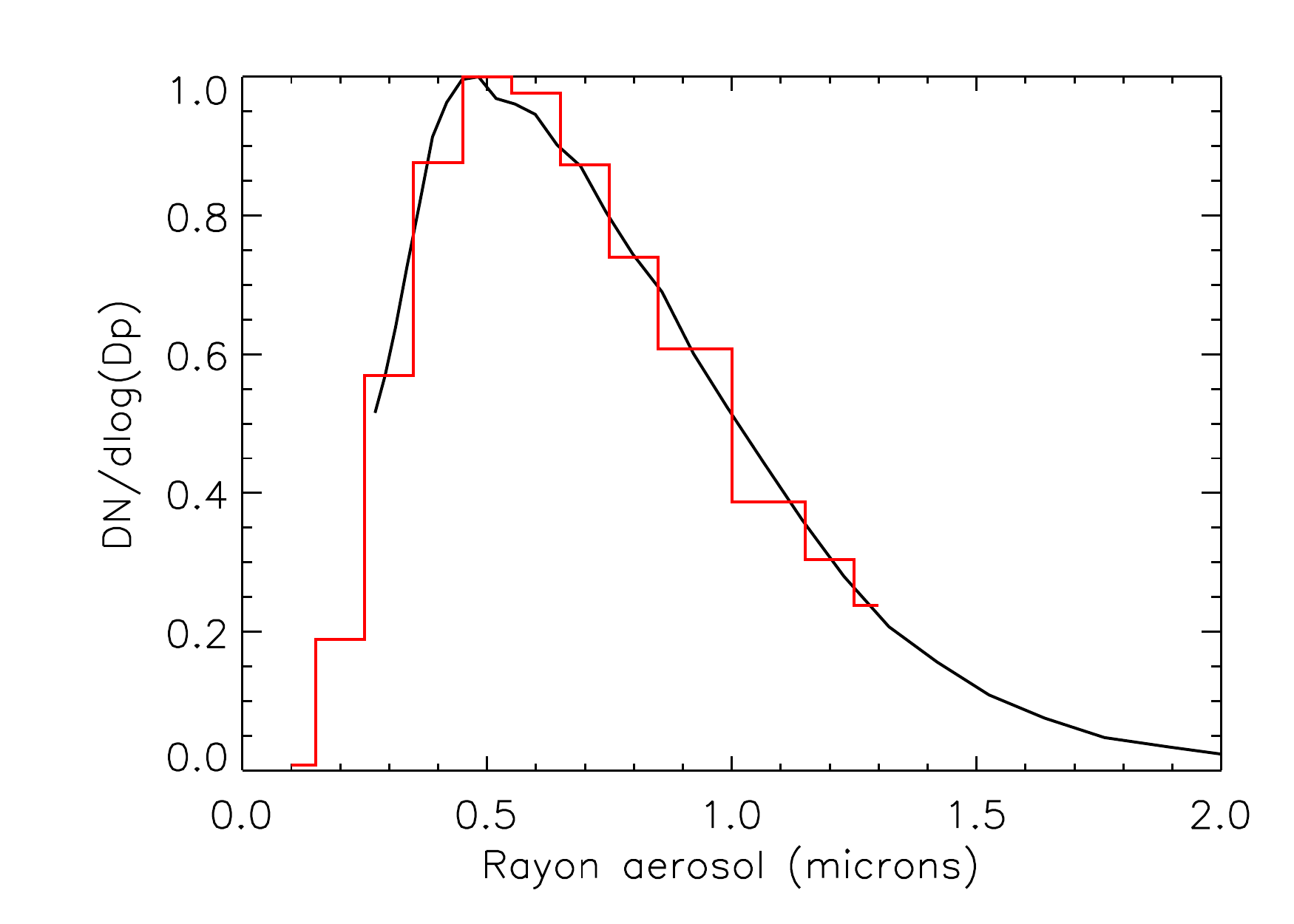}
\caption{Sample sizes and shapes. {\it Left panel}: Image of the glassy Mg-rich sample taken with a scanning electron microscope, the scale bar in the lower right is $1\,\mu$m. {\it Middle panel}:  Image of the solgel Fe-rich sample E20 taken with a transmission electron microscope, the scale bar is 100\,nm.  {\it Right panel}: Black line shows the normalised size distribution measured for sample E30, Mg$_{0.7}$Fe$_{0.3}$SiO$_3$.  The red histogram-like curve shows the size distribution adopted in order to study the impact of the grain size distribution on the optical constant calculation (See Sect.~\ref{sect_error}).}
    \label{TEM_shape}% label for figure
\end{figure*}

\section{Calculation of the optical constants}
\label{sect_calculs}

Here we derive the optical constants ($m = n + ik$) of our cosmic dust analogues from their measured mass absorption coefficients (MACs, designated ${\kappa}_{\rm{exp}}$, cm$^2$\,g$^{-1}$). The MAC is related to the extinction, absorption, and scattering cross sections ($C_{\rm{ext}}$,  $C_{\rm{abs}}$, $C_{\rm{sca}}$, cm$^2$)  and efficiencies ($Q_{\rm{ext}}$,  $Q_{\rm{abs}}$, $Q_{\rm{sca}}$) by the relations: 
\begin{equation}
\label{kappa}
{\kappa}_{\rm{exp}} = {\kappa}_{\rm{abs}} + {\kappa}_{\rm{sca}} = \frac{C_{\rm{abs}} + C_{\rm{sca}}}{\upsilon \rho}  = \frac{G}{{\upsilon \rho}}\left({Q}_{\rm{abs}} + {Q}_{\rm{sca}}\right), 
\end{equation}
where $\upsilon$ is the grain volume (cm$^3$), $G$ its geometrical cross section (cm$^2$), and $\rho$ the density of the grain material (g cm$^{-3}$). The efficiency factors $Q_{\rm{abs}}$ and $Q_{\rm{sca}}$ as a function of the material optical constants depend on the shape, structure, and size distribution of the sample particles. In this study, we are dealing with a collection of grains of undefined shape, size, and shape distribution. We therefore need to make some assumptions about these parameters.

For ellipsoidal grains that are small compared to the wavelength, the absorption cross section for a collection of identical, non-spherical, randomly oriented particles is given by \citep{bohren1998}:
\begin{equation}
        \label{kappa_abs_ellipsoid}     
 <{\kappa}_{\rm{abs}}>  = \frac{2\pi N}{\upsilon \rho \lambda}  \mathrm{Im}\left( \frac{{\alpha}_1 + {\alpha}_2 + {\alpha}_3}{3} \right), 
\end{equation}
where $N$ is the refractive index of the surrounding medium and ${\alpha}_i$ is the polarisability along each axis of the ellipsoid: 
\begin{equation}
        \label{alpha_ellipsoid} 
{ \alpha}_{\rm{i}}  = \upsilon \frac{\epsilon - {\epsilon}_{\rm{m}}}{ {\epsilon}_{\rm{m}} + L_{\rm{i}} \left(\epsilon - {\epsilon}_{\rm{m}} \right) }, 
 \end{equation}
where $\epsilon = \epsilon' + \epsilon'' = m^2 $ is the dielectric function of the grain material and ${\epsilon}_{\rm{m}}$ that of the surrounding medium (${\epsilon}_{\rm{m}} = N^2$). The geometrical factors, $L_i$, can take any value between 0 and 1 with $L_1 + L_2 + L_3= 1$ and for a spherical grain $L_1 = L_2 = L_3= 1/3$. For spheroids, a special case of ellipsoids, where two of the three rotational axes ($a$, $b$, $c$) are equal, two of the geometrical factors are equal and adopting $b = c$ we have $L_2 = L_3 = (1 - L_1)/2$. For oblate grains of semi axis $ a < b = c $, the geometrical factor $L_1$ is expressed by \citep{vandehulst1957}: 
\begin{equation}
\begin{aligned}
        \label{L_oblate}        
 L_{\rm{1}}  = \frac{1 + f^2}{f^2} \left(1 - \frac{1}{f} \rm{arctan}(f) \right)  \; \rm{where} \; f^2 =  \frac{b^2}{a^2} -1   \\
  \end{aligned}
,\end{equation}
and for prolate grains of semi axis $ a > b = c $, the geometrical factor $L_1$ is: 
\begin{equation}
        \label{L_prolate}       
 L_{\rm{1}}  = \frac{1 - e^2}{e^2} \left( -1 + \frac{1}{2e}\rm{ln}\left( \frac{1+e}{1-e}\right)  \right)  \; \rm{where} \; e^2 = 1 - \frac{b^2}{a^2}.   \\
 \end{equation}
Setting $L = L_1$, the absorption mass coefficient of identical spheroidal grains, which are small compared to the wavelength and randomly oriented, is given by: 
\begin{equation}
\label{eq:kappa_abs}
<{\kappa}_{\rm{abs}}>  = \frac{2\pi N}{3 \rho \lambda}  \mathrm{Im} \left( \frac{\epsilon - \epsilon_m}{\epsilon_m + L(\epsilon -\epsilon_m)} + \frac{4(\epsilon -\epsilon_m)}{2 \epsilon_m + (1-L)(\epsilon -\epsilon_m)} \right).  
\end{equation}

The optical constants $n$ and $k$ as well as the real and imaginary parts of the dielectric function $\epsilon$ are linked by the Kramers-Kronig relations \citep{bohren1998} which allow us to derive $n$ (or $\epsilon'$), knowing $k$ (or $\epsilon''$). Defining the functions $\zeta$ and $\xi$ such that: 
\begin{equation}
\label{eq:xi_exp}
\xi = \frac{ \lambda \rho}{2 \pi } \kappa_{\rm{exp}} = \mathrm{Im}(\zeta)  = \zeta''. 
\end{equation}
From Eq.~(\ref{eq:kappa_abs}), for a collection of identical spheroidal particles that are randomly oriented, we have: 
\begin{equation}
\label{eq:zeta}
\zeta  = \frac{N}{3 }  \left( \frac{\epsilon - \epsilon_m}{\epsilon_m + L(\epsilon -\epsilon_m)} + \frac{4(\epsilon -\epsilon_m)}{2 \epsilon_m + (1-L)(\epsilon -\epsilon_m)} \right).  
\end{equation}
The function $\zeta$ must satisfy the Kramers-Kronig relations and can be expressed as: 
\begin{equation}
 \label{eq:zeta_xi}
\zeta =  \zeta' + \rm{i} \zeta''  = \zeta_{\infty} + \mathrm{KK}(\xi) + \rm{i} \xi, 
\end{equation}
where $\zeta_{\infty}$ is real and calculated from Eq.~(\ref{eq:zeta}) adopting $\epsilon = \epsilon_{\infty} = n_{\rm{vis}}^2$, ($n_{\rm{vis}}$ is the value of the refractive index of the material in the visible domain at $\sim$ 0.5-0.7 $\mu$m) and $ \mathrm{KK}(\xi)$ is calculated from the Kramers-Kronig integral as follows ($P$ is the Cauchy principal value of the integral): 
 \begin{equation}
 \label{eq:kk}
 \mathrm{KK}(\xi) =  \frac{2}{\pi} P \int_{\rm{vis}}^{\infty} \frac{\lambda^2}{\Lambda} \frac{\xi''(\Lambda)}{\lambda^2 - \Lambda^2} d\Lambda.
\end{equation}
 
To derive the optical constants from the experimental data, we proceeded as follows. We first calculated $\xi$ from the experimental data using Eq.~(\ref{eq:xi_exp}). We then used $\xi$ to calculate $\zeta'$ from Eqs.~(\ref{eq:zeta_xi}) and (\ref{eq:kk}). We used the publicly available code kktrans\footnote{https://hera.ph1.uni-koeln.de/$\sim$ossk/Jena/pubcodes.html} from V. Ossenkopf to calculate the integral. Knowing the real and imaginary parts of $\zeta$ from the experimental data, we then used Eq.~(\ref{eq:zeta}) to calculate the dielectric function $\epsilon$. To do this, we expressed Eq.~(\ref{eq:zeta}) as a second-order polynomial in $\epsilon$: 
\begin{equation}
\label{eq:pol}
\begin{aligned}
& A {\epsilon}^2 + B \epsilon + C = 0 \\
& \rm{where:} \\
&A = \zeta L (1-L) - \frac{\sqrt{{\epsilon}_{\rm{m}}}}{3}  (1 + 3L)\\
&B = \left( \zeta (1 + L - 2L(1-L) ) - \frac{\sqrt{{\epsilon}_{\rm{m}}}}{3}  (4 - 6L) \right){\epsilon}_{\rm{m}} \\
&C= \left( \zeta (1-L) (1+L) + \frac{\sqrt{{\epsilon}_{\rm{m}}}}{3}  (5 - 3L) \right) {\epsilon}_{\rm{m}}^2. \\
\end{aligned}
\end{equation}
Equation~(\ref{eq:pol}) can then be solved analytically with $\epsilon$ given by:
\begin{equation}
\begin{aligned}
&\epsilon' = \mathrm{Re}{ \left( \frac{-B - \sqrt{B^2 - 4AC}}{2A} \right) } \\
&\epsilon'' = \mathrm{Im}{ \left(\frac{-B - \sqrt{B^2 - 4AC}}{2A} \right) }. \\
\end{aligned}
\end{equation}

The optical constants $n$ and $k$ were calculated from $\epsilon$ via the relations: 
\begin{equation}
\begin{aligned}
& \epsilon = m^2\\
&n =  \sqrt{ \frac{ \sqrt{{\epsilon'}^2 + {\epsilon''}^2} + \epsilon'}{2} } \\
&k =  \sqrt{ \frac{ \sqrt{{\epsilon'}^2 + {\epsilon''}^2} - \epsilon'}{2} }. \\
\end{aligned}
\end{equation}

\section{Optical constants in the MIR/FIR domain at selected temperatures}
\label{sect_results}

The optical constants of the 12 samples studied by \cite{demyk2017a,demyk2017b} were calculated using the method described above for each temperature measurement (10, 100, 200, and 300\,K) and over the spectral range from 5\,$\mu$m to 800\,$\mu$m or $1000 \,\mu$m, depending on the sample, where the longest wavelength depends on the sample. The optical constants were also calculated at 30\,K for the eight Fe-rich samples, which are close to those at 10 K and therefore not shown. The MAC of the Mg-rich samples at 30\,K are identical to those at 10\,K and we therefore did not calculate the optical constants at 30\,K for these samples. The optical constants are calculated from the MACs obtained from transmission measurements for grains embedded in a polyethylene (PE) matrix and we therefore had to make assumptions about the grain size and shape distributions and the refractive indices in the visible $n_{\rm{vis}}$. 

To calculate the optical constants, we set the refractive index of the medium, $N$,  to be 1.52, the value for polyethylene \citep{mennella1998}. The refractive indices of our samples in the visible, $n_{\rm{vis}}$, which depend upon the sample composition, are unknown. In order to estimate their values, we use the HJPDOC database,\footnote{Heidelberg - Jena - St.Petersburg - Database of Optical Constants, http://www.mpia.de/HJPDOC/} which gathers optical constant studies for a range of materials, including amorphous silicates. We find that, in the spectral range from 0.5 to 1\,$\mu$m,  the refractive indices of samples similar to those studied here are in the range $1.5 - 1.7$. The values vary slightly with the composition of the samples: for glassy, pyroxene-like silicates, $n_{\rm{vis}} \simeq 1.55$ for MgSiO$_3$, $\leqslant 1.68$ for a sample containing 60\% iron \citep{jaeger1994,dorschner1995}, while solgel Mg-rich silicates show values from 1.55 (MgSiO$_3$) up to 1.64 (Mg$_{2.4}$SiO$_{4.4}$) \citep{jaeger2003}. Glassy and solgel MgSiO$_3$ have the same value for  $n_{\rm{vis}}$ of 1.55 and we therefore assume that it does not depend on the structure of the material. We chose $n_{\rm{vis}}$ values closest to those of the equivalent database samples. The values for $n_{\rm{vis}}$ adopted for the different samples are indicated in Table~\ref{table:extrapol}.

The density of the grain material is needed in order to compute the optical constants. In the absence of information on the density of the samples, we adopted the values for similar materials. For the glassy samples, we set the density to 2.7 g\,cm$^{-3}$, corresponding to a glassy MgSiO$_3$ \citep{boudet2005}. For the solgel Fe-rich samples, we set the density to values in the range $2.8 - 3.1$\,g\,cm$^{-3}$, which were interpolated from the values measured by \cite{jaeger1994} and \cite{dorschner1995} for solgel samples containing 5\%, 30\%,\ and 50\% iron (see Table~\ref{table:extrapol}).

\begin{figure*}[!h]
\centering
  \includegraphics[scale=.3, trim={0 1cm 0 1.5cm}, clip]{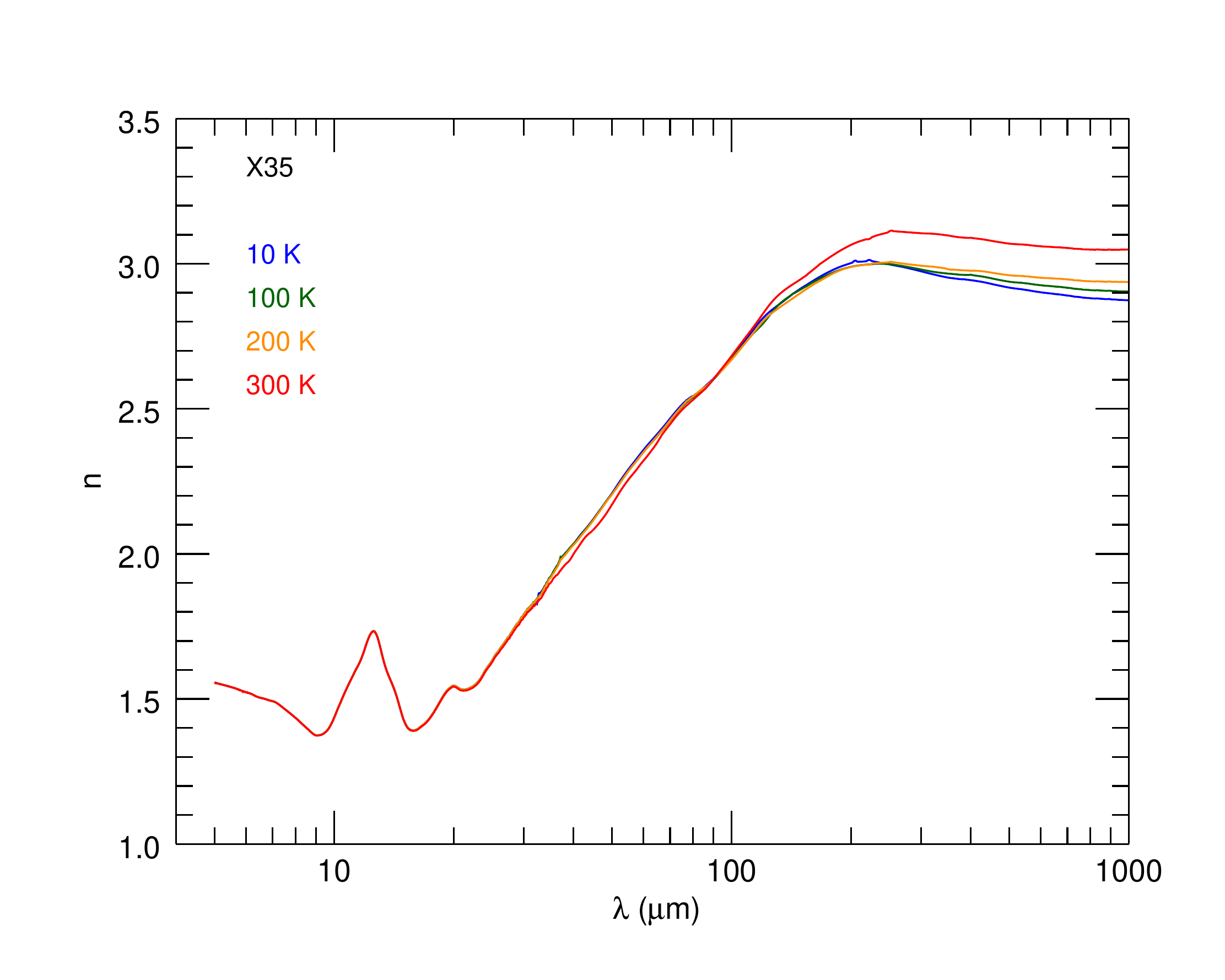}
  \includegraphics[scale=.3, trim={0 1cm 0 1.5cm}, clip]{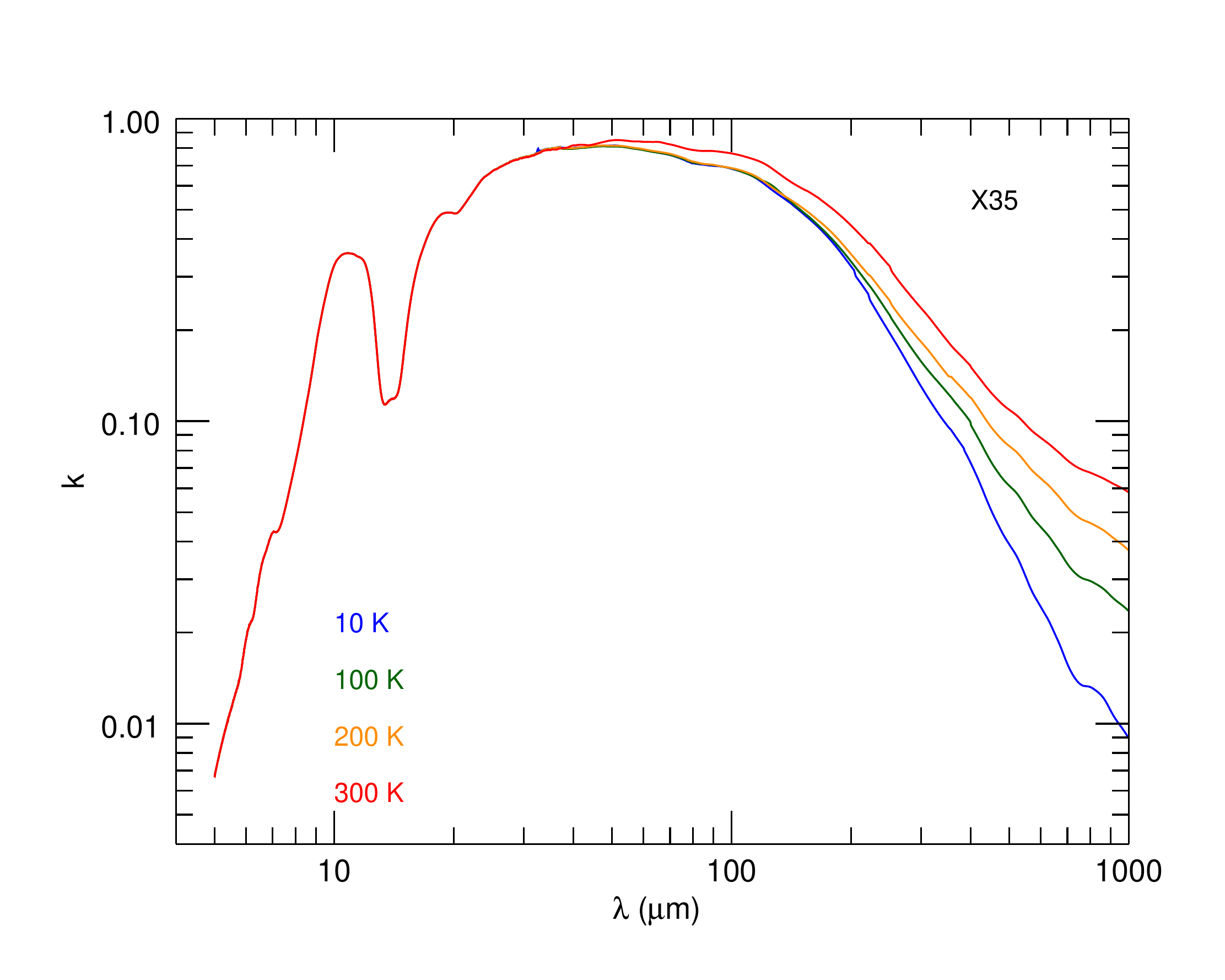}
  \includegraphics[scale=.3, trim={0 1cm 0 1.5cm}, clip]{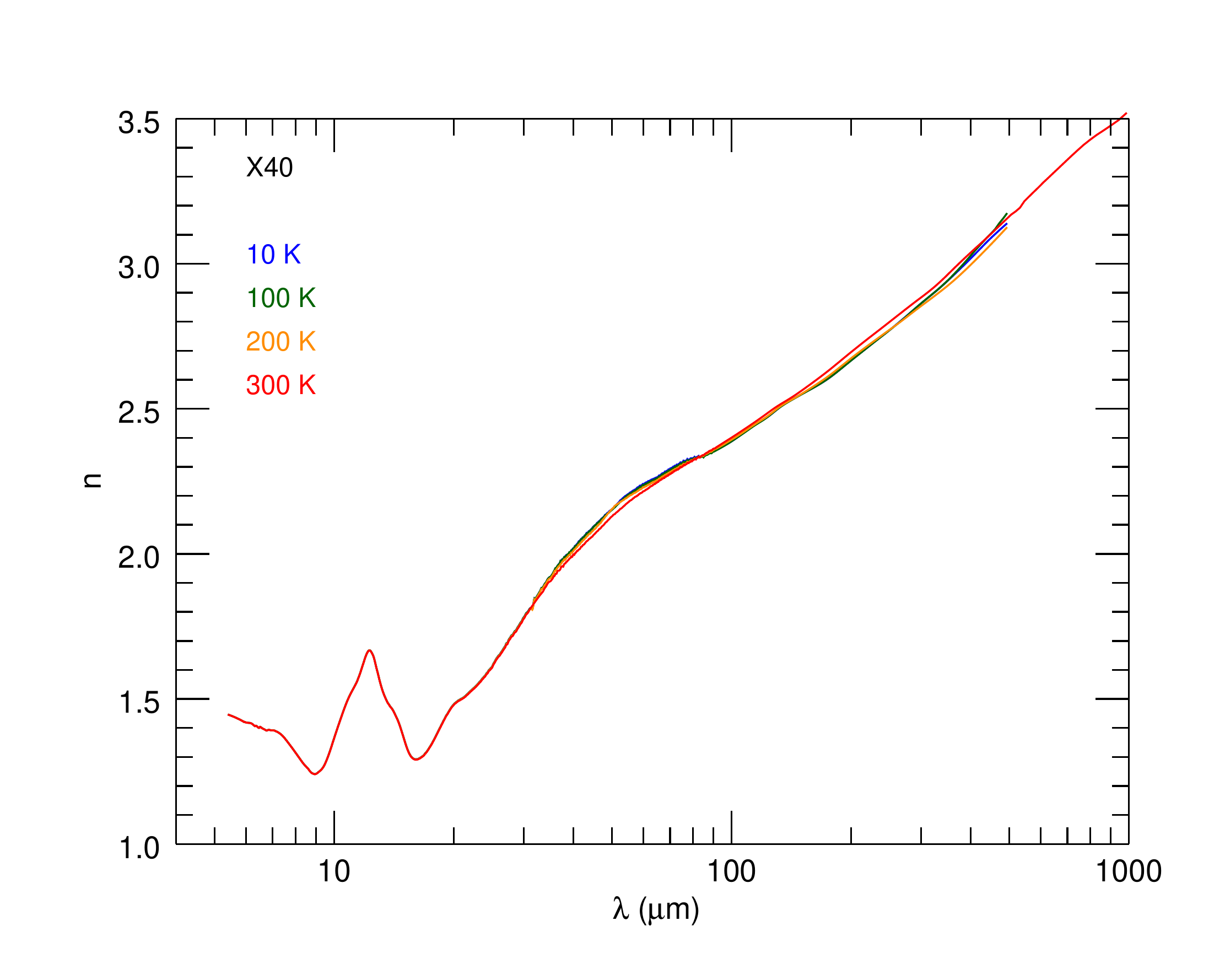}
  \includegraphics[scale=.3, trim={0 1cm 0 1.5cm}, clip]{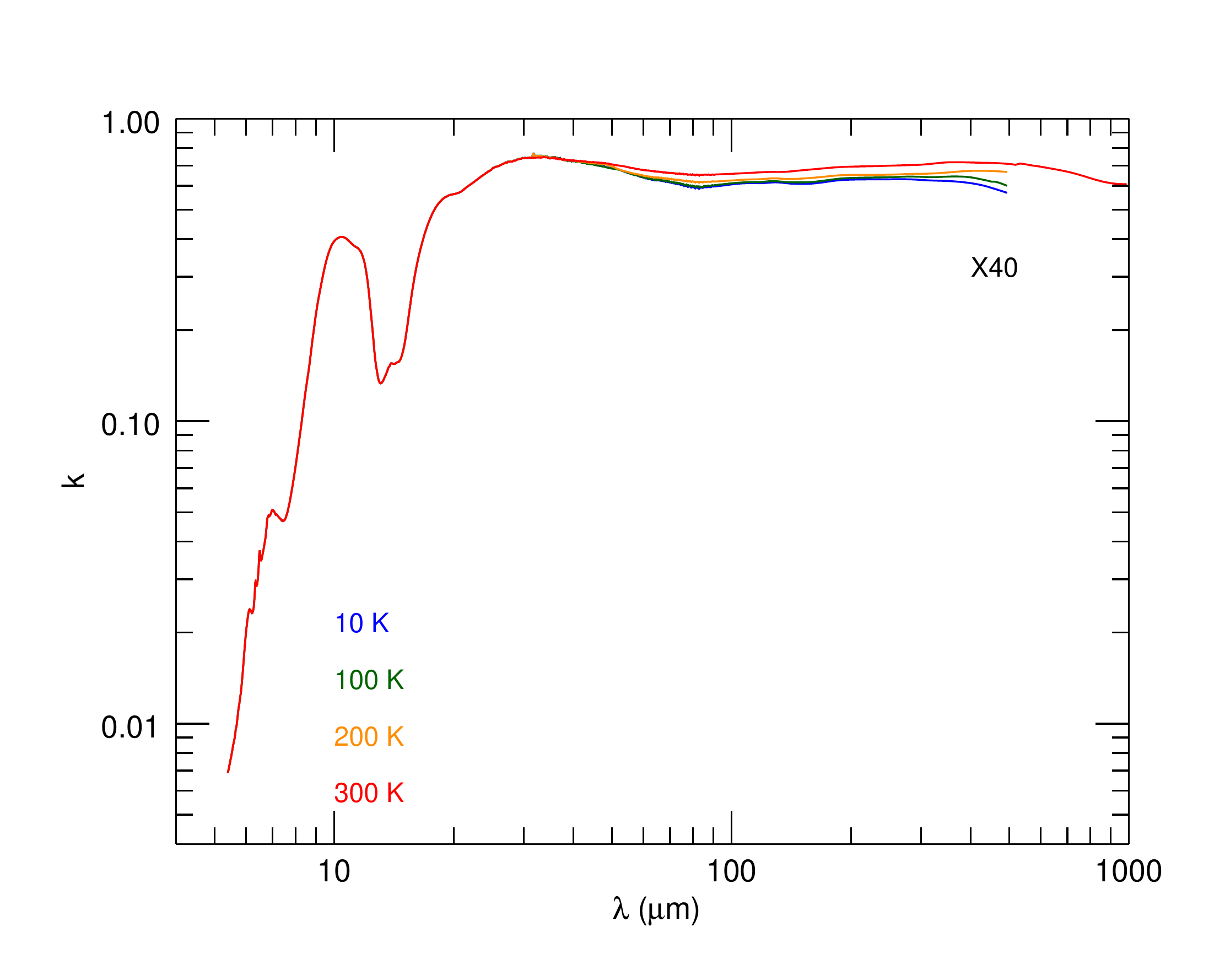}
  \includegraphics[scale=.3, trim={0 1cm 0 1.5cm}, clip]{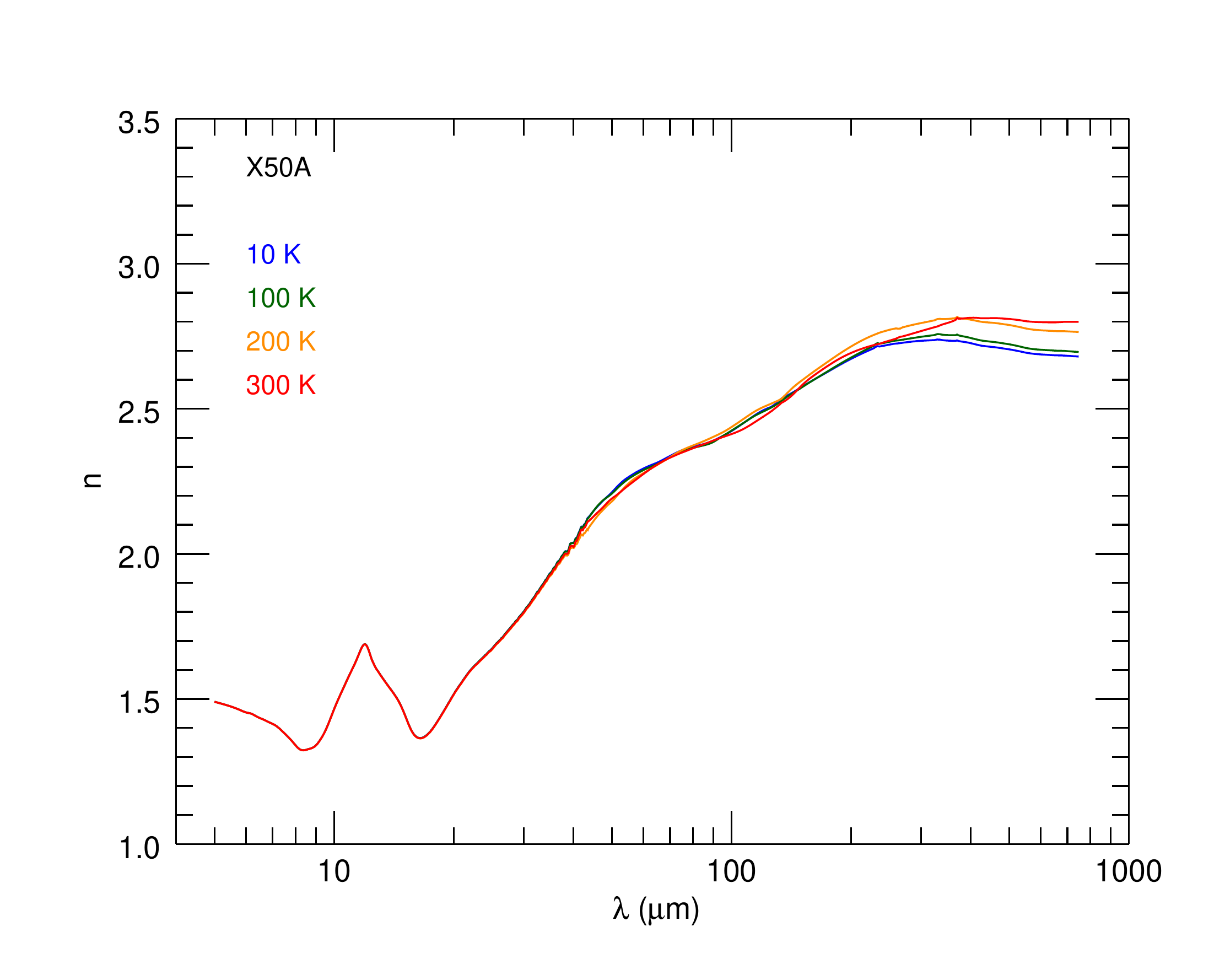}
  \includegraphics[scale=.3, trim={0 1cm 0 1.5cm}, clip]{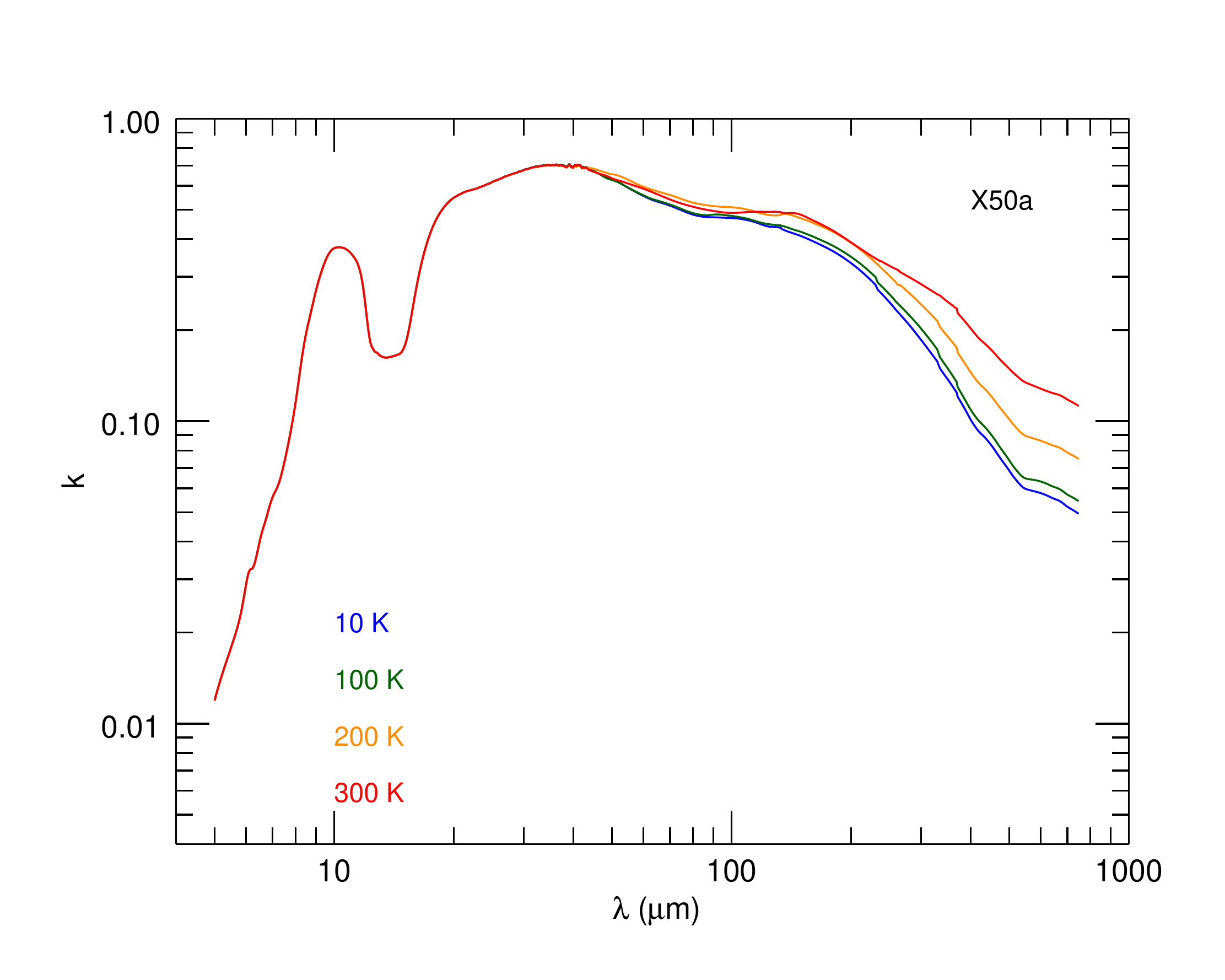}
  \includegraphics[scale=.3, trim={0 1cm 0 1.5cm}, clip]{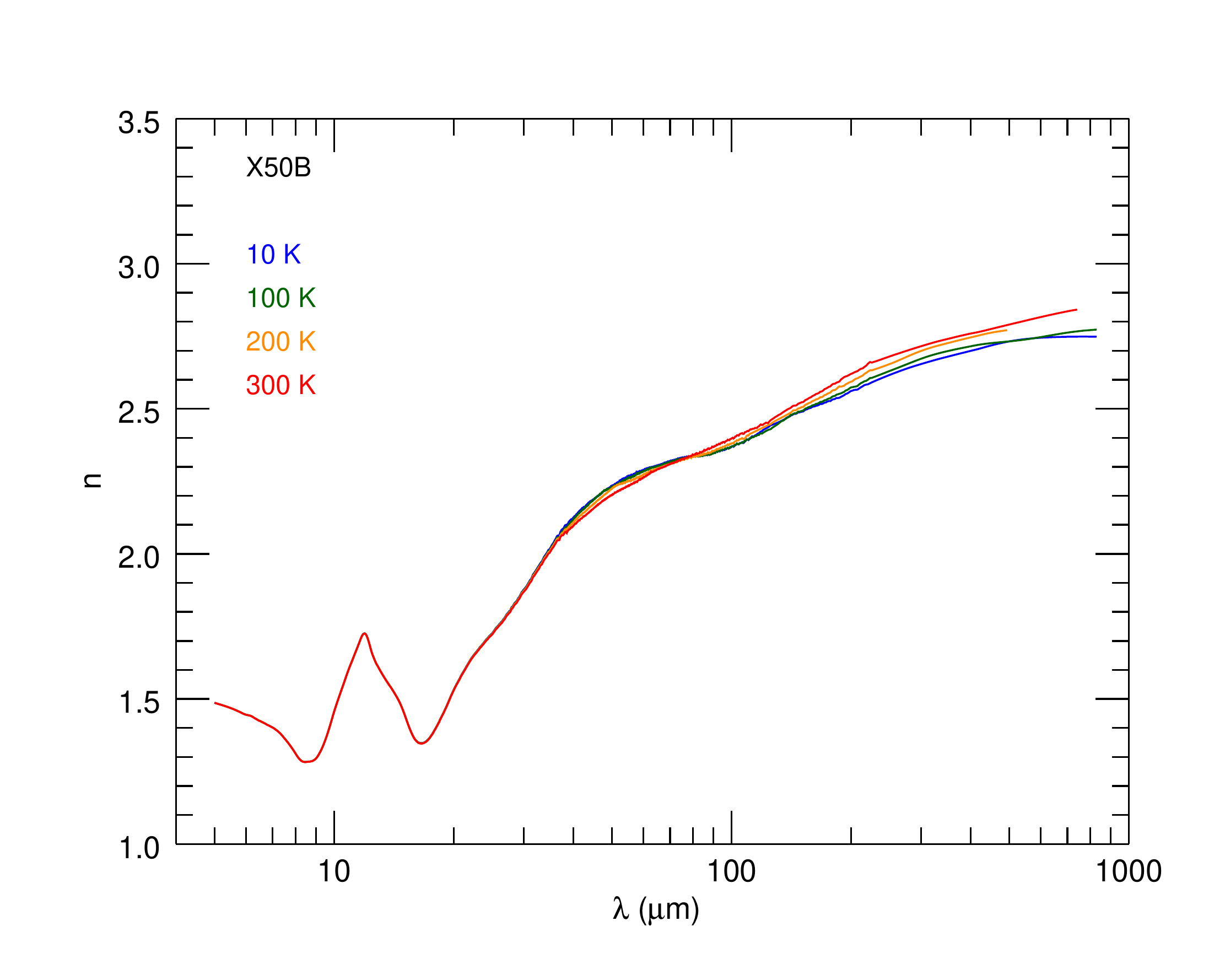}
  \includegraphics[scale=.3, trim={0 1cm 0 1.5cm}, clip]{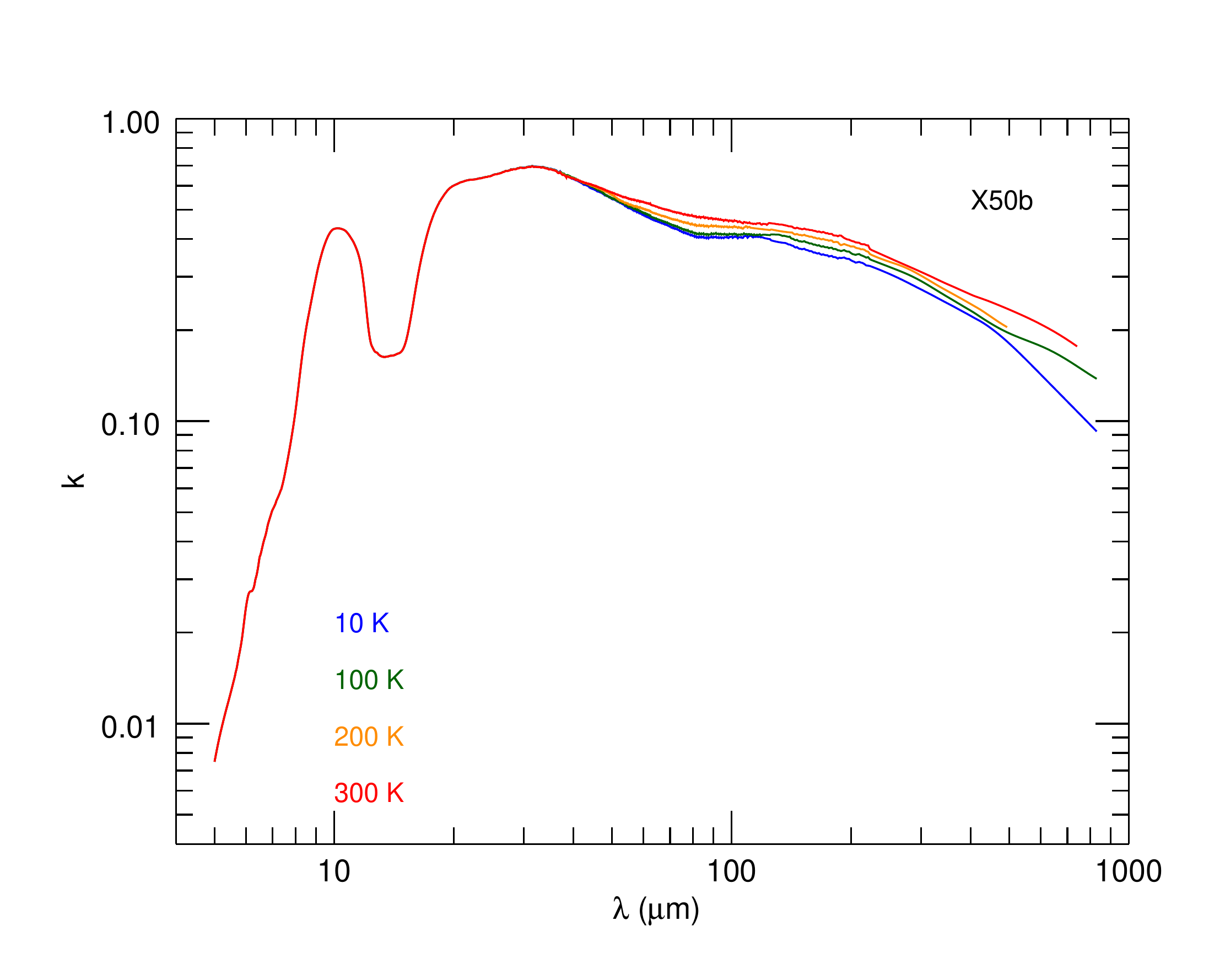}
          \caption{Optical constants for the Mg-rich silicate samples at temperatures of 10 K (blue), 100 K (green), 200K (orange), and 300 K (red).  }
    \label{netk_mgrich}% label for figure
\end{figure*}

\begin{figure*}[!h]
\centering
  \includegraphics[scale=.3, trim={0 1cm 0 1.5cm}, clip]{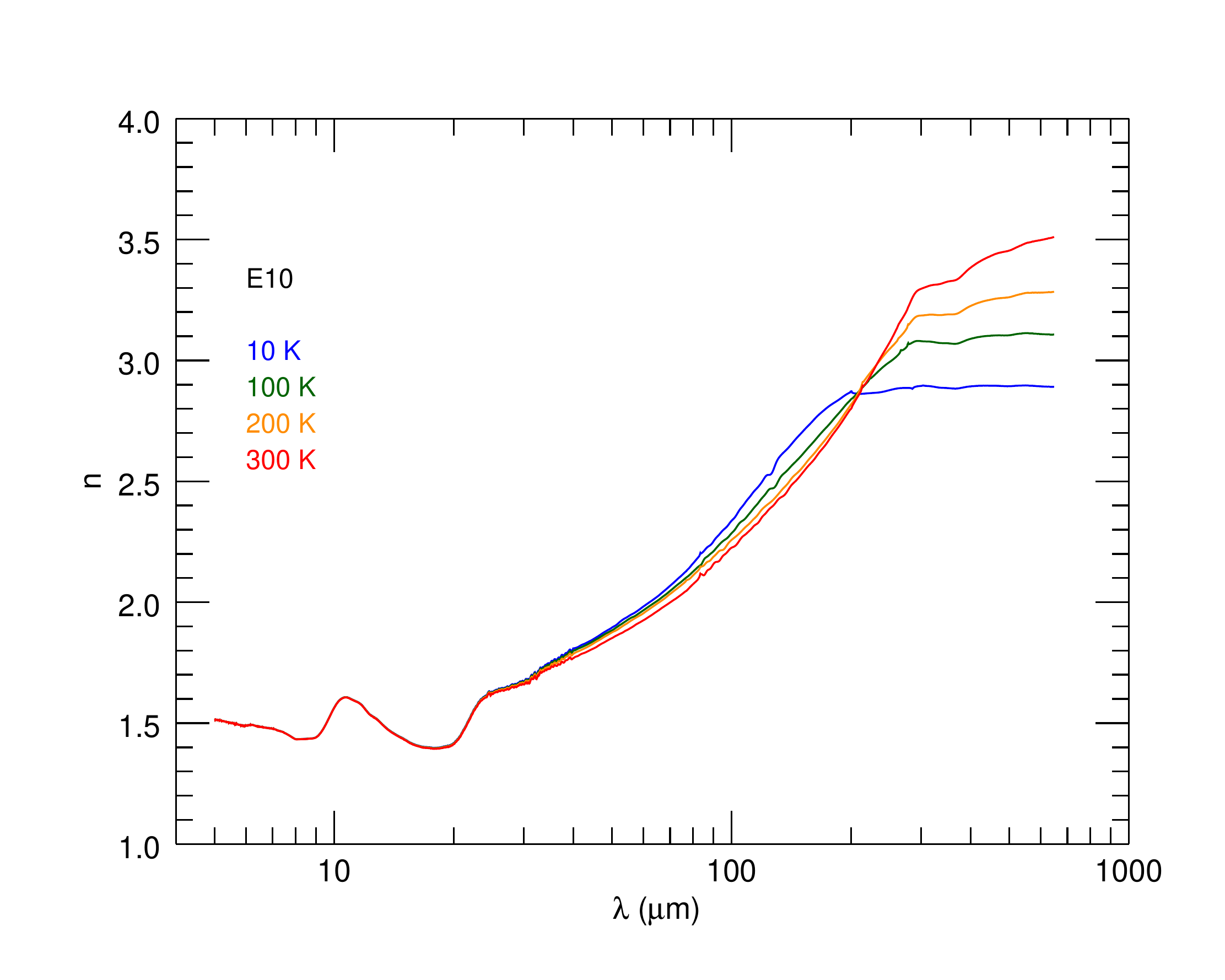}
  \includegraphics[scale=.3, trim={0 1cm 0 1.5cm}, clip]{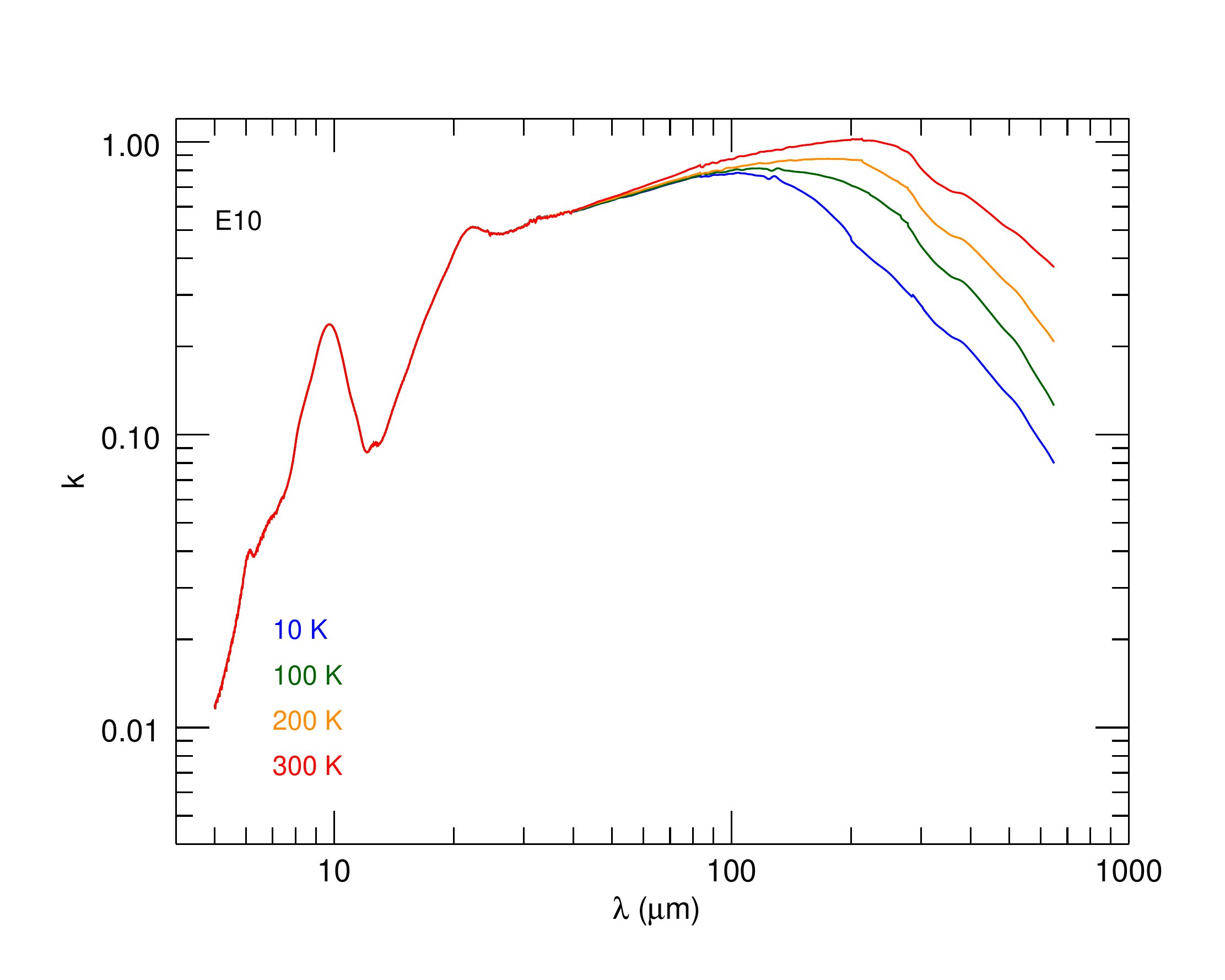}
  \includegraphics[scale=.3, trim={0 1cm 0 1.5cm}, clip]{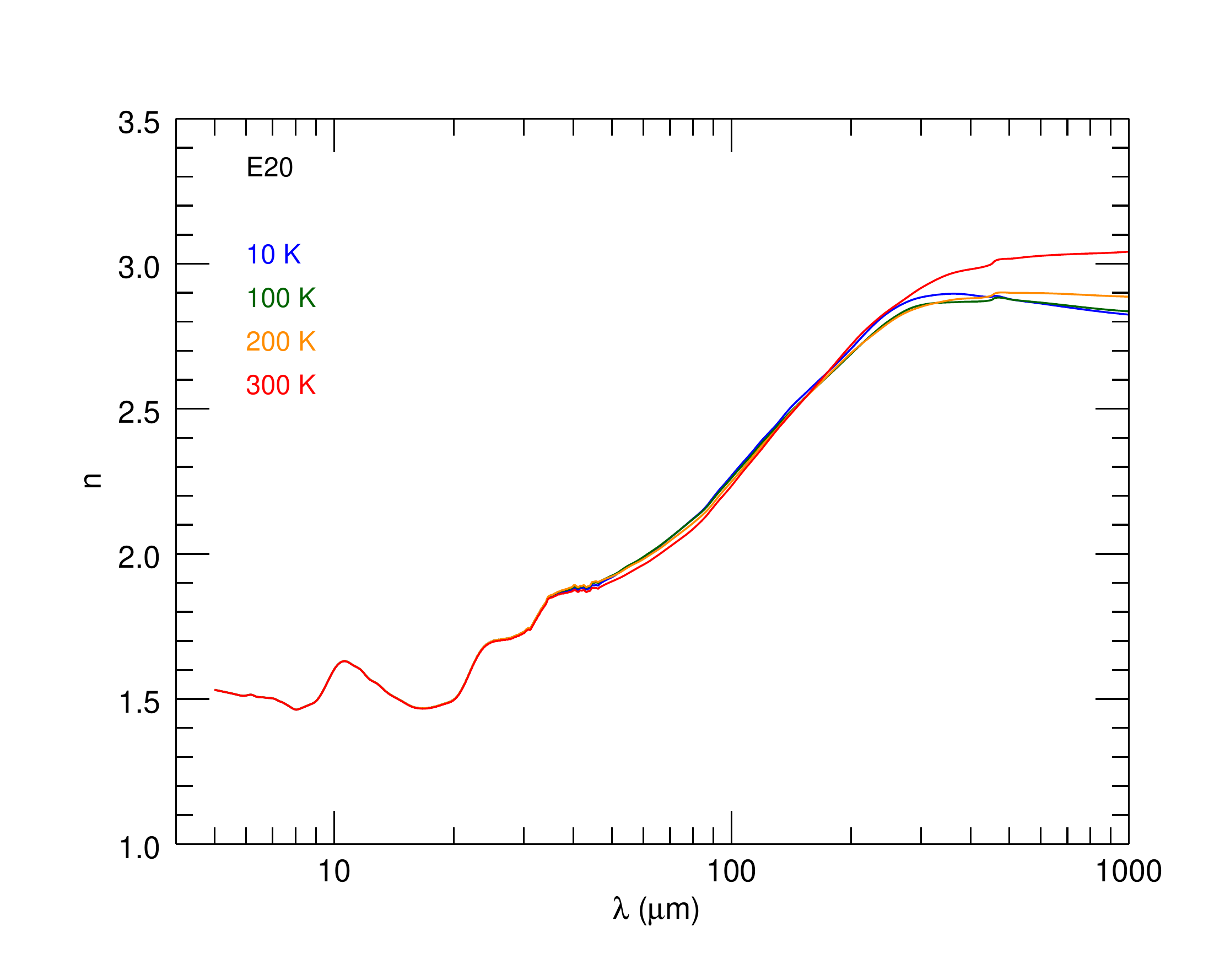}
  \includegraphics[scale=.3, trim={0 1cm 0 1.5cm}, clip]{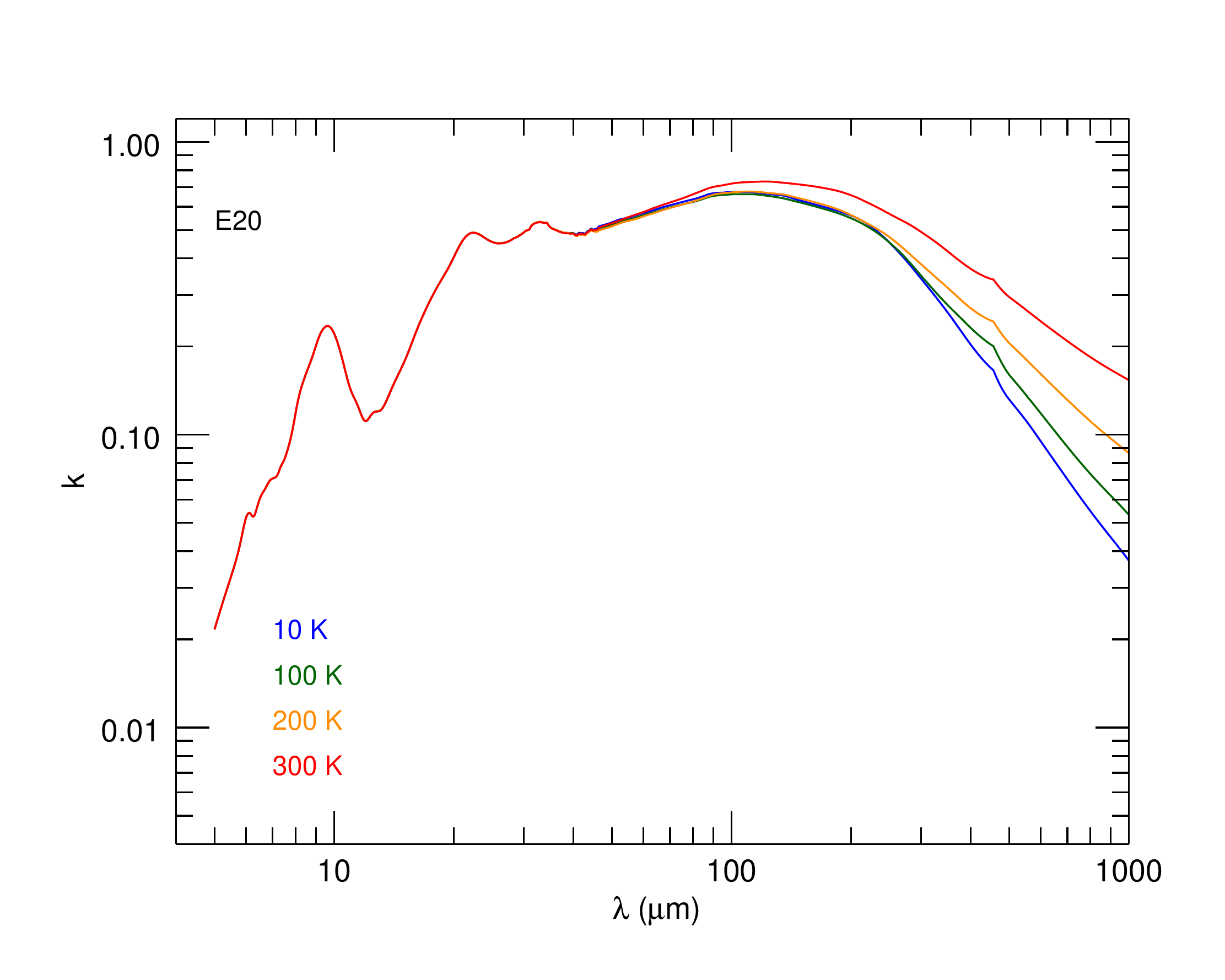}
  \includegraphics[scale=.3, trim={0 1cm 0 1.5cm}, clip]{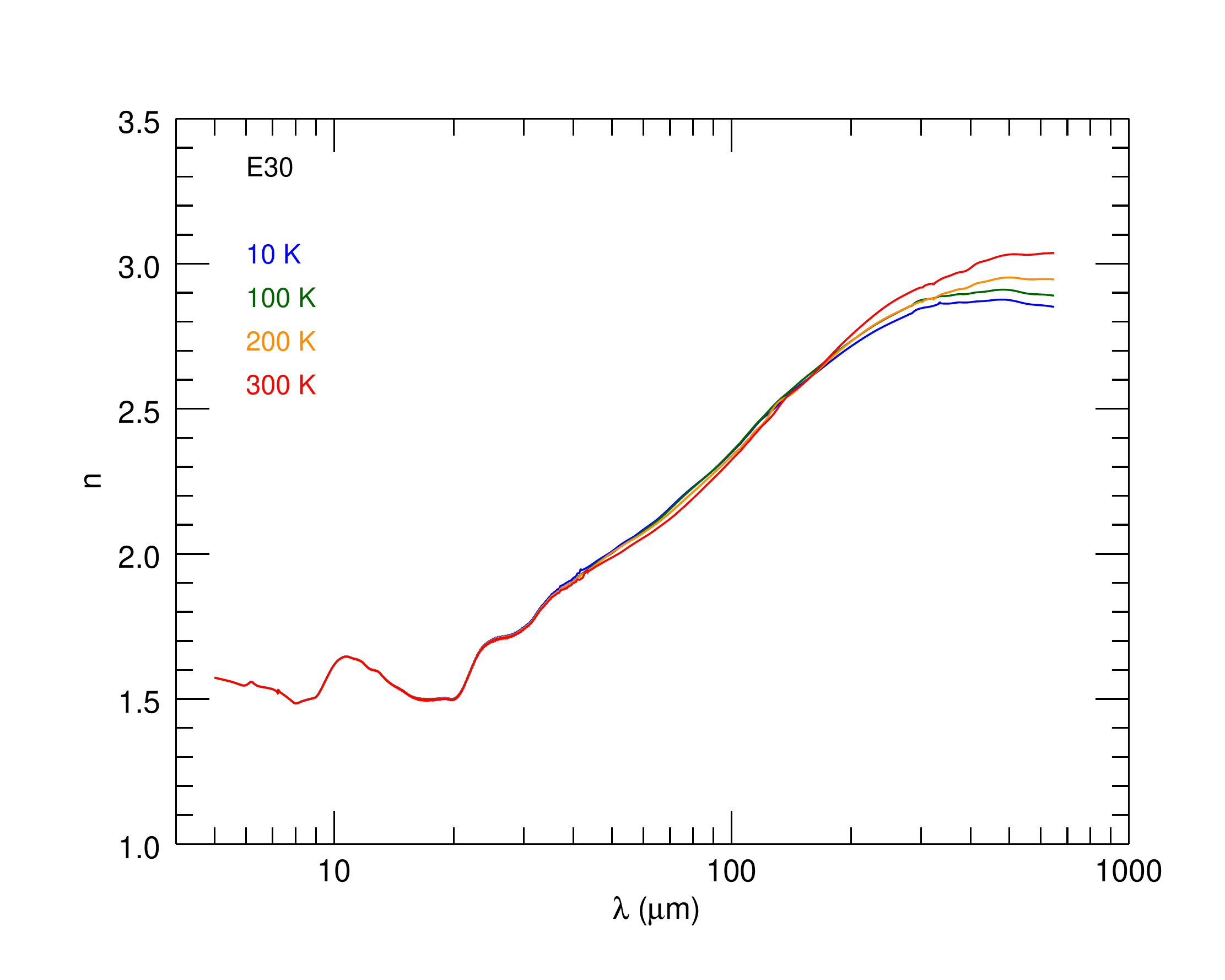}
  \includegraphics[scale=.3, trim={0 1cm 0 1.5cm}, clip]{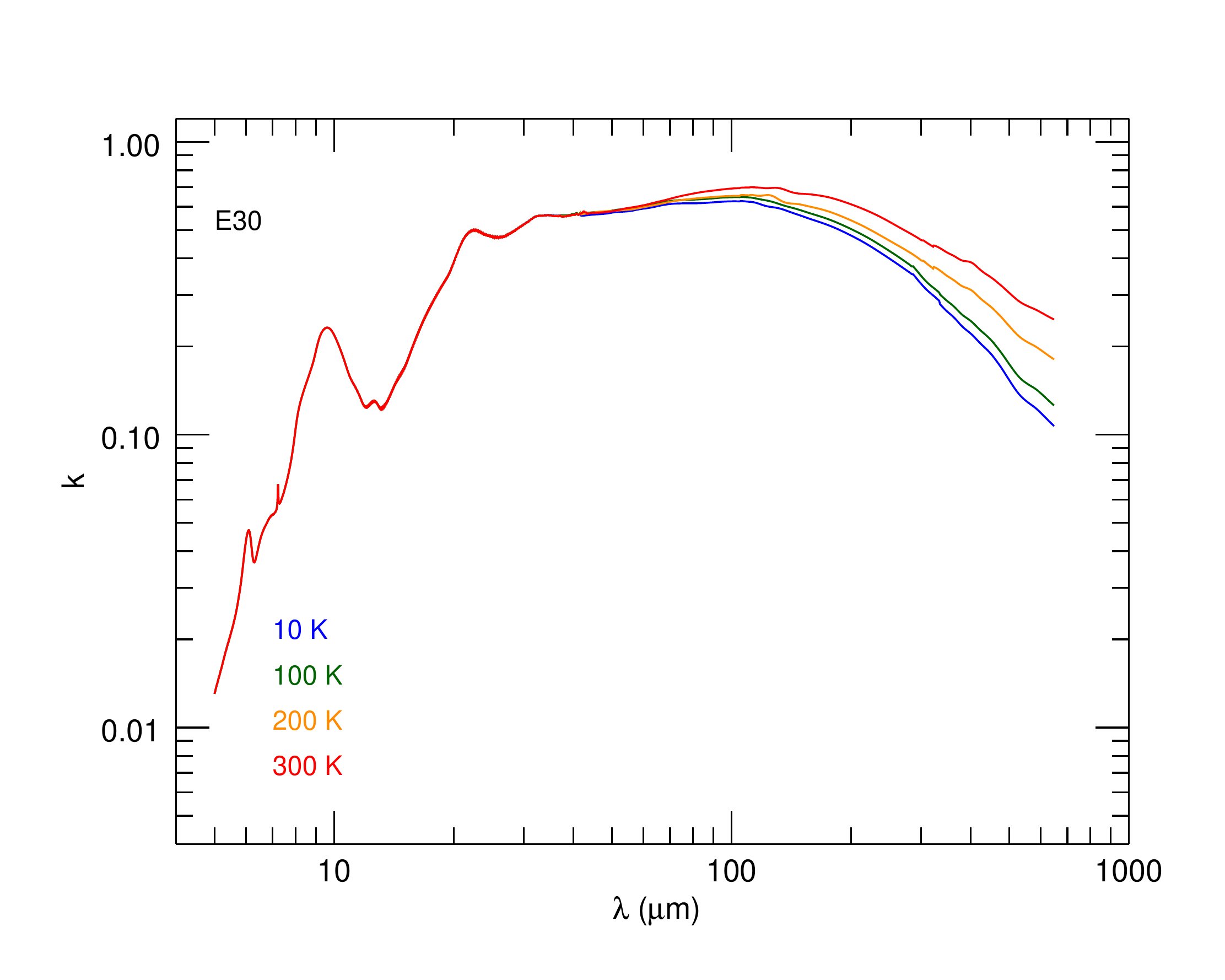}
  \includegraphics[scale=.3, trim={0 1cm 0 1.5cm}, clip]{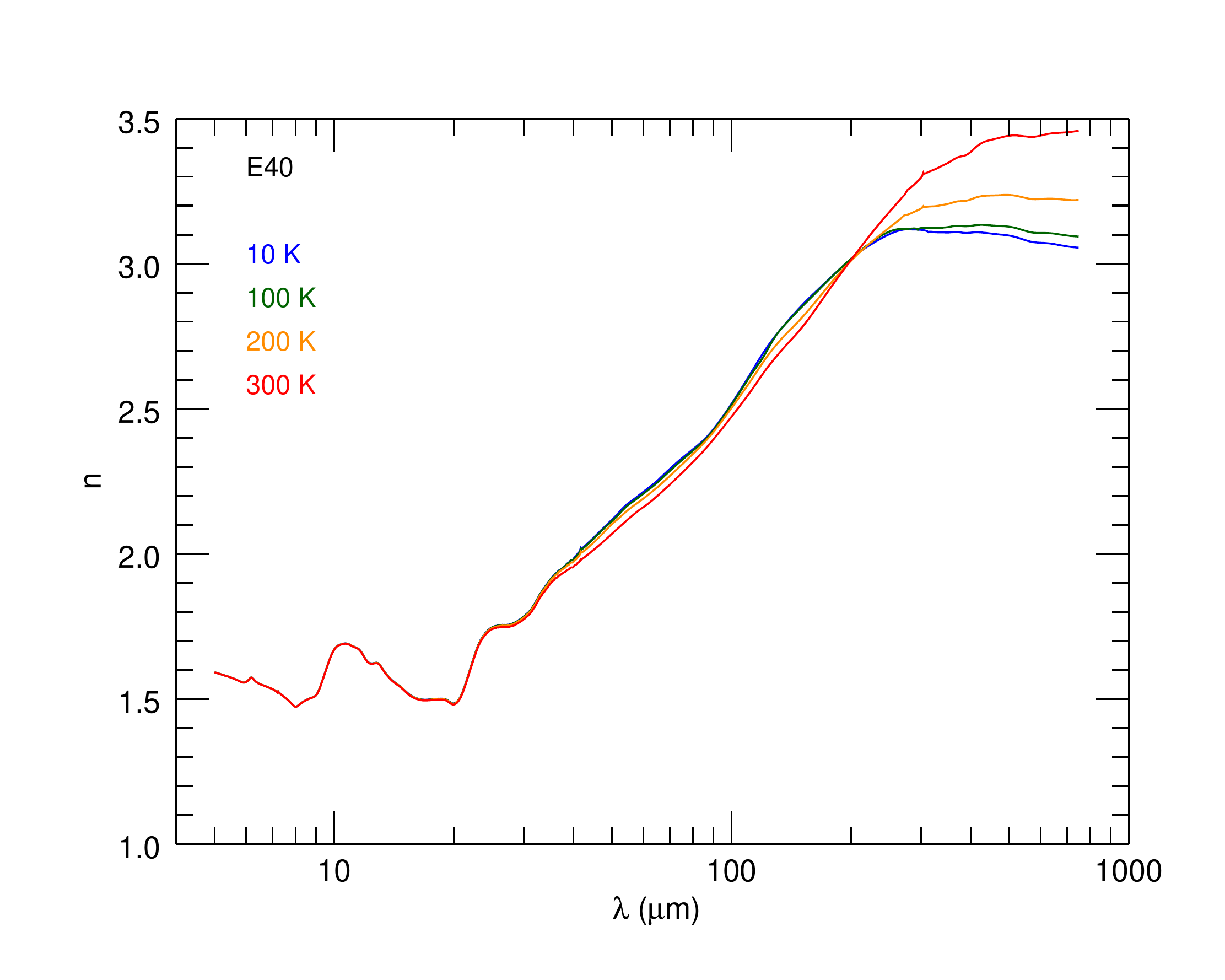}
  \includegraphics[scale=.3, trim={0 1cm 0 1.5cm}, clip]{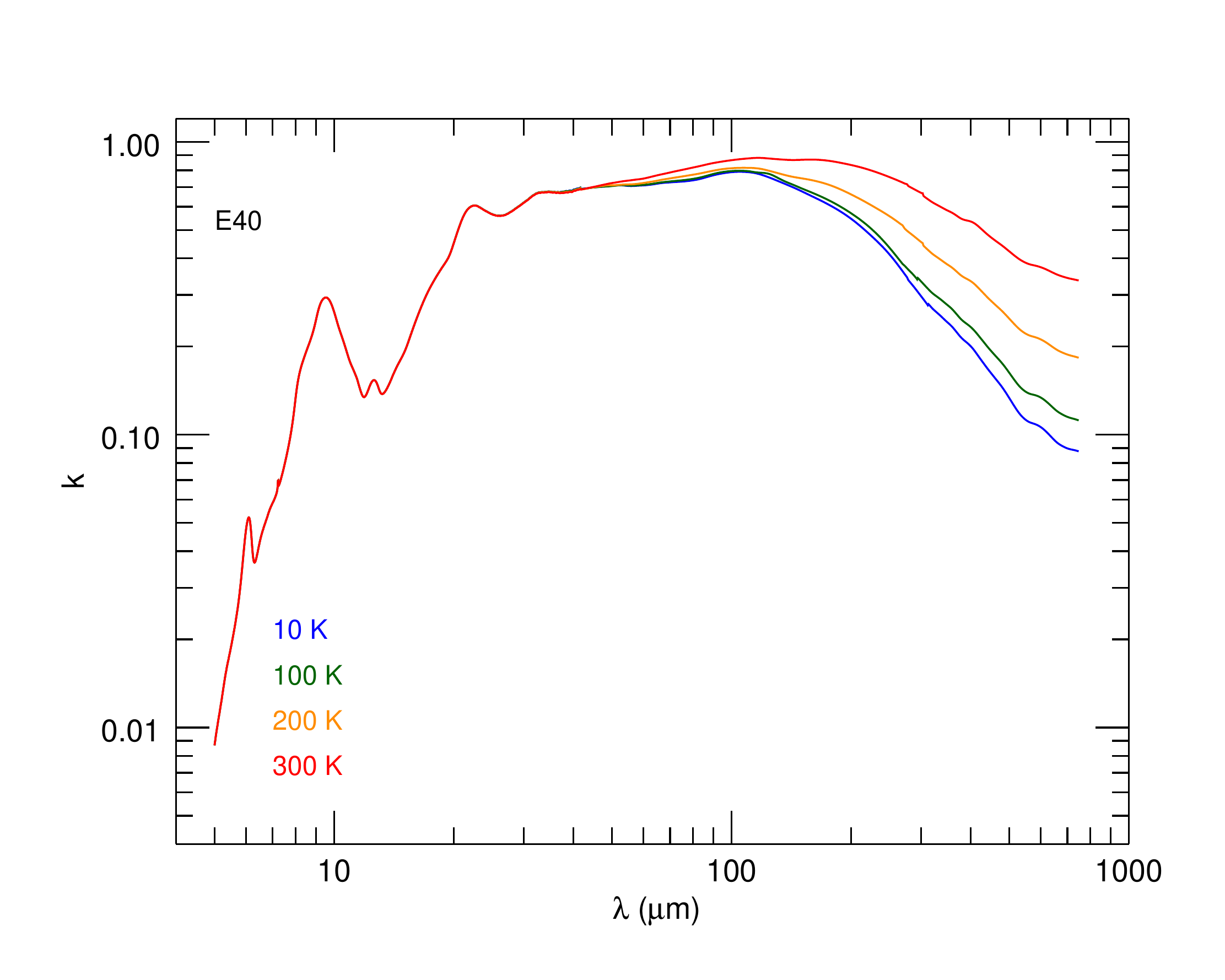}
          \caption{Optical constants for the Fe-rich silicate samples at temperatures of 10 K (blue), 100 K (green), 200 K (orange), and 300 K (red).  }
    \label{netk_ferich}% label for figure
\end{figure*}

\begin{figure*}[!h]
\centering
  \includegraphics[scale=.3, trim={0 1cm 0 1.5cm}, clip]{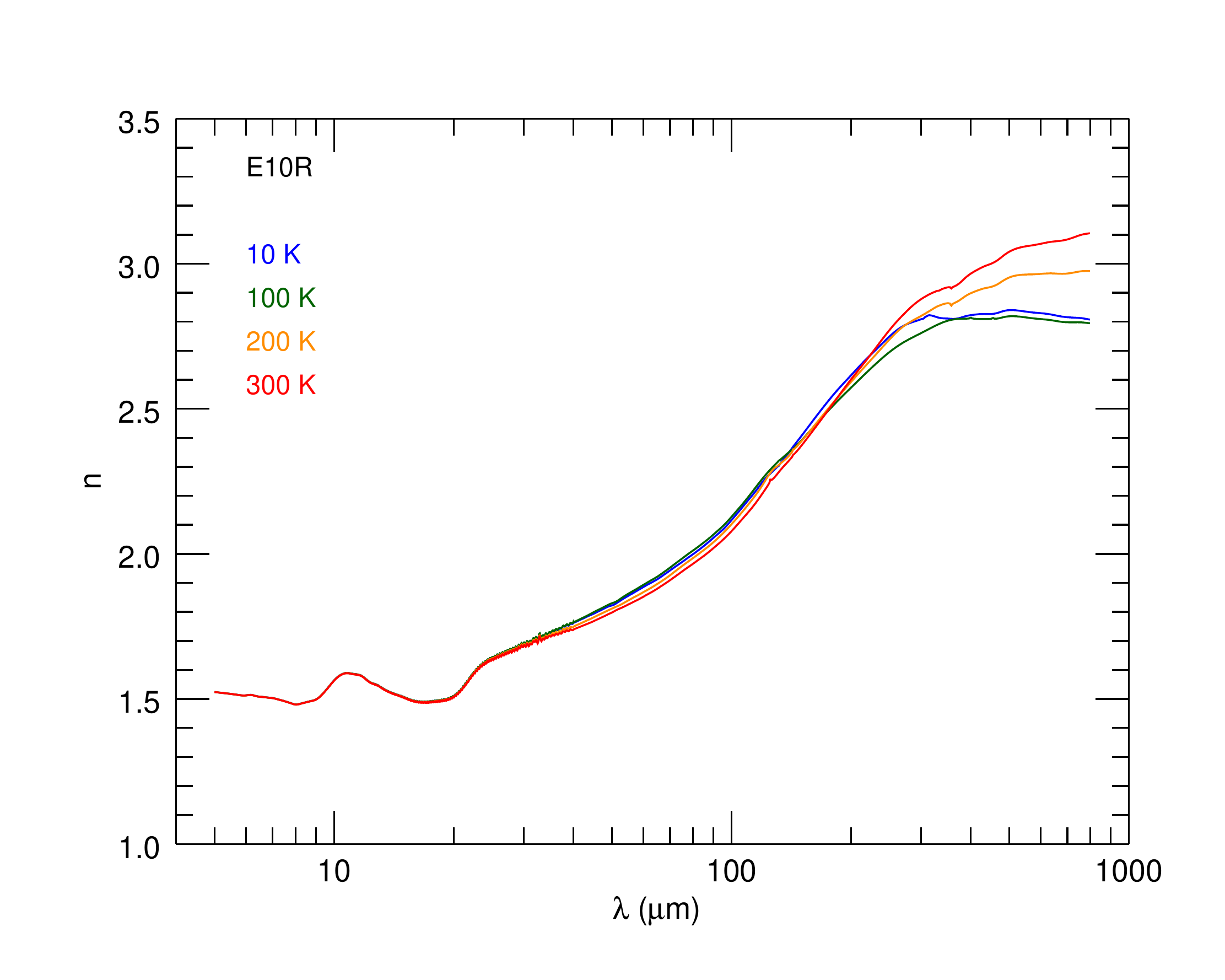}
  \includegraphics[scale=.3, trim={0 1cm 0 1.5cm}, clip]{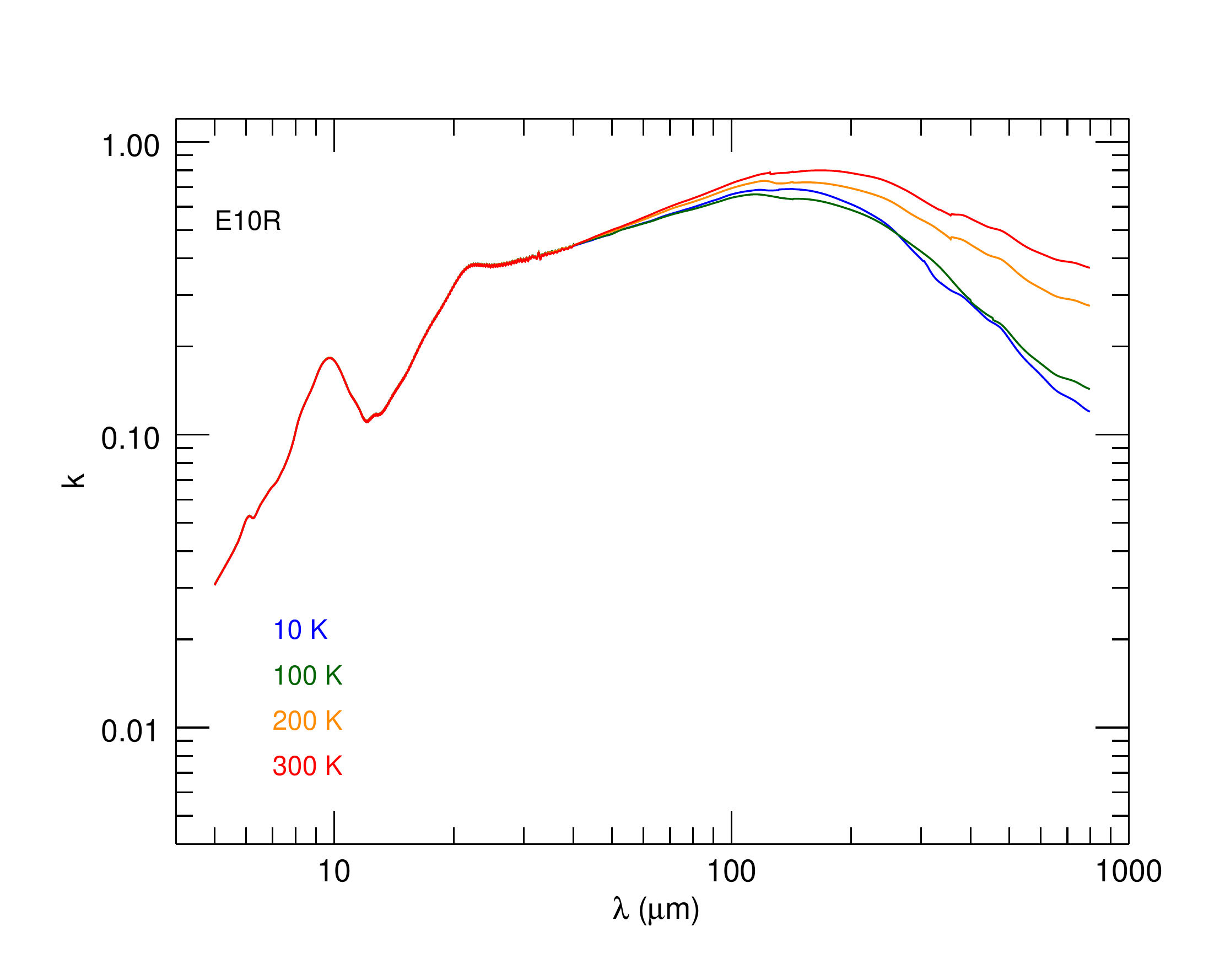}
  \includegraphics[scale=.3, trim={0 1cm 0 1.5cm}, clip]{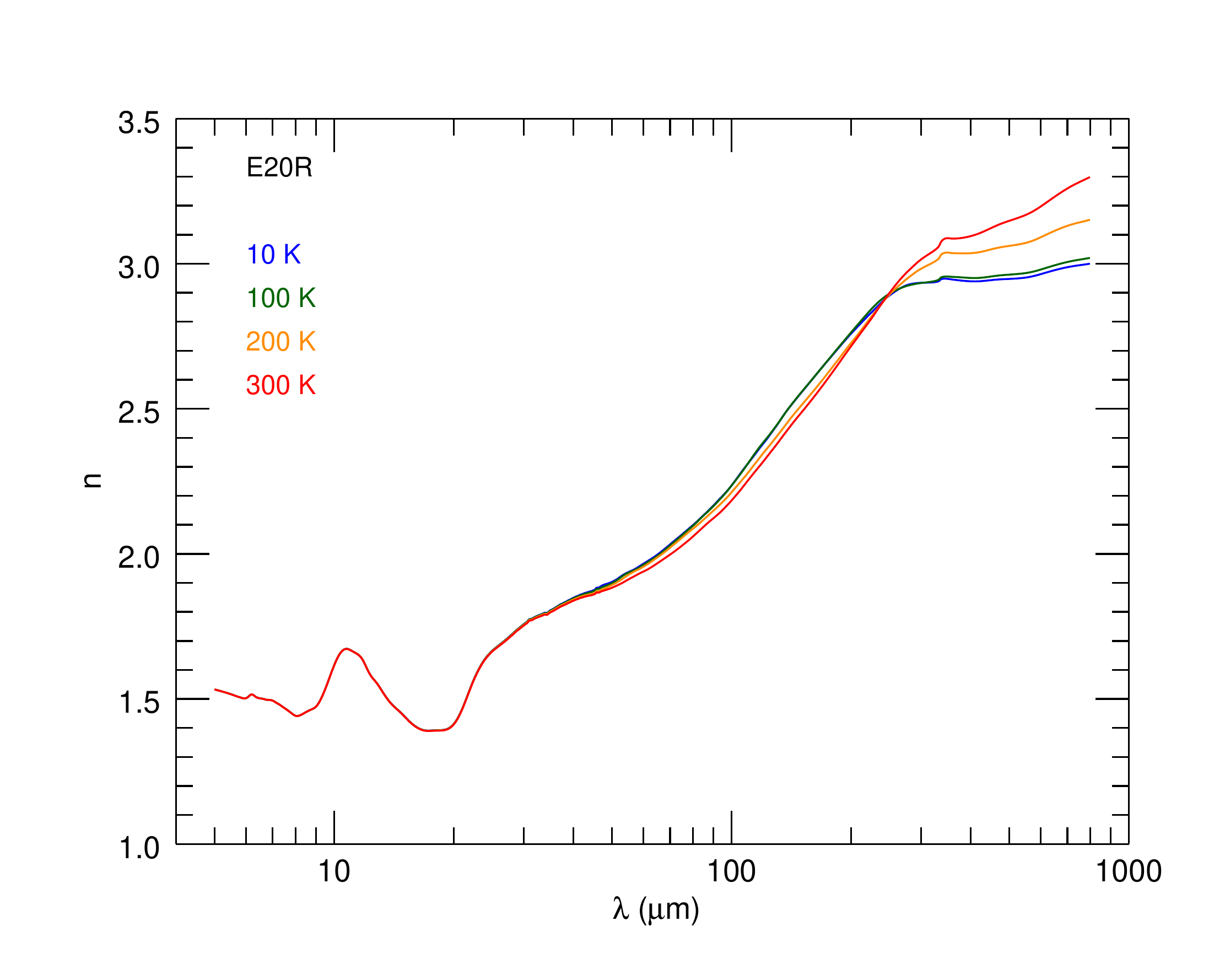}
  \includegraphics[scale=.3, trim={0 1cm 0 1.5cm}, clip]{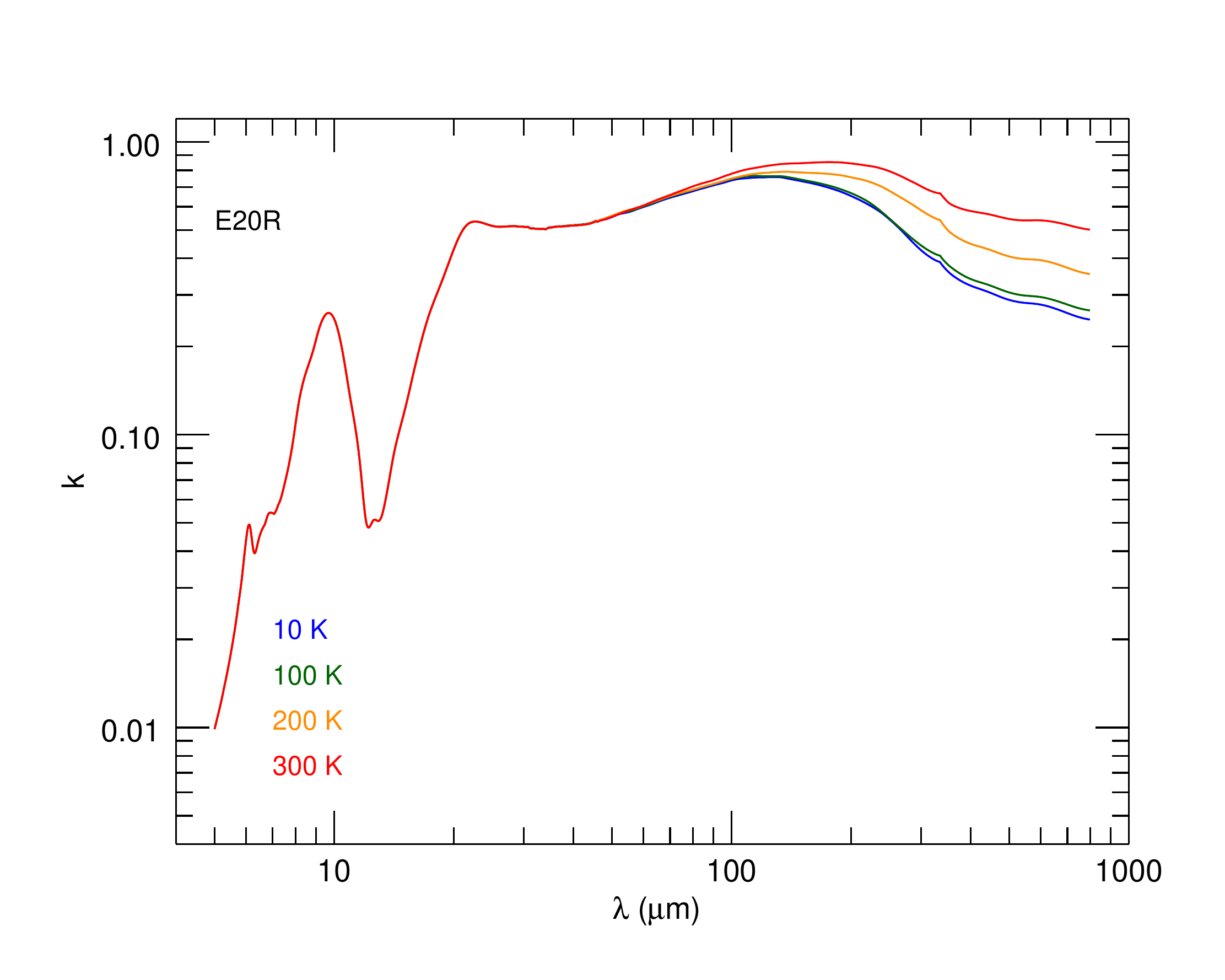}
  \includegraphics[scale=.3, trim={0 1cm 0 1.5cm}, clip]{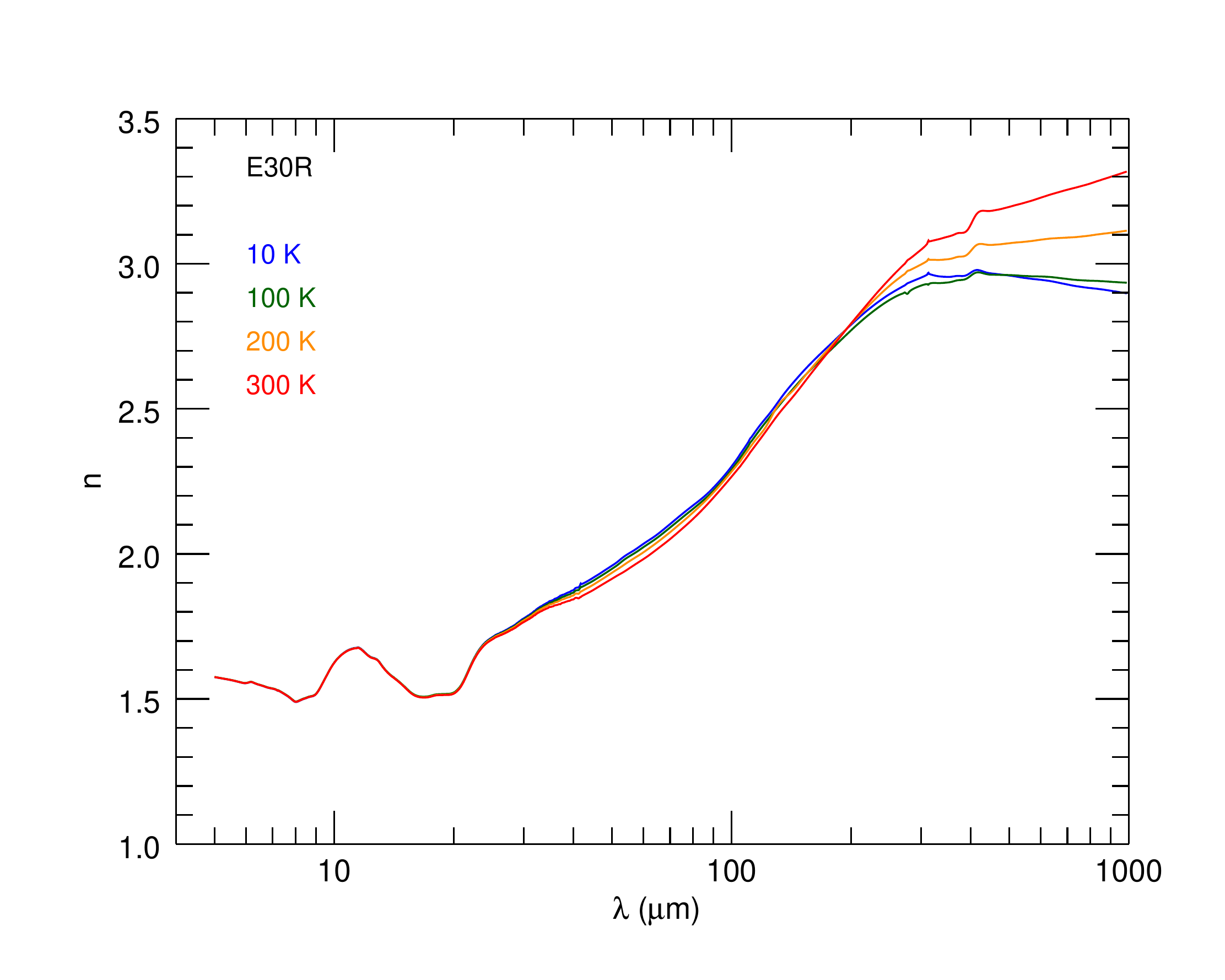}
  \includegraphics[scale=.3, trim={0 1cm 0 1.5cm}, clip]{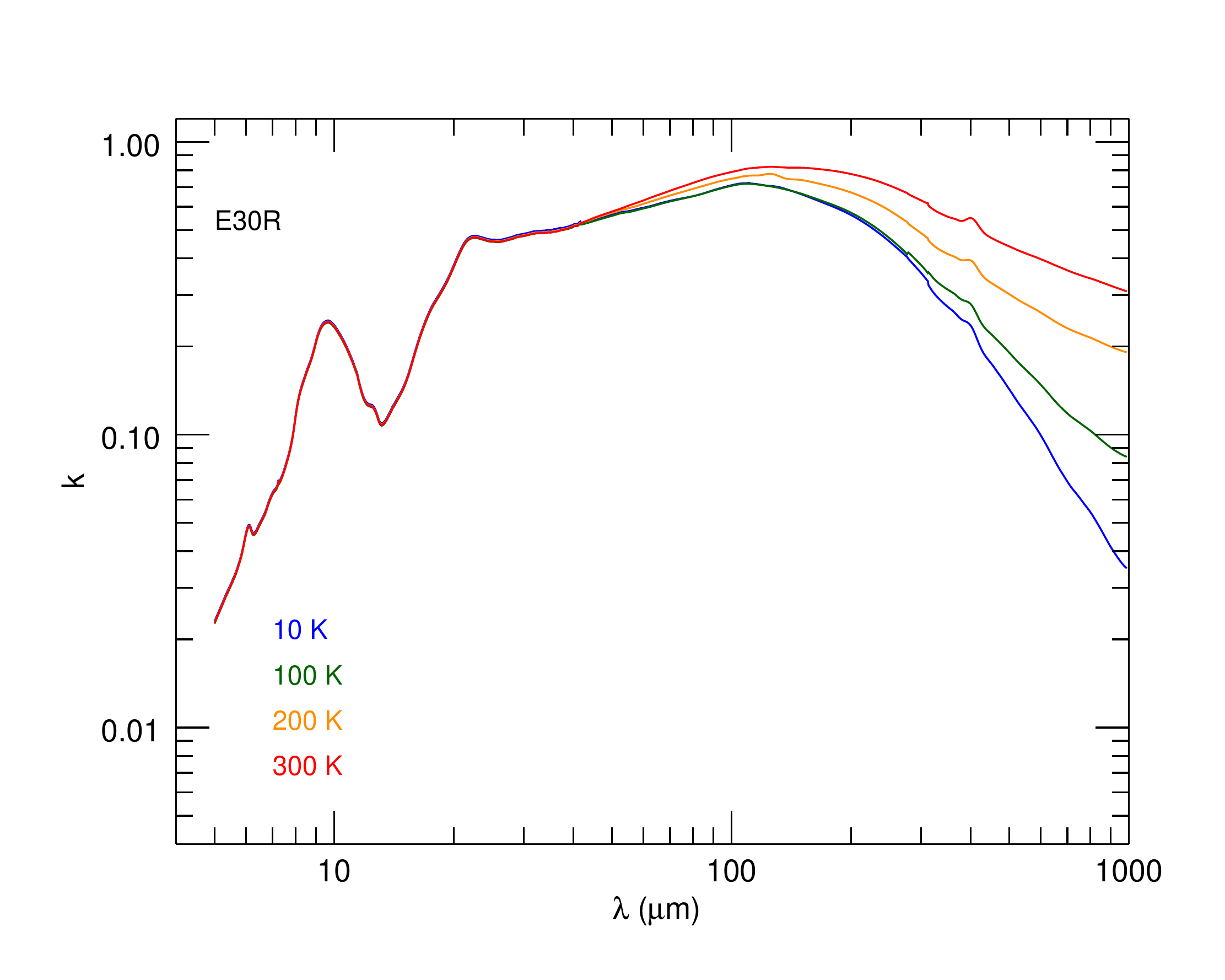}
  \includegraphics[scale=.3, trim={0 1cm 0 1.5cm}, clip]{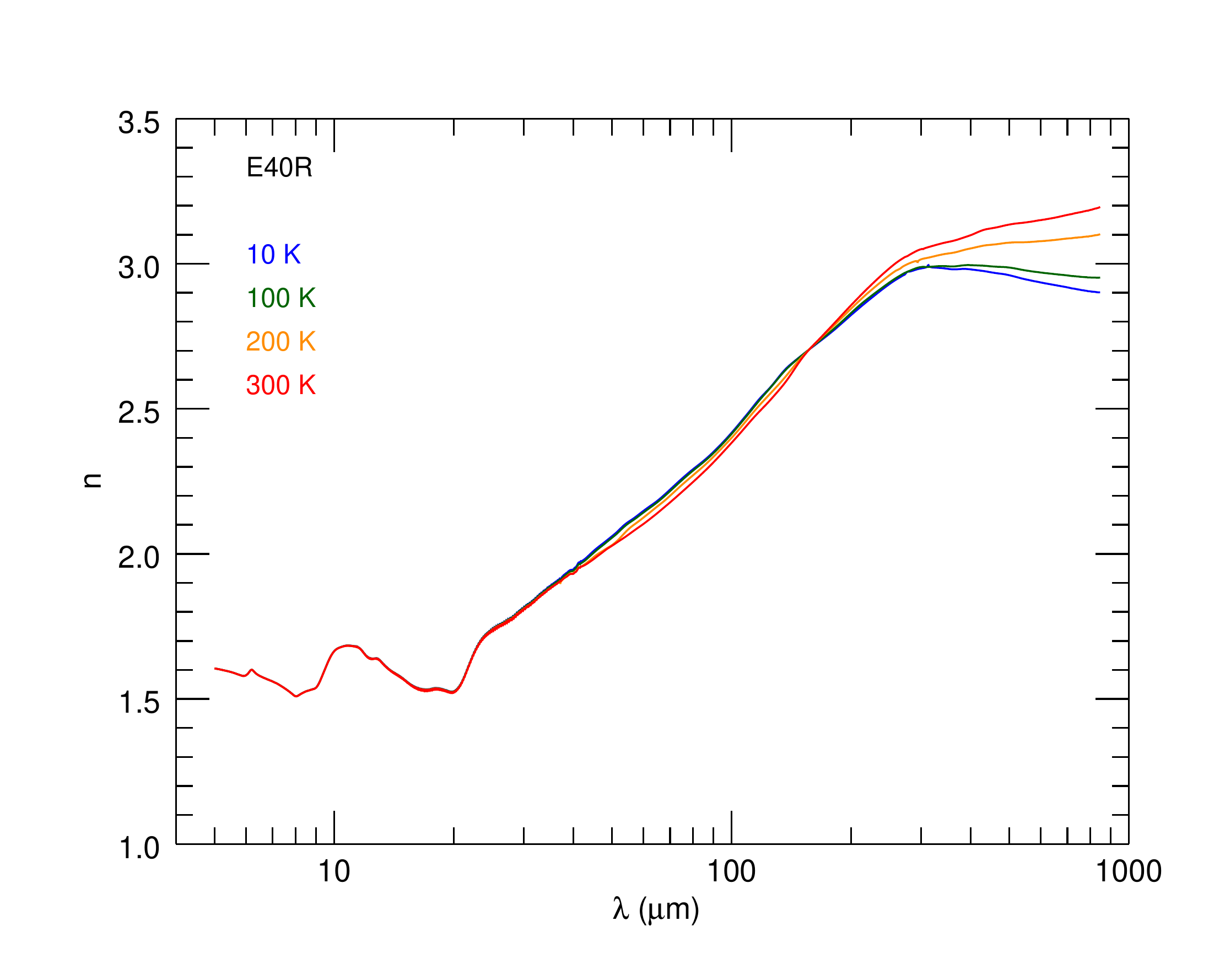}
  \includegraphics[scale=.3, trim={0 1cm 0 1.5cm}, clip]{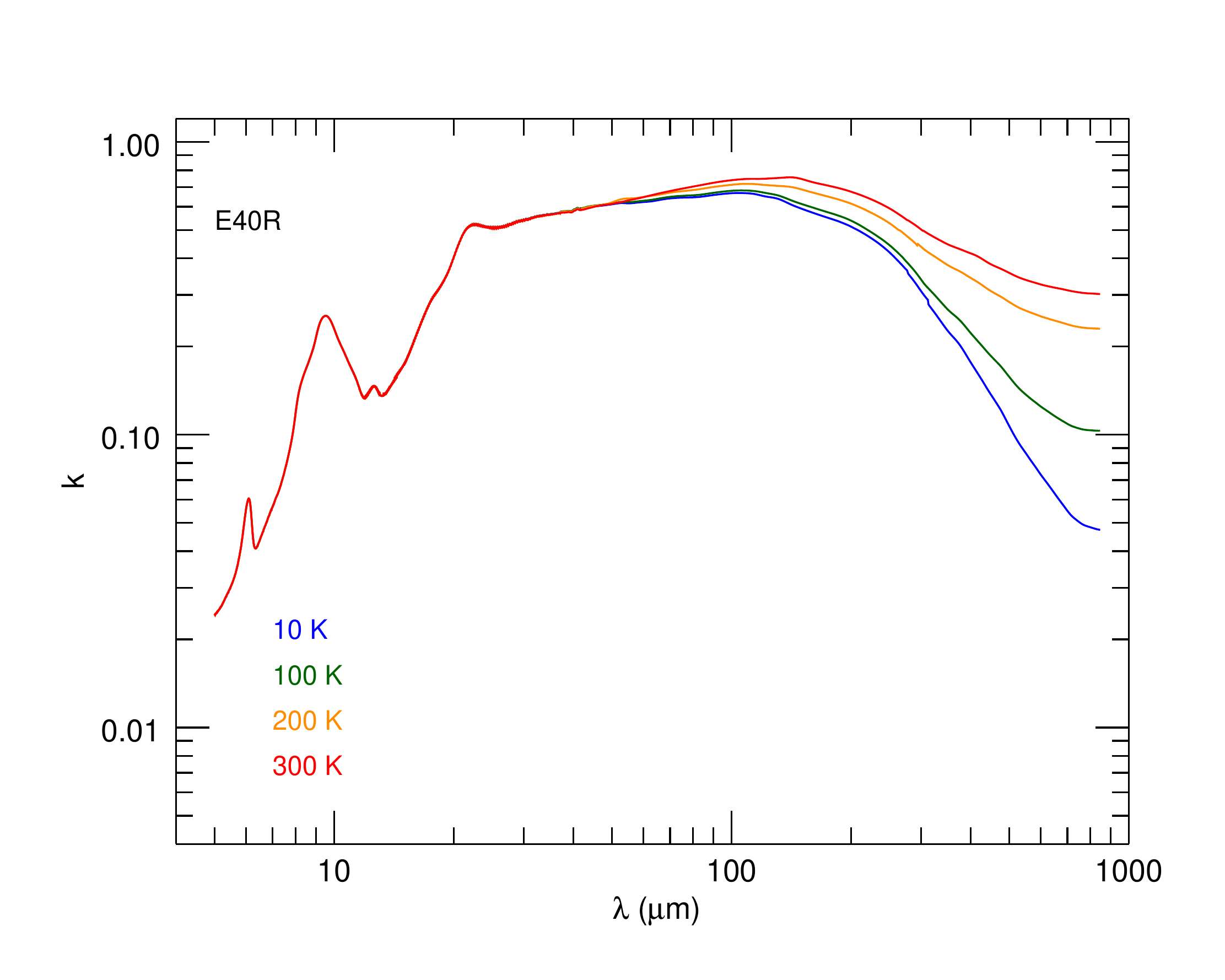}
          \caption{Optical constants for the reduced Fe-rich silicate samples at temperatures of 10 K (blue), 100 K (green), 200 K (orange), and 300 K (red).  }
    \label{netk_ferich_reduced}% label for figure
\end{figure*}

The studied samples consist of grains of different sizes and shapes. Electronic transmission microscopy (TEM, Fig.~\ref{TEM_shape}) images of the grains show that, even though they may be partially agglomerated, they are mostly submicron in size. This was confirmed by the measurement of the size distribution of one sample (Fig.~\ref{TEM_shape}) which peaks at a radius of $\sim 0.5 \,\mu$m, with only a minority of grains in the $1 - 2 \,\mu$m range. The grains are therefore small compared to the wavelength and are in the Rayleigh limit. 

The TEM images of the samples show that the grains have irregular shapes (Fig.~\ref{TEM_shape}), with the majority of the grains being reasonably well approximated by spheroids, most often of prolate shape with an axis ratio, $a/b$, in the range $1.5 - 2$; a few have more elongated or spherical shapes. The glassy samples (X35, X40, X50A and X50B) appear to be more roundish than the solgel samples (E10, E20, E30, E40, E10R, E20R, E30, E40R). We therefore adopted a population of spheroidal grains of prolate shape with an axis ratio $a/b$ = 1.5 for the glassy silicates and $a/b$ = 2 for the solgel Fe-rich silicates. 

The optical constants computed with the preceding assumptions, gathered in Table~\ref{table:extrapol}, are presented in Figs.~\ref{netk_mgrich}, \ref{netk_ferich}, and \ref{netk_ferich_reduced} for the Mg-rich, Fe-rich, and reduced Fe-rich samples, respectively, for temperatures 10, 100, 200, and 300 K (the optical constants at 30 K are not shown for clarity). 

%\subsection{Error related to the refractive index at short wavelength}

\section{Error on the optical constant determinations}
\label{sect_error}

We investigated the implications of the assumptions made on  $n_{\rm{vis}}$ on the shape and size of the particles. To do so, for each sample, we calculated the optical constants by adopting different values of $n_{\rm{vis}}$, the grain shape (prolate or oblate), and the axis ratio. In the following, we designate $n_{\rm{ref}}$ and $k_{\rm{ref}}$ the reference optical constants ---that is, those presented in Figs.~\ref{netk_mgrich}, \ref{netk_ferich}, and \ref{netk_ferich_reduced}--- calculated for prolate grains, and use the $n_{\rm{vis}}$ and $a/b$ values given in Table~\ref{table:extrapol}.

\subsection{Influence of the adopted parameters}
\label{subsect_error}

To estimate the effect of an error made on $n_{\rm{vis}}$, we calculated $n_i$ and $k_i$ for each sample for values of $n_{\rm{vis}} = n_{\rm{vis}}^{\rm{ref}} \pm$ 5~\%, $n_{\rm{vis}}^{\rm{ref}}$ being the value given in Table~\ref{table:extrapol}, and adopting the same grain shape as that used to derive $n_{\rm{ref}}$ and $k_{\rm{ref}}$, that is prolate grains with an axis ratio of 1.5 for the glassy silicates and 2 for the solgel samples. We define the resulting optical constant uncertainty as: 
\begin{equation}
\label{eq:delta_n_k_nvis}
\begin{aligned}
{\delta}n_{\rm{nvis}} =  \frac{ ( n_{\rm{ref}} -  n_i ) }{ n_{\rm{ref}} }  ;  \ \ \ \ \ {\delta}k_{\rm{nvis}} =  \frac{ ( k_{\rm{ref}} -  k_i ) }{ k_{\rm{ref}} }. 
\end{aligned}
\end{equation}
The uncertainties on $n$ and $k$ result in an uncertainty on the MAC (computed for each sample with the grain shape and axis ratio from Table~\ref{table:extrapol}), which is expressed as: 
\begin{equation}
\label{eq:delta_kappa_nvis}
\begin{aligned}
{\delta}{\kappa}_{\rm{nvis}} =  \frac{ {\kappa}(n_{\rm{re}f}, k_{\rm{ref}}) - {\kappa}(n_i, k_i)  }{  {\kappa}(n_{\rm{re}f}, k_{\rm{ref}}) }.  
\end{aligned}
\end{equation}

A look at measurements of the refractive index of Mg- and Fe-rich amorphous silicates in the visible domain in the HJPDOC database shows that $n_{\rm{vis}}$ is in the range $\sim1.5 - 1.7$. We therefore consider that the error associated with our choice of $n_{\rm{vis}}$ does not exceed 5\%. As can be seen in Fig.~\ref{test_n0}, for all samples,  this results in an error on $n$ of $\sim$ 5\% over the whole spectral range and an error on $k$ of less than $\sim 3\%-4$\%  for $\lambda \leqslant 20 \,\mu$m and up to $\sim 8$\% on $k$ for $\lambda \geqslant 30 \,\mu$m, the resulting error on the simulated MAC being negligible. Test cases assuming larger values of the error on $n_{\rm{vis}}$ showed that the uncertainties on $n$ and $k$ are proportional to that on $n_{\rm{vis}}$.

%\subsection{Error related to grain shape }

The calculation of optical constants also requires an assumption about grain shape and an error on the assumed shape results in an uncertainty on the derived optical constants. We estimate this by calculating, for each sample, the optical constants, noted $n_{a/b}$ and $k_{a/b}$, adopting the value of $n_{\rm{vis}}$ given in Table~\ref{table:extrapol}, and different assumptions for the axis ratio. The uncertainty on the optical constants is then given by: 
\begin{equation}
\label{eq:delta_n_k_shape}
\begin{aligned}
{\delta}n_{\rm{shape}} =  \frac{ ( n_{\rm{ref}} -  n_{a/b} ) }{ n_{\rm{ref}} }  ;  \ \ \ \ \ {\delta}k_{\rm{shape}} =  \frac{ ( k_{\rm{ref}} -  k_{a/b} ) }{ k_{\rm{ref}} }.  
\end{aligned}
\end{equation}
\begin{figure*}[!th]
\centering
\includegraphics[angle=90, scale=0.26, trim={0 1cm 0 0.cm}, clip]{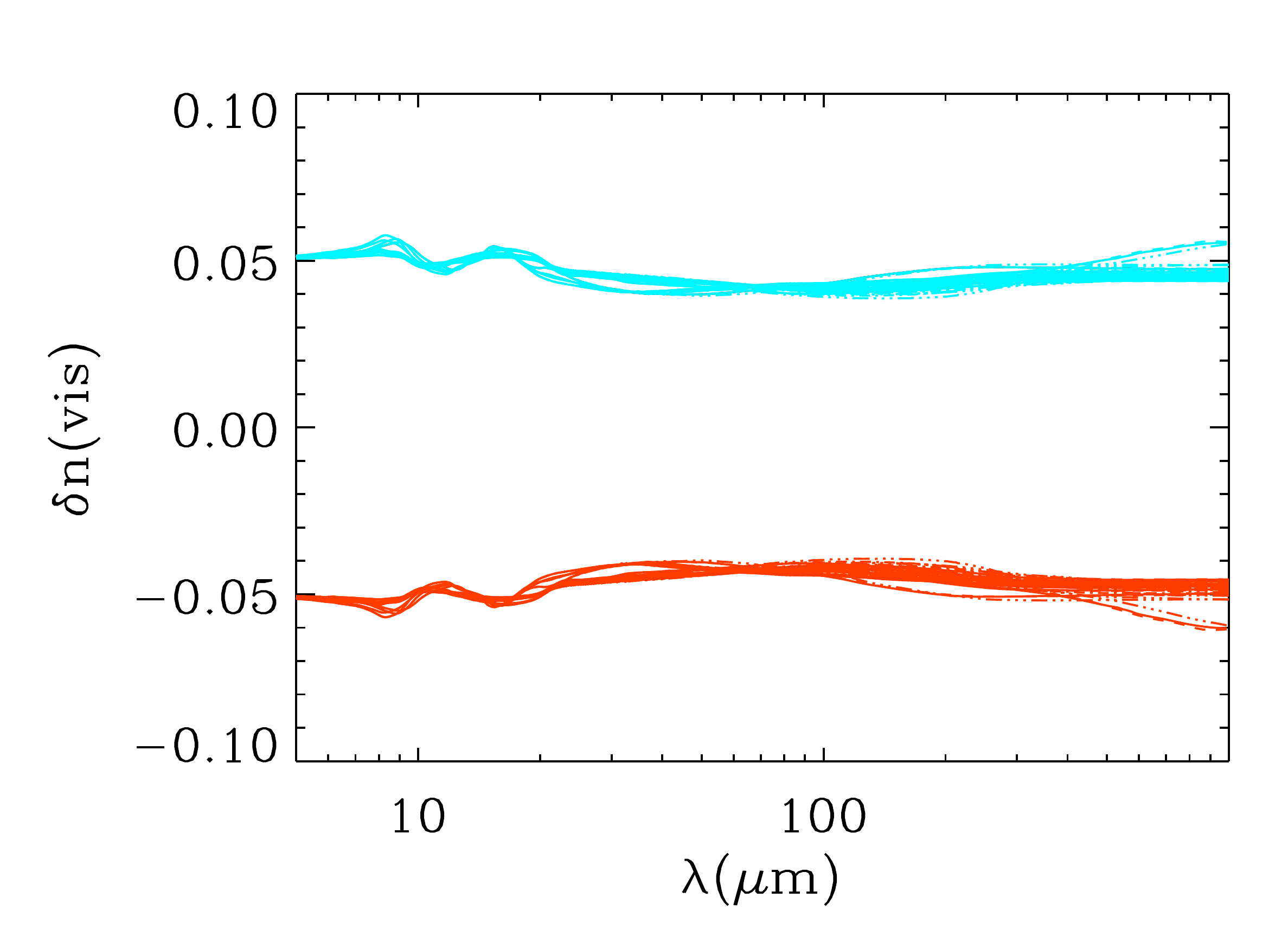}
\includegraphics[angle=90, scale=0.26, trim={0 1cm 0 0.cm}, clip]{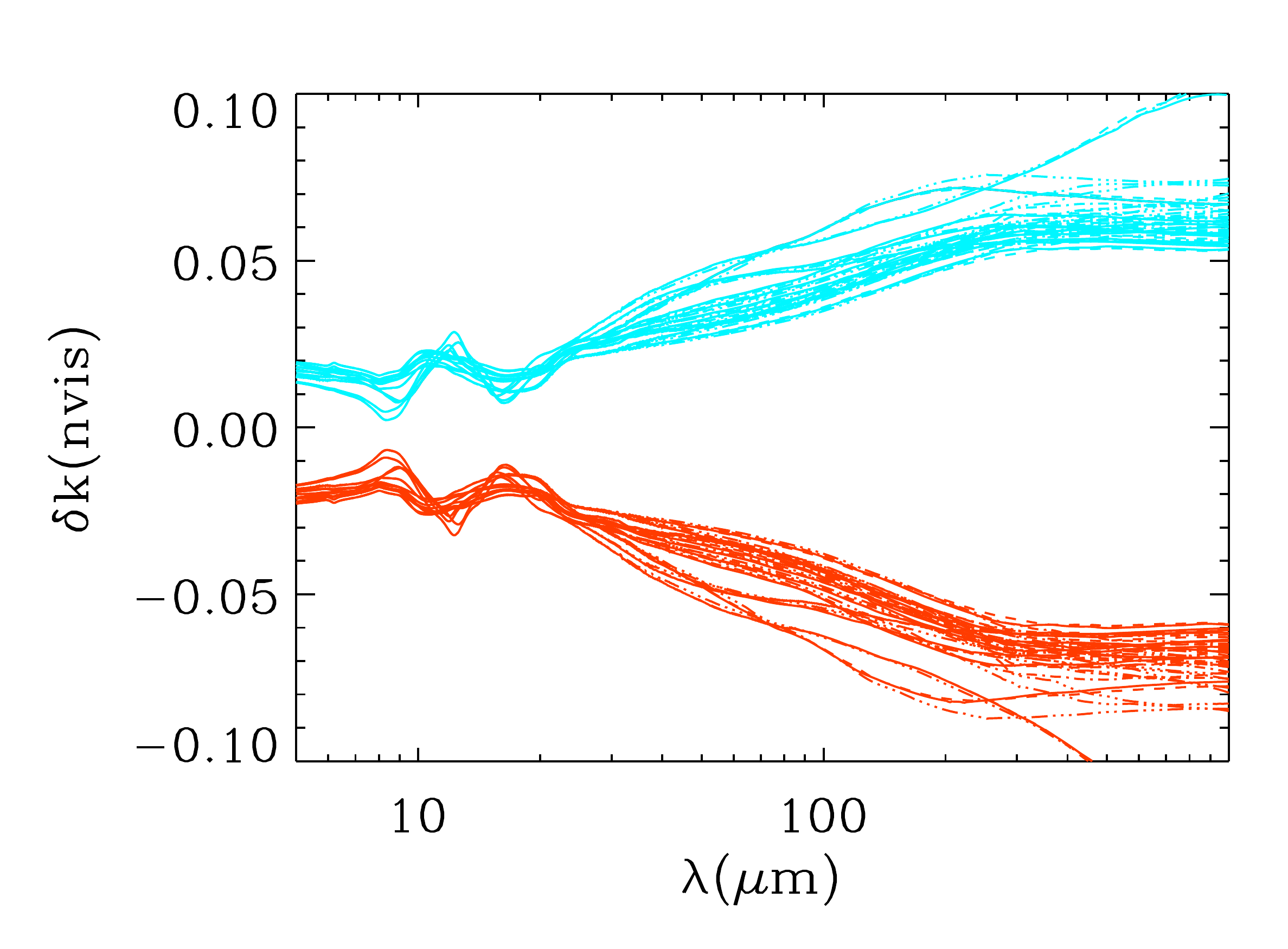}
\includegraphics[angle=90, scale=0.26, trim={0 1cm 0 0.cm}, clip]{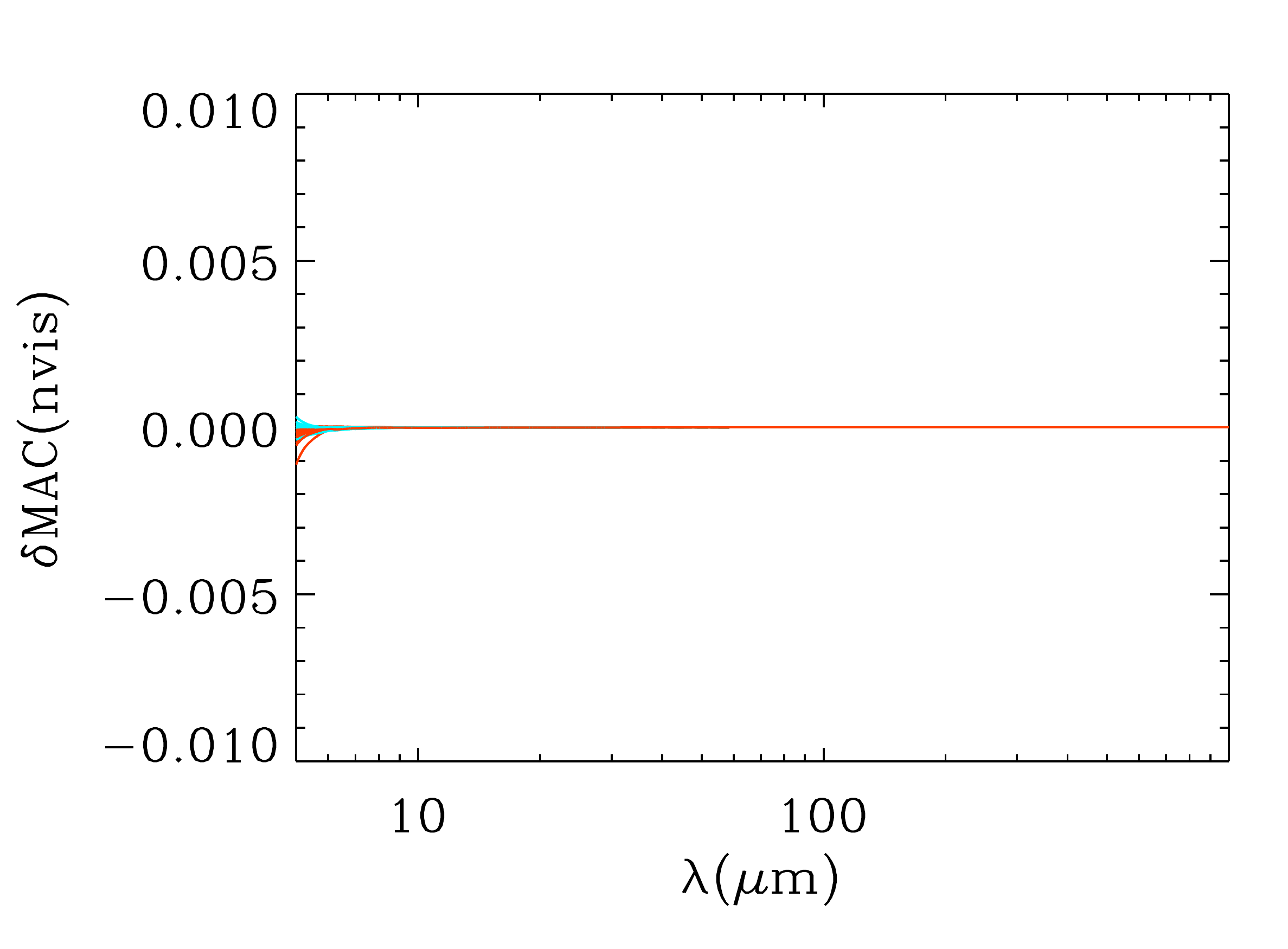}
          \caption{Influence of the assumed value of the refractive index at short wavelengths ($n_{\rm{vis}}$) on the calculated optical constants (left and middle panels) and on the computed MAC (right panel) for all samples and temperatures (continuous lines: 10K, dashed line: 100K, dotted-dashed lines : 200K, triple dot-dashed lines: 300K).  The assumed refractive indices in the visible are: $n_{\rm{vis}} = n_{\rm{vis}}^{\rm{ref}} -  5$\% (blue lines) and $n_{\rm{vis}} = n_{\rm{vis}}^{\rm{ref}} +  5$\% (red lines).}
    \label{test_n0}% label for figure
\end{figure*}
\begin{figure*}
\centering
\includegraphics[angle=90, scale=0.27, trim={0 1cm 0 1.cm}, clip]{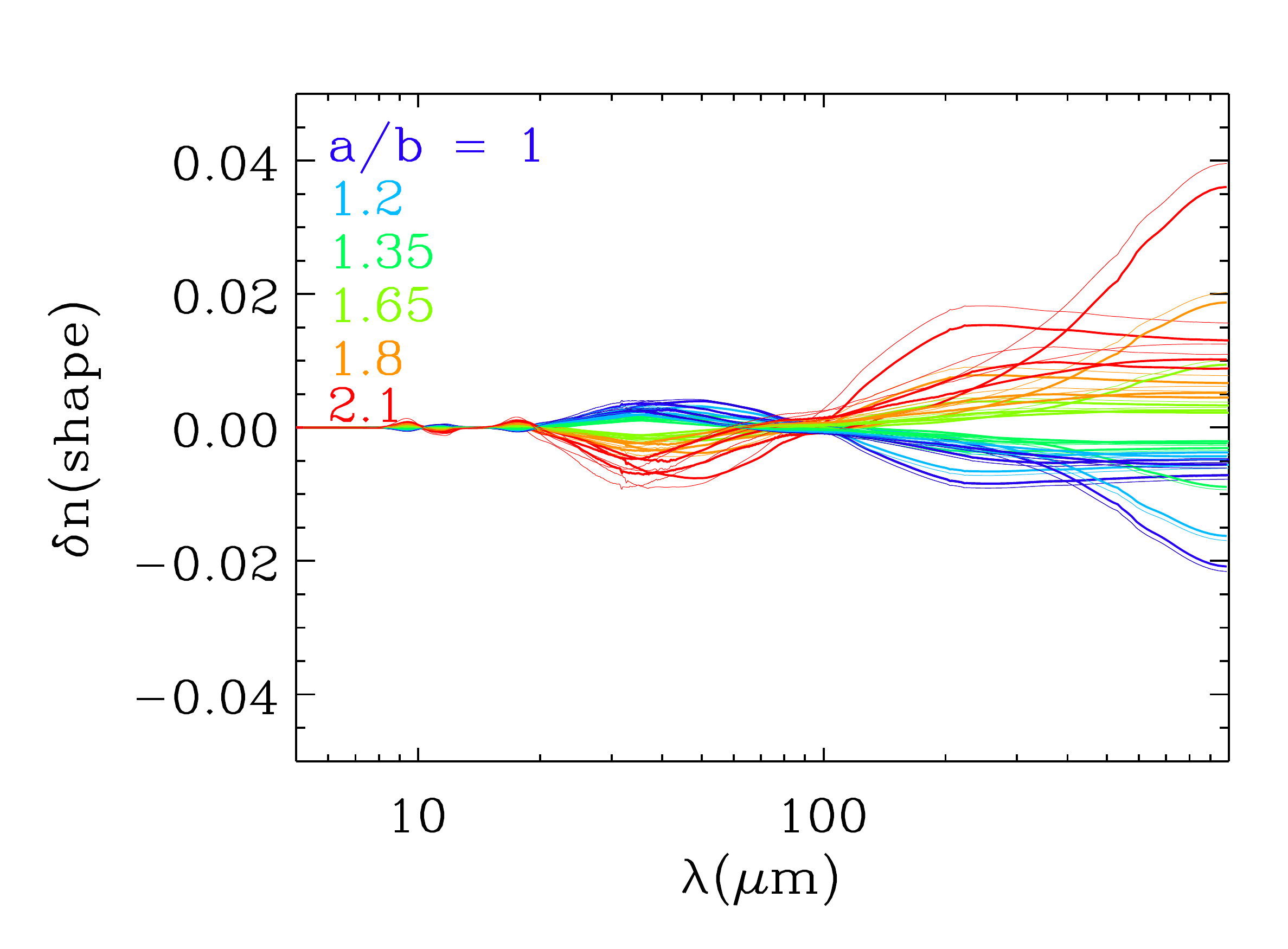}
\includegraphics[angle=90, scale=0.27, trim={0 1cm 0 1.cm}, clip]{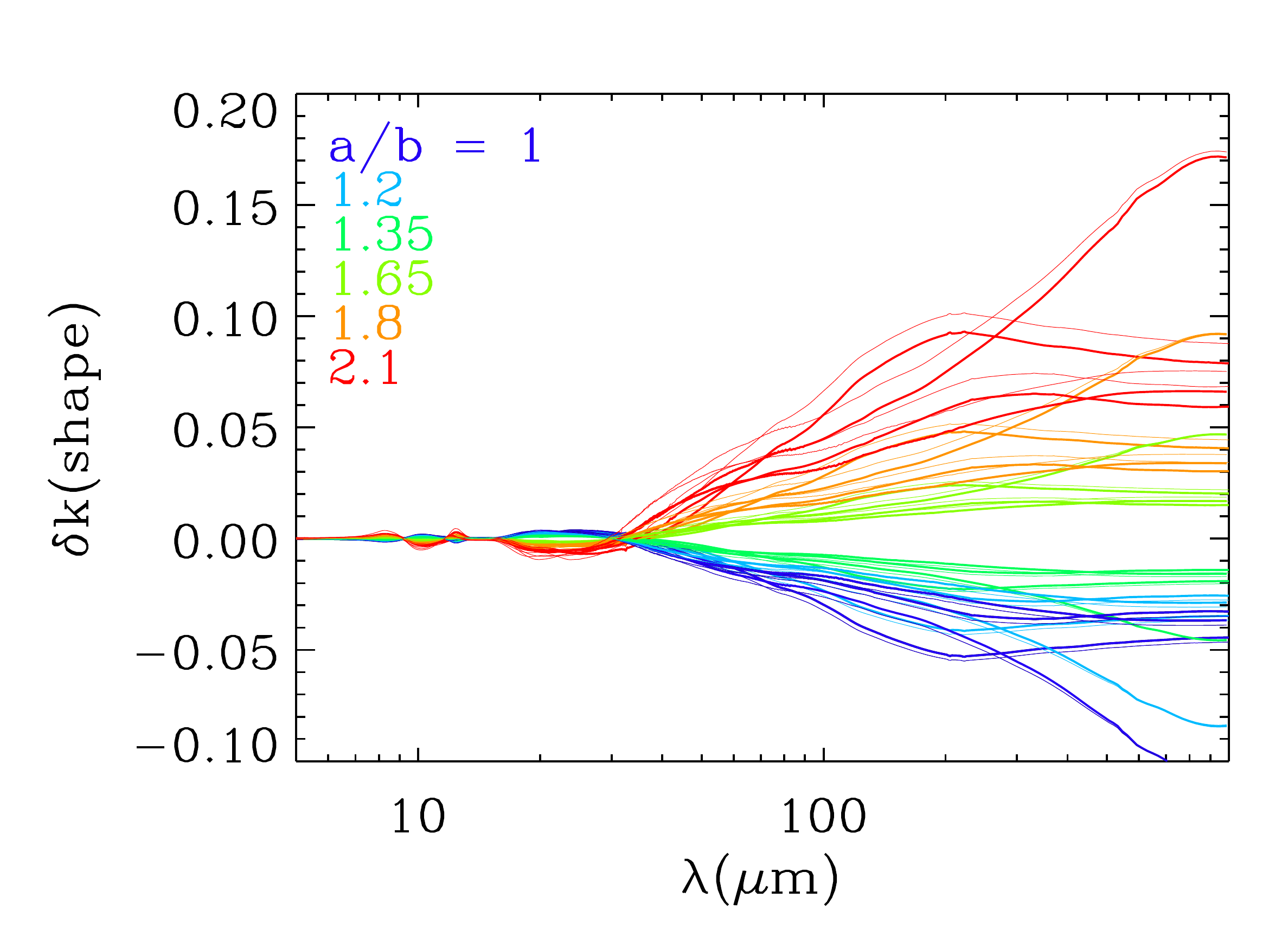}
\includegraphics[angle=90, scale=0.27, trim={0 1cm 0 1.cm}, clip]{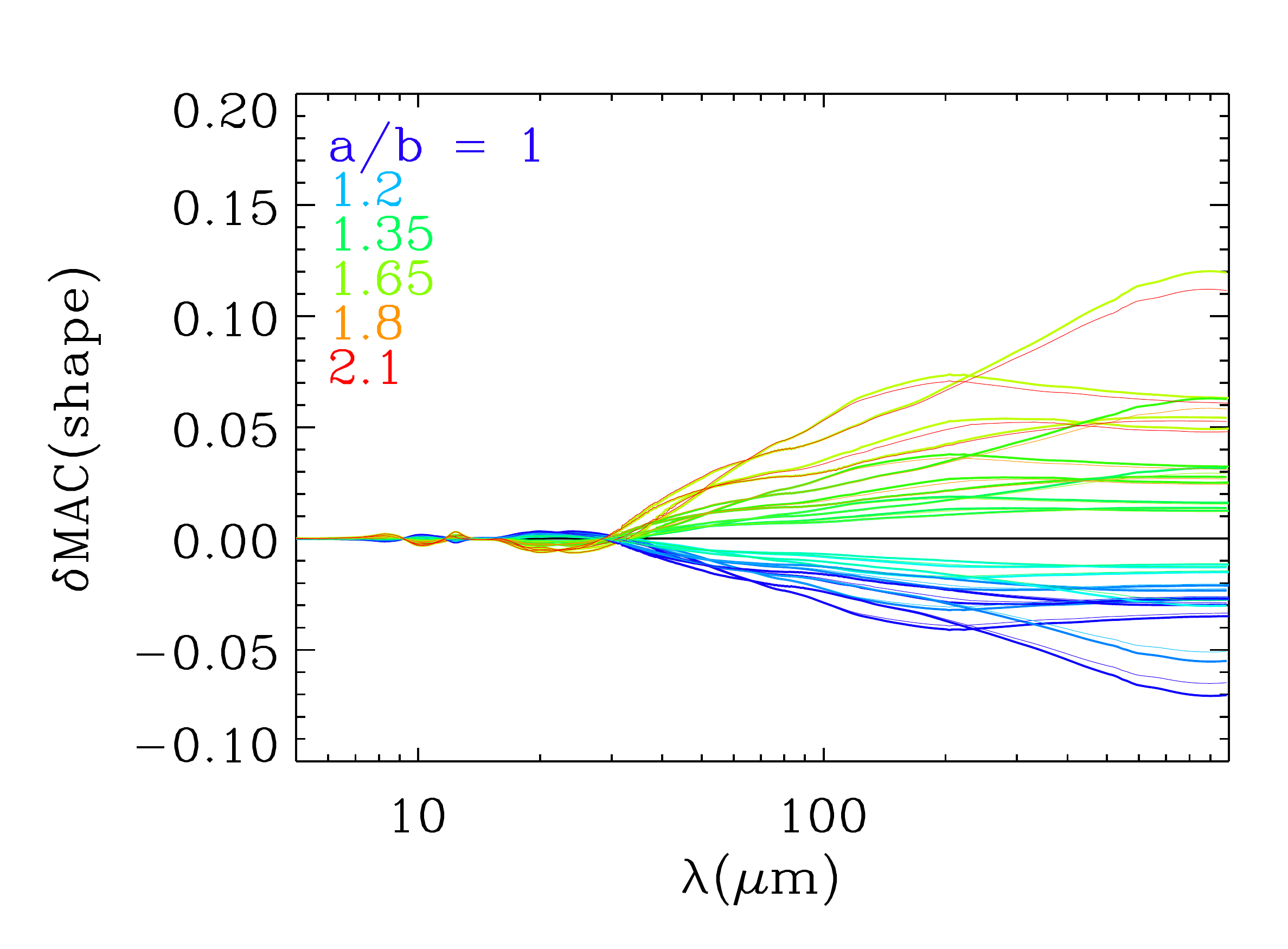}
\caption{Effect of an error on the grain shape assumption on the calculation of the optical constants for glassy Mg-rich samples at 10\,K and on the resulting MAC. The results are shown for different grain shapes: spherical (dark blue),  prolate (thick lines), and oblate (thin lines) with axis ratios, $a/b$, of 1.2, 1.35, 1.65, 1.8, and 2.1. From left to right, the figures show the resulting errors on $n$, $k,$ and the MAC.}
    \label{test_shape_X}% label for figure
\end{figure*}
\begin{figure*}
\centering
\includegraphics[angle=90, scale=0.27, trim={0 1cm 0 1.cm}, clip]{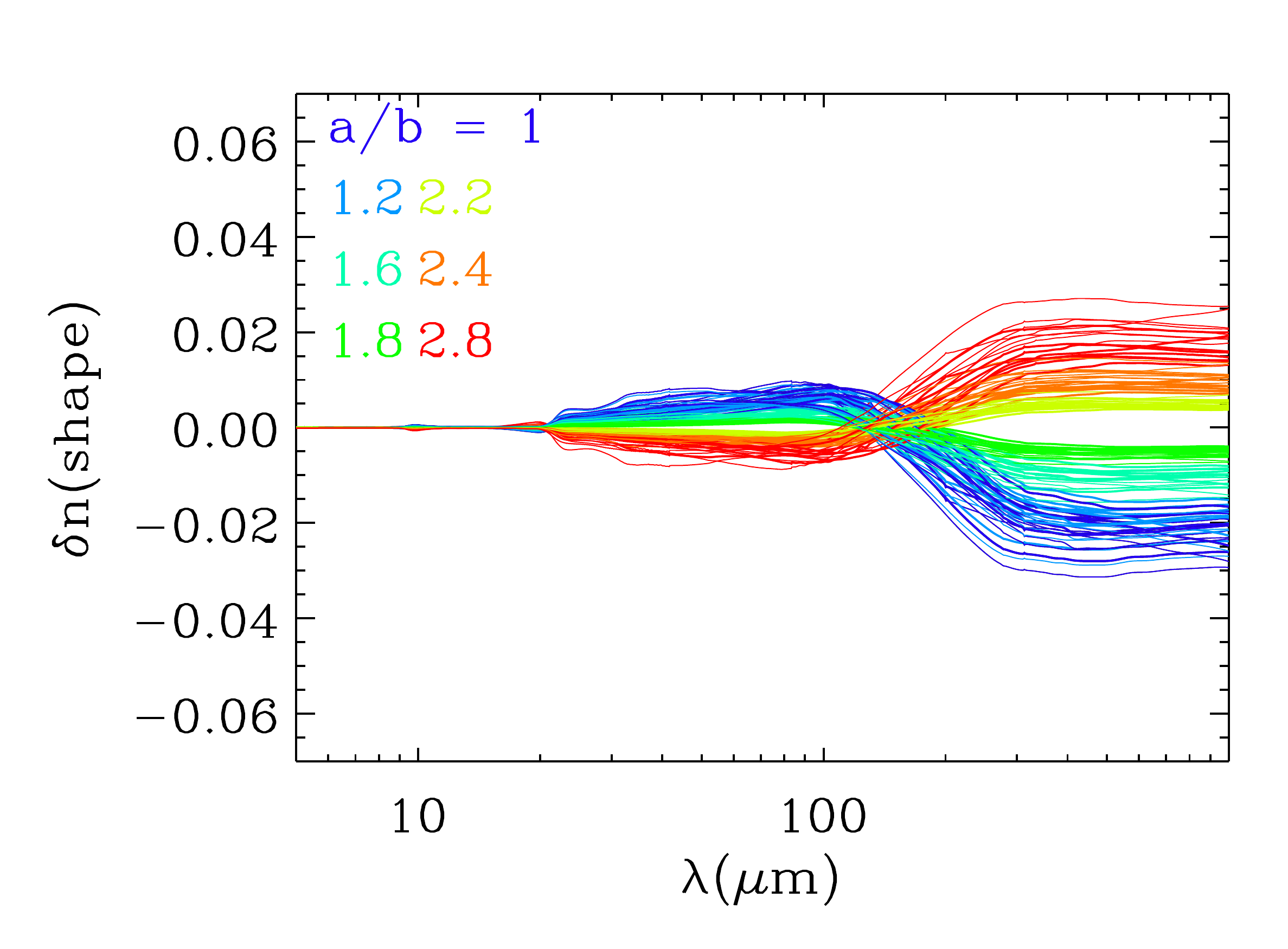}
\includegraphics[angle=90, scale=0.27, trim={0 1cm 0 1.cm}, clip]{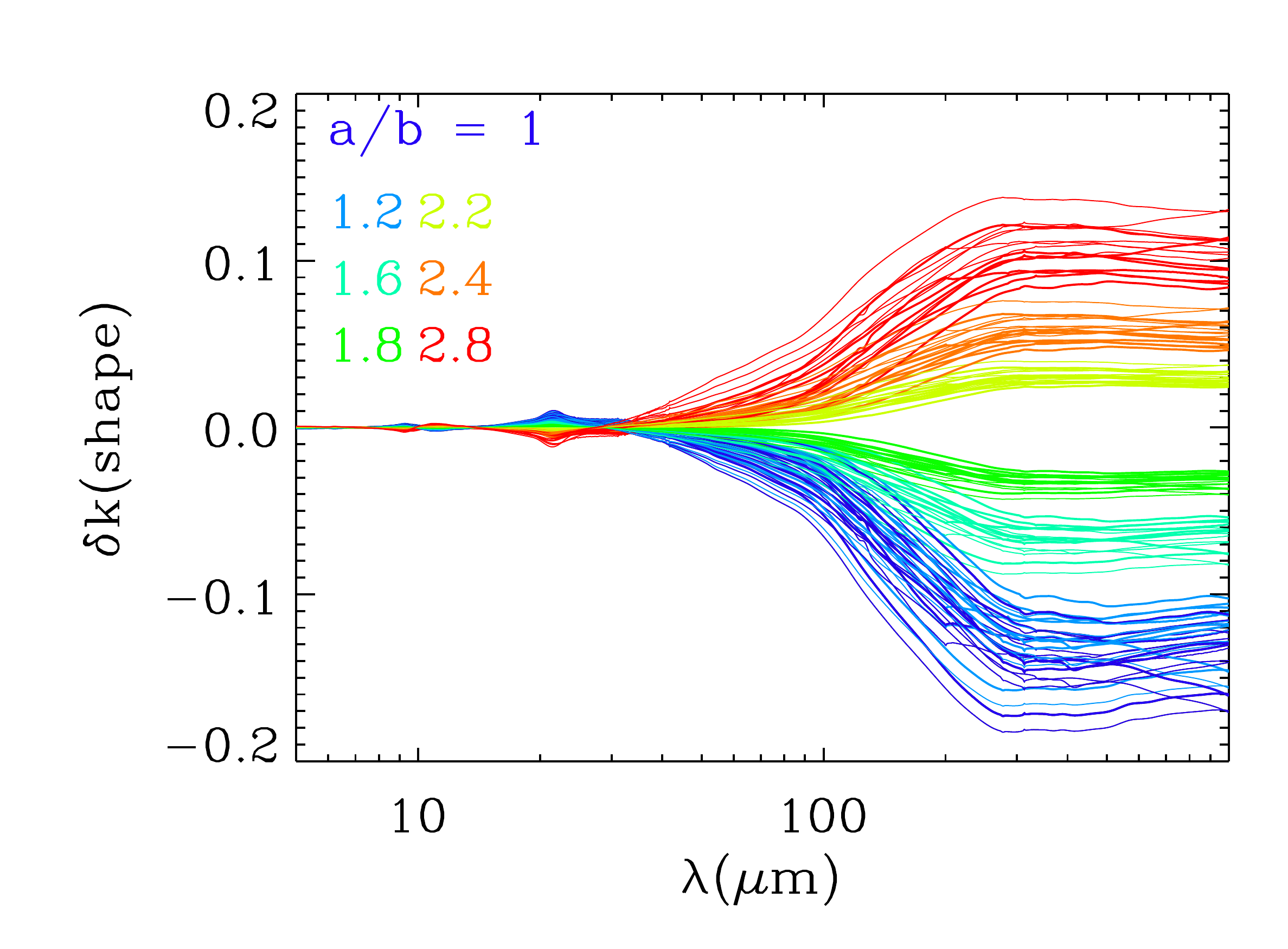}
\includegraphics[angle=90, scale=0.27, trim={0 1cm 0 1.cm}, clip]{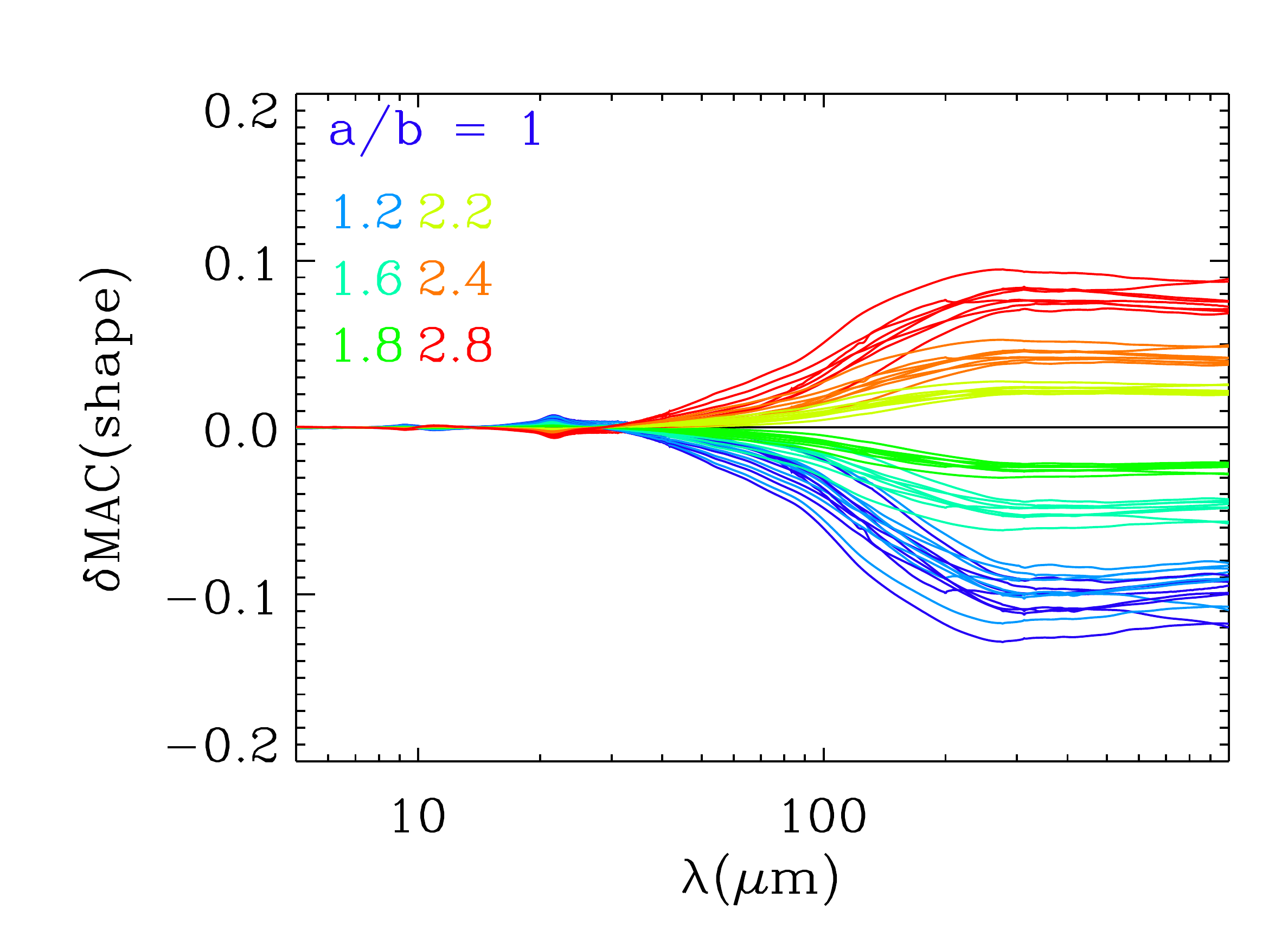}
\caption{Effect of an error on the grain-shape assumption on the calculated optical constants for Fe-rich samples at 10\,K and on the resulting MAC. The results are shown for different grain shapes: spherical (dark blue),  prolate (thick lines), and oblate (thin lines) with  axis ratios, $a/b$, of 1.2, 1.6, 1.8, 2.2 , 2.4, and 2.8. From left to right, the figures show the resulting errors on $n$, $k,$ and the MAC.}
    \label{test_shape_E}% label for figure
\end{figure*}
%
% propagation de l'erreur à kappa
%
The uncertainties on $n$ and $k$ resulting from an error made on the assumption on the grain shape will propagate onto the MAC for grain populations with different shapes. Adopting, for the computation of $\kappa$, the grain shapes and axis ratios from Table~\ref{table:extrapol}, the uncertainty on the MAC calculated for a given grain shape is expressed as: 
\begin{equation}
\label{eq:delta_kappa_shape}
\begin{aligned}
{\delta}{\kappa}_{\rm{shape}} =  \frac{ {\kappa}(n_{\rm{re}f}, k_{\rm{ref}}) - {\kappa}(n_{a/b}, k_{a/b})  }{  {\kappa}(n_{\rm{re}f}, k_{\rm{ref}}) }  \\
\end{aligned}
.\end{equation}
\begin{figure*}[!h]
\centering
\includegraphics[scale=0.4]{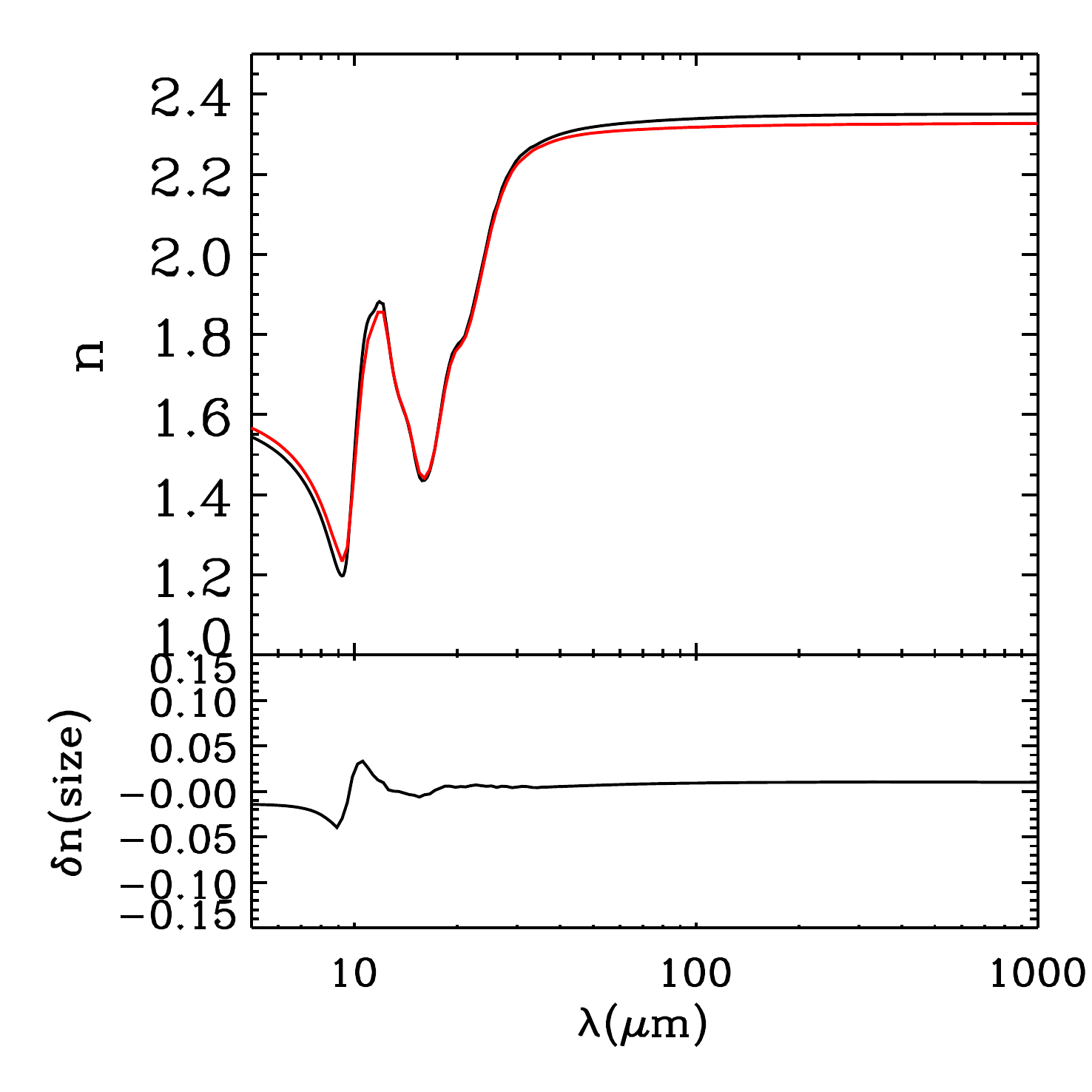}
\includegraphics[scale=0.4]{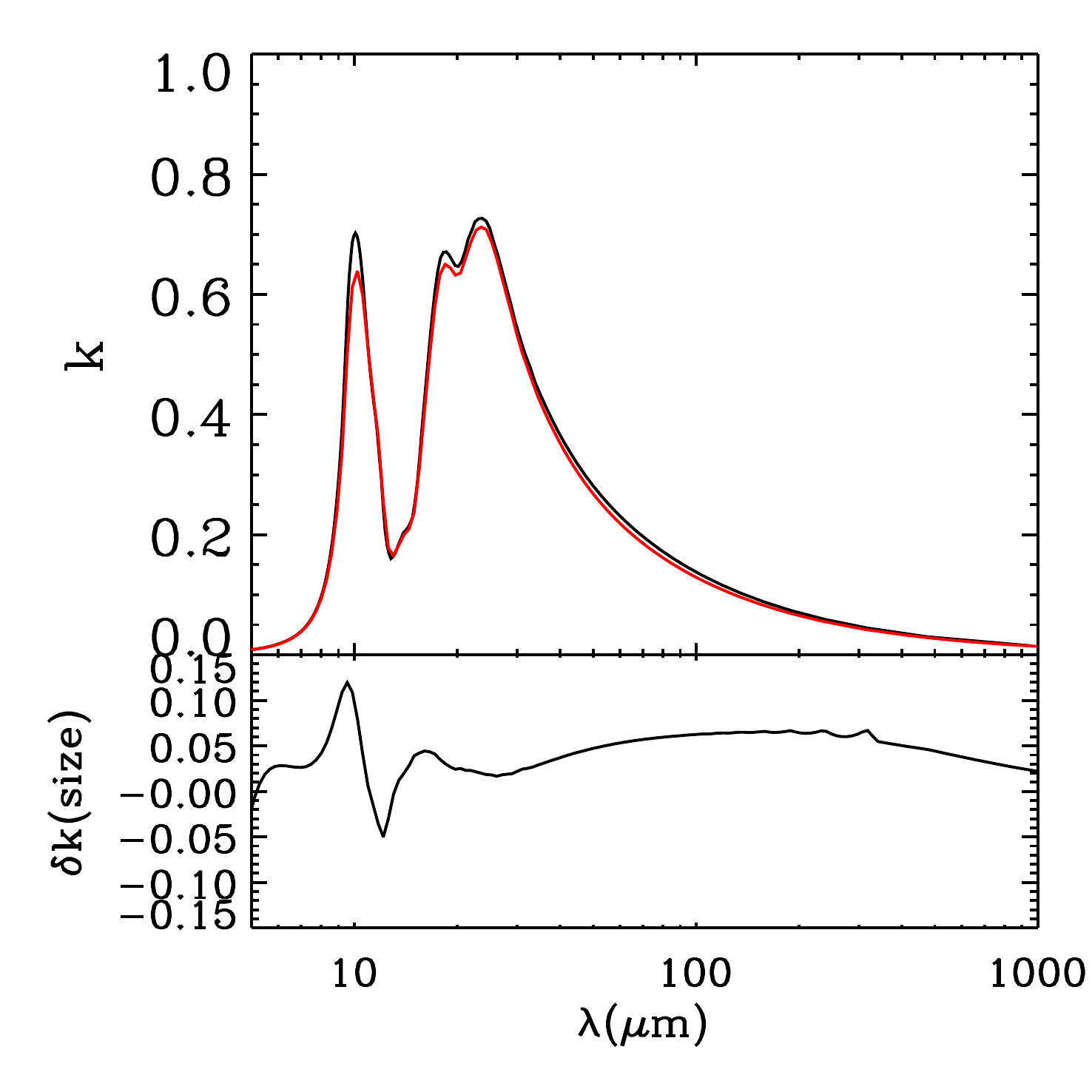}
\includegraphics[scale=0.4]{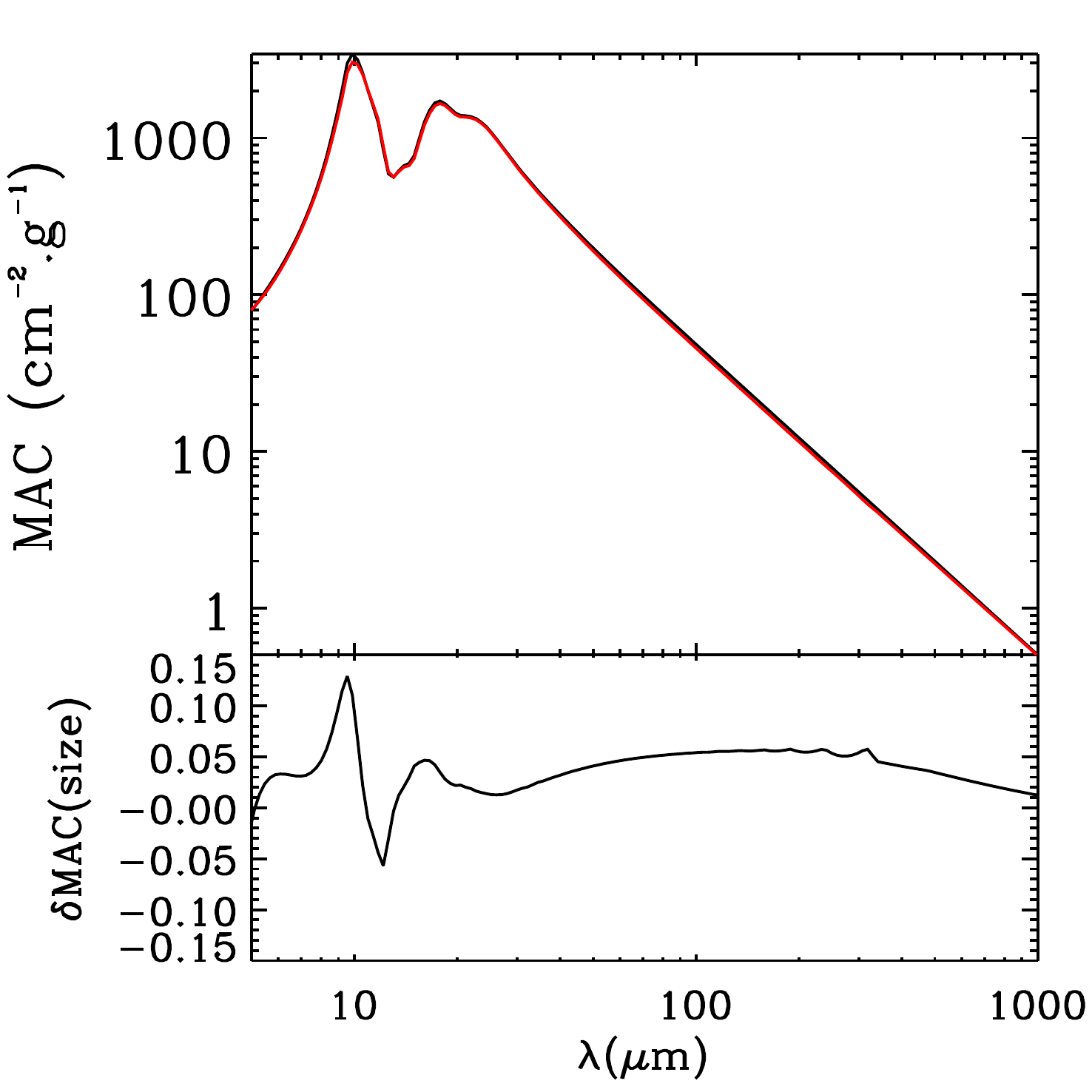}
\caption{Effect of neglecting the size distribution in the calculation of the optical constants. Calculations are made on a spectrum simulated using optical constants of amorphous forsterite silicates from \cite{jaeger2003} assuming a prolate grain shape with an axis ratio of 2 and a size distribution similar to the one measured experimentally. The top panels of the two left figures show the refractive index, $n$, and the absorption coefficient, $k$, from the database (black lines) and those derived from the simulated MAC (red lines); the bottoms panels show the error made on $n$ and $k$. The right figure shows the simulated MAC (black line) and the new MAC calculated from the $n$ and $k$ derived from the simulated MAC (top panel) and the error made on the MAC (bottom panel).}    
\label{test_size}% label for figure
\end{figure*}

Looking at the TEM images of the samples, it can be seen that only a minority of grains in the samples have very elongated or flattened shapes. We therefore assumed that the grains in the samples are characterised by axis ratios of $a/b = (a/b)_{\rm{ref}} \pm 40$\%. To investigate the effects of the choice of grain shape and elongation, we derived the optical constants assuming oblate and prolate grains with values of the axis ratio ranging from 1 (spherical grains) to 1.4 times the value adopted to compute the optical constants of each sample ($a/b = 1.5$ and 2 for the Mg-rich and Fe-rich samples, respectively). The results are presented in Figs.~\ref{test_shape_X} and ~\ref{test_shape_E} for the Mg- and Fe-rich samples, respectively. The errors for the 10\,K measurements are superimposed on the figures, allowing us to investigate the overall effect of the grain shape. For a given axis ratio, the shape of the grain, prolate or oblate, has little effect on the optical constants. On the contrary, for a given grain shape, the axis ratio has a strong impact on the calculated optical constants, in particular in the FIR. The derived uncertainty on the refractive index, $n$, is negligible at short wavelengths ($\lambda \leqslant 20 \, \mu$m) for all values of ($a/b$). It is smaller than $\sim 2$\% in the range $\sim 20 - 200 \, \mu$m and never exceeds 4\% in the FIR for $1 \leqslant a/b \leqslant 2.1-2.8$, depending on the sample. The uncertainty on $k$ is negligible for $\lambda \leqslant 30 \, \mu$m but increases with the wavelength and can be up to $\sim 10\%-15$\% for the Mg-rich samples for $a/b = 2.1$ and  up to $\sim 20\%-25$\% for the Fe-rich samples for $a/b = 2.8$, depending on the sample. The uncertainty on the calculated MAC follows the same trend as that for $k$. However, it is smaller than the uncertainty on $k$: for $\lambda \geqslant  30 \, \mu$m, ${\delta}{\kappa}_{\rm{shape}} \leqslant 5\%-10$\% for the Mg-rich samples and $a/b \leqslant 2.1$, and ${\delta}{\kappa}_{\rm{shape}} \leqslant 10\%-15$\% for $a/b \leqslant 2.8$ for the Fe-rich samples, depending on the sample.

The samples studied here consist of collections of grains of different sizes. In our calculation of the optical constants, we assume that the grains have a single size in the Rayleigh limit. We performed tests to investigate the effect of such an assumption. Using the optical constants of the amorphous forsterite-type silicates from \cite{jaeger2003} we simulated the MAC of a population of grains characterised by the measured size distribution. The absorption cross sections for each size are calculated with the discrete dipole approximation \citep[DDA,][]{purcell1973}, using version 7.3.3  of the ddscat routine described in \cite{flatau2012}, assuming prolate grains with axis ratios of 1.5 and 2 for the Mg- and Fe-rich samples, respectively. We calculated the optical constants from this simulated MAC and compared them with the original values (Fig.~\ref{test_size}). The errors on $n$, $k$, and the MAC were estimated in the same way as for the errors on $n_{\mathrm{vis}}$ and grain shape. For a prolate grain with $(a/b) = 2$, we find that the error induced by neglecting a size distribution effect is less than 4\% for the refractive index over the entire spectral range. It is larger for the extinction coefficient, $k$, and the MAC for which the error is $\sim 5\%\ - 13$\% in the $8-15 \, \mu$m range and $\leqslant 3\%-5$\% for $\lambda \geqslant 15 \, \mu$m. For a prolate grain shape with $a/b = 1.5$, the error follows the same trend with  wavelength but is slightly smaller ($\sim 1$\% less over the whole spectral domain).

Another approximation made in the calculation of the optical constants is that we neglect the fact that the grains can be agglomerated in the pellets. The main change induced by coagulation is an increase in dust extinction efficiency in the FIR. \cite{ysard2018} undertook a detailed study of the effects of grain coagulation on the extinction efficiencies from the visible to FIR domain as a function of monomer composition, size, shape, number of monomers, and aggregate geometry. Studies of grain coagulation generally assume that the grains are in vacuum, and the magnitude of the effect observed in vacuum may be different when the grains are embedded in a matrix, as is the case in this work. We therefore  calculated the extinction efficiency of coagulated grains embedded in a polyethylene matrix. Following the work of \cite{ysard2018}, we calculated the optical properties of ten different geometry aggregates, each made of six spherical monomers of   100\,nm  radius. We considered coagulated grains and also grains in close proximity that are not in contact,  the latter being representative of grains in a matrix. Finally, to mimic the fact that the grains have an irregular surface, we varied the size of the dipoles constituting the monomers. The calculations were performed at 250 and $500 \,\mu$m with the DDA \citep[][]{purcell1973} using the optical constants of the amorphous forsterite-type silicates from \cite{jaeger2003}. As can be seen in Fig.~\ref{test_coagulation},  the effect of coagulation is much weaker in PE than in vacuum. In the worst case of tightly bound aggregates (with half-radius overlapped grains), the extinction efficiency is 1.1 times greater than that of isolated grains while it is only 1.07 times larger for close-proximity grains and 1.04 for non-smooth grains. The effect of monomer shape should not change these results because, as noted by  \cite{ysard2018}, averaging over a shape distribution gives results close to those for spherical monomers. Furthermore, calculations show that the effect of grain shape is attenuated in the matrix as compared to vacuum. Considering that the error made on the opacity is at most 10\%, the uncertainty on the calculated optical constants is smaller than this factor. We therefore do not expect it to be greater than 10\% and, considering that the filling factor of the grains in the pellets is smaller than 0.15 and that we do not expect strong aggregation, it is most likely of the order of 7\%.

\begin{figure}[t]
\centering
\includegraphics[scale=0.32]{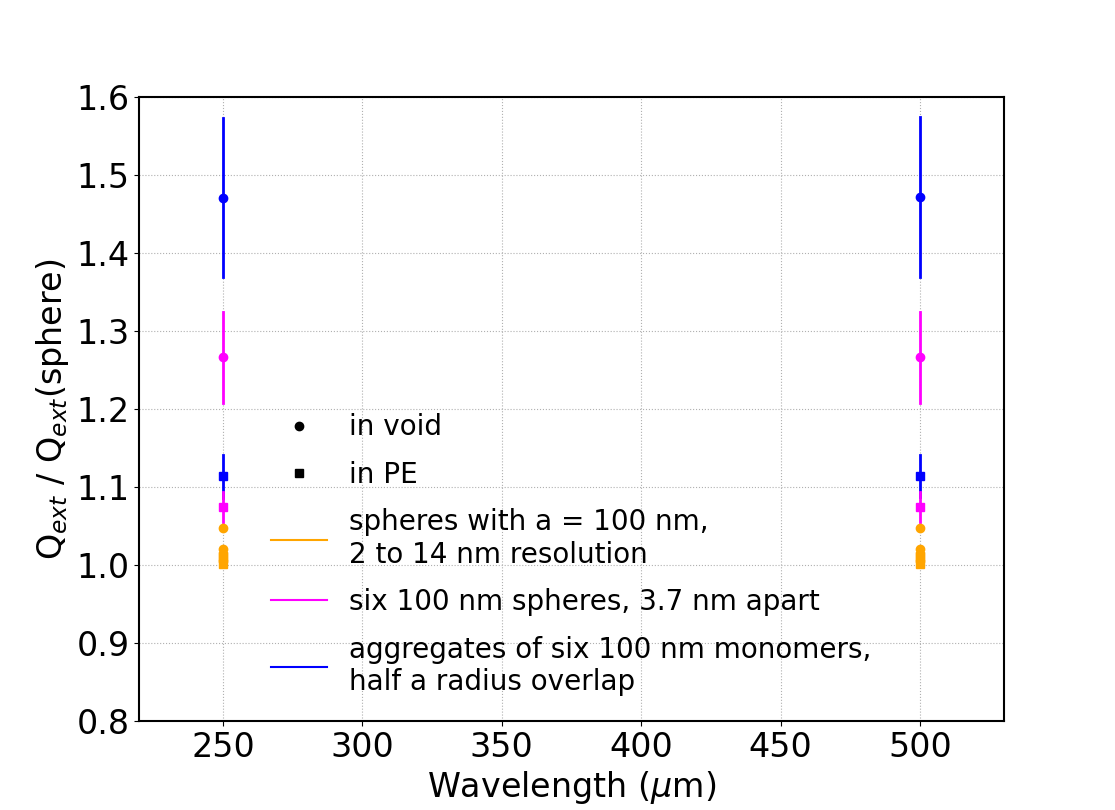}
\caption{Effect of coagulation on the extinction efficiency as a function of the surrounding medium. The extinction efficiency ratios of different types of aggregates and irregular spheres to that of isolated smooth spherical grains are plotted for grains in vacuum (filled circles) and in polyethylene (squares). The vertical bars indicate the dispersion of the results for the ten aggregate geometries. }    
\label{test_coagulation}% label for figure
\end{figure}

\subsection{Resulting total uncertainties}
\label{sect_uncertainties}

The quadratic sum of the three sources of error (on $n_{\rm{vis}}$, and the shape and size of the grains) gives an estimation of the uncertainty on the derived refractive index, absorption coefficient, and mass absorption coefficient. We  calculated the uncertainties for each sample, adopting an error of $\pm 5$\% on $n_{\rm{vis}}$ and of $\pm 40$\% on the shape (i.e., grains with axis ratio $a/b = 1 - 2.1$ for the Mg-rich glassy samples and  $a/b = 1 - 2.8$ for the Fe-rich solgel samples). The uncertainty originating from an error on the size distribution was not estimated from the measured samples but from a MAC simulated with the optical constants of the Mg$_2$SiO$_4$ sample from \cite{jaeger2003}. The wavelength dependence of the uncertainty is therefore characteristic of this specific sample and cannot be used for our measured samples as it induces an artificial spectral dependence into the uncertainty. This is indeed the case for the uncertainty on $k$ and on the MAC; that on $n$ is small enough to hide such an effect. To overcome this problem, we parametrised the uncertainty on $k$ and on the MAC as follows: for $\lambda \leqslant 8\,\mu$m we adopted an uncertainty of 5\%, for $8\,\mu$m $< \lambda \leqslant 15\,\mu$m an uncertainty of 12\%, and for $\lambda > 15\,\mu$m an uncertainty of 5\%. The total uncertainty was calculated for each sample and is shown in the Appendix (Figs.~\ref{netk_mgrich_error}, \ref{netk_ferich_error} and \ref{netk_ferich_reduced_error}). We find that, for all samples, the total uncertainty on $n$, ${\delta}n_{tot}$,  is $\sim 4\%-6$\% for the studied wavelength range and that it is dominated by the likely errors on the adopted values of $n_{\mathrm{vis}}$. The total uncertainty on $k$, ${\delta}k_{tot}$, is dominated by the error on the grain size in the MIR and by the error on the grain shape in the FIR. It is comparable for the Fe-rich samples and for the Mg-rich samples in the MIR: ${\delta}k_{tot} \sim 5$\% for $\lambda \leqslant 30\,\mu$m, with errors of 13\% at most in the 10 $\mu$m spectral feature. For wavelengths greater than $30\,\mu$m, ${\delta}k_{tot}$ is larger for the Fe-rich samples than for the Mg-rich samples. For the Fe-rich samples, ${\delta}k_{tot}^{Fe}$ $\sim 5-15$\% for  $30\,\mu$m $\leqslant \lambda \leqslant 200\,\mu$m and ${\delta}k_{tot}^{Fe}$ reaching 25\% at most for $\lambda > 200\,\mu$m, depending on the sample. For the Mg-rich samples, ${\delta}k_{tot}^{Mg}\leqslant 5\%-13$\% for $ \lambda \geqslant 30\,\mu$m, except for sample X40 for which it reaches 17\% at 1\,mm. As discussed above, in the FIR, an additional uncertainty from grain aggregation  in the pellets could be added, but as it has not been estimated over the entire spectral range, we have not included it in the calculation of the total uncertainties. However, we estimate that this will, at most, increase the total error in the FIR by up to 27\% for Fe-rich samples and by up to 13\%-16\% for the Mg-rich samples.

\section{Extrapolation to short and long wavelengths}
\label{sect_extrapolation}

For use in astronomical models, the optical constants ---calculated in the wavelength range from 5 to $1000\,\mu$m--- need to be extrapolated to shorter and longer wavelengths. To do so, we first extrapolate $k$ in the range $0.01 - 10^5\,\mu$m and then calculate $n$ on this spectral range using the Kramers-Kronig relations. 

The absorption band in the UV range, because of the interaction of the radiation field with electrons of the Si-O bonds, varies slightly with the structure and composition of the material (see \cite{siegel1974,kitamura2007} for the case of silica). \cite{nitsan1976} measured the optical constants of crystalline olivine for $\lambda \leqslant 0.2\,\mu$m. These data were used by \cite{scottduley1996} to provide optical constants of silicates in the UV to MIR range, data that are used in the THEMIS model. As no other relevant data are available in this domain, we use the value of $k$ from \cite{scottduley1996} for all samples. For $0.2\,\mu$m $\leqslant \lambda \leqslant 5.6\,\mu$m, we use $k$ values from Mg- and Fe-rich silicates from \citet{jaeger2003}. The absorption coefficient adopted in the UV domain closely matches  that adopted above $0.2\,\mu$m. 

At long wavelengths, we extrapolate $k$ with a power law whose spectral index is determined from $k$ calculated from the measurements in the range $500-1000\,\mu$m, depending on the sample. The details of the extrapolation for each sample are summarised in Table~\ref{table:extrapol}. We verified that the extrapolation on $k$  has no effect on $n$ or the MAC in the experimental spectral range. The difference in the extrapolated $n$ and MAC compared to the non-extrapolated $n$ and MAC never exceeds 1\% and 2\%, respectively. The extrapolated optical constants are shown in the Appendix (Fig.~\ref{netk_mgrich_extrapolated}, \ref{netk_ferich_extrapolated}, and ~\ref{netk_ferich_reduced_extrapolated}).

\begin{figure*}
\centering
\includegraphics[scale=0.45]{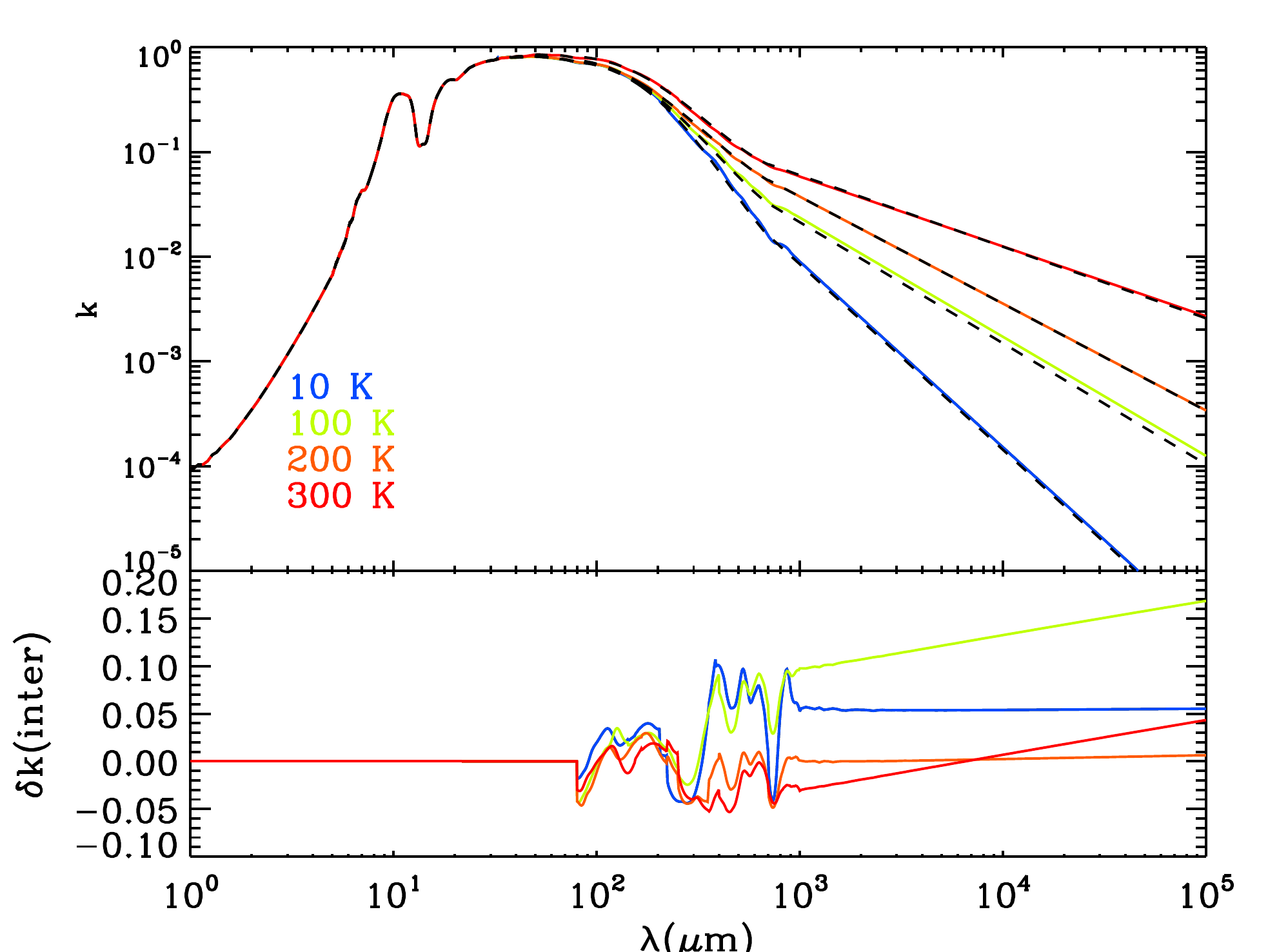}
\includegraphics[scale=0.45]{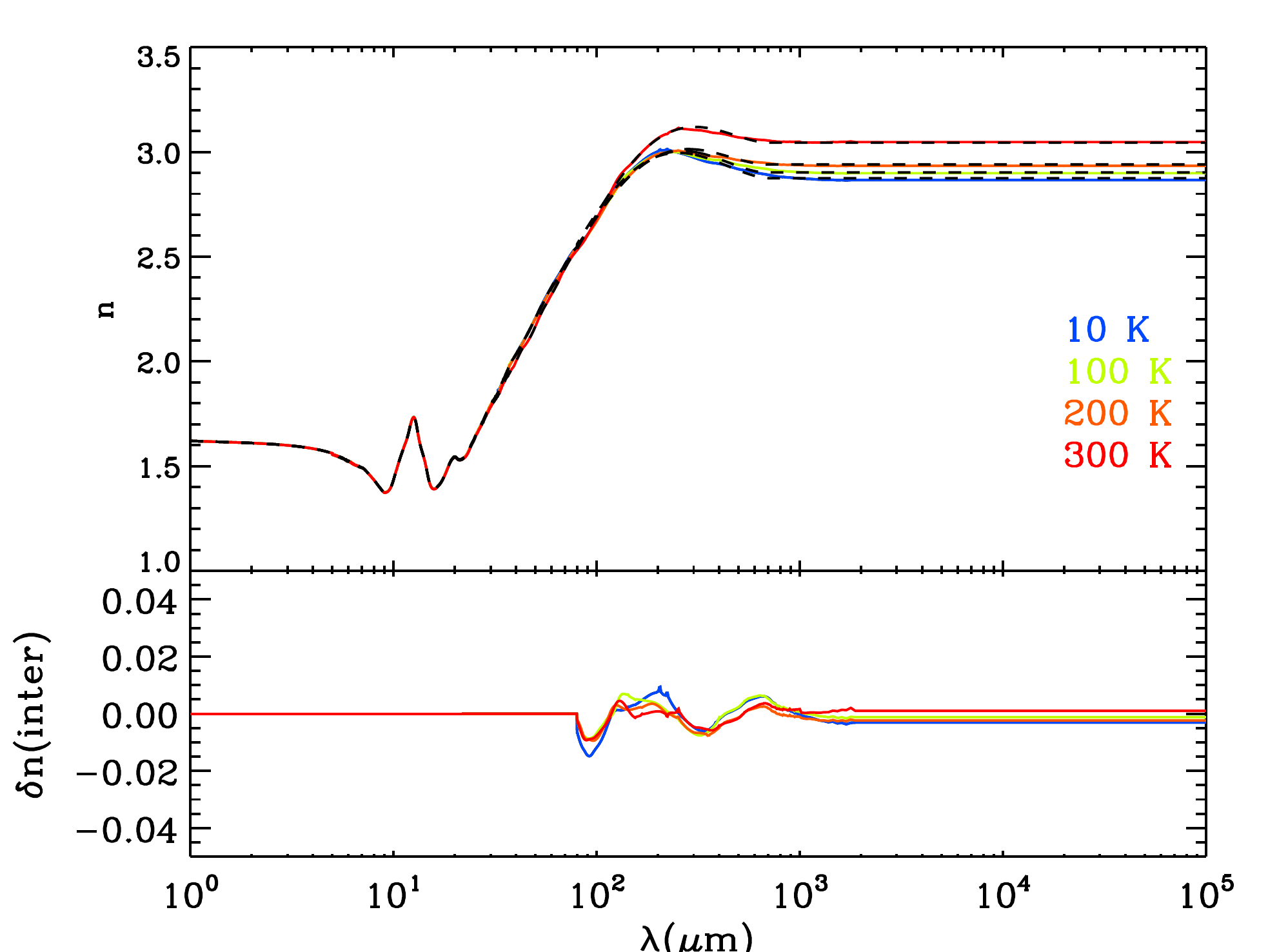}
\includegraphics[scale=0.39]{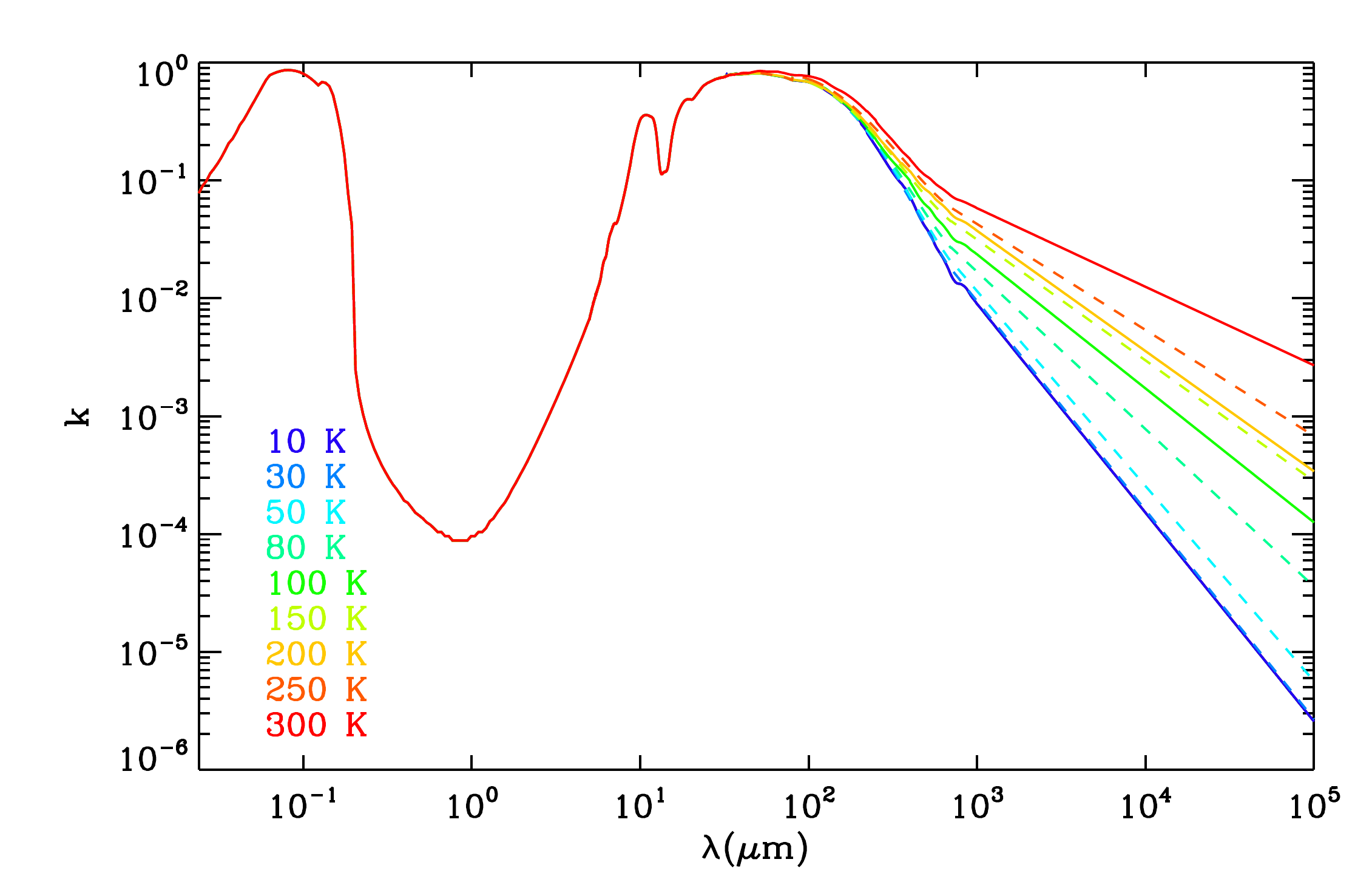}
\includegraphics[scale=0.39]{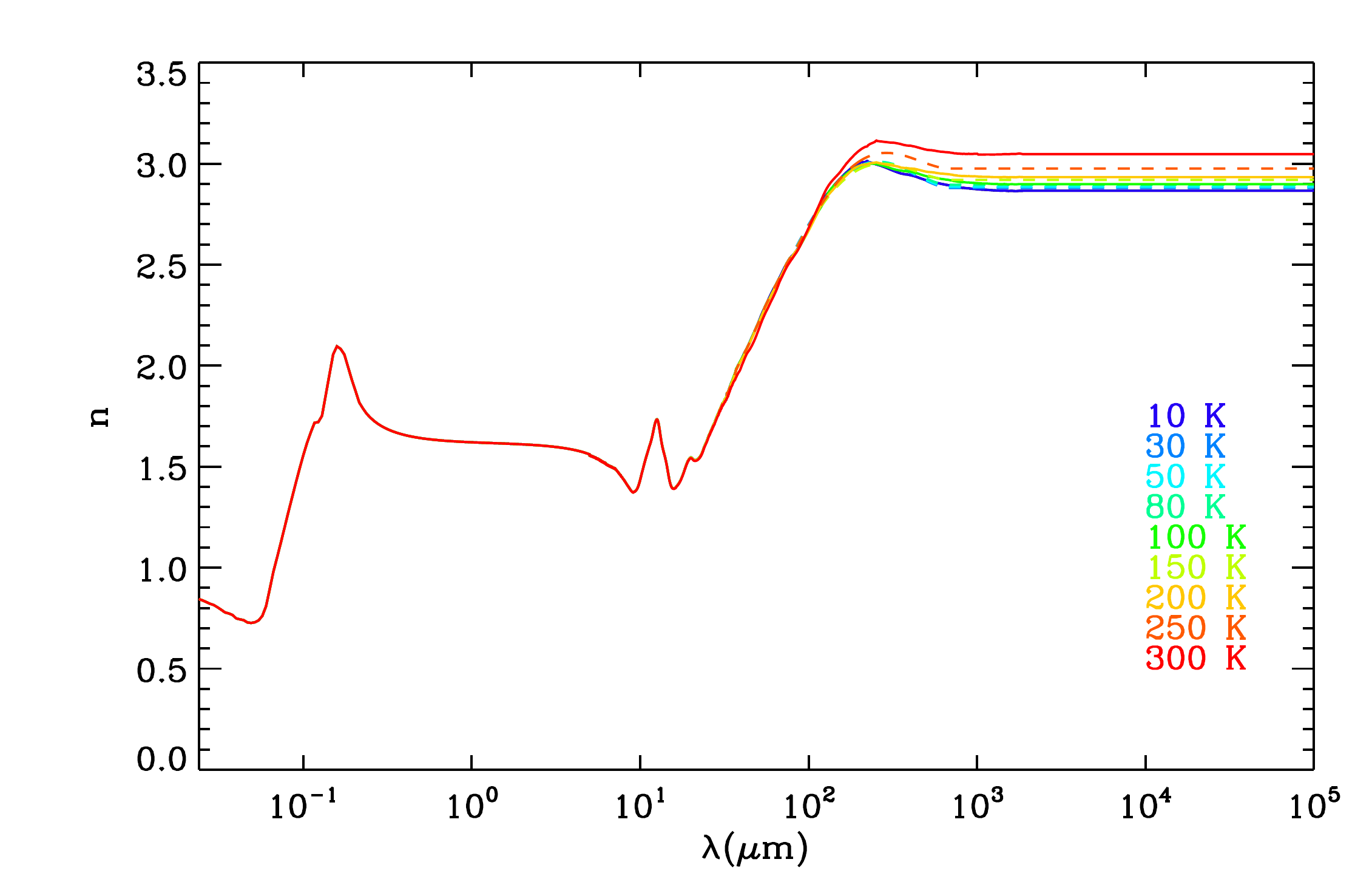}
\caption{Temperature interpolation of the optical constants for sample X35. The two upper panels show the comparison of $k$ (upper left) and $n$ (upper right) calculated from the measurements at 10, 100, 200, and 300 K (blue, green, orange, and red curves, respectively) with the interpolated ones (black dashed lines) together with their associated error. The two lower panels show $k$ (lower left) and $n$ (lower right) at temperatures from 10 K to 300K. The interpolated optical constants are shown by the dashed lines and those calculated from measurements are the continuous lines. }    
\label{fig_interpol_T}% label for figure
\end{figure*}

\section{Interpolation to any temperature}
\label{sect_interpol_T}

The optical constants of the silicate dust analogues were calculated for the temperatures at which the MACs were measured: 10, 30, 100, 200, and 300K.  However, it is of interest to interpolate the optical constants at temperatures other than those studied in the 10 - 300 K range, to enable their use in radiative transfer models, for example. We therefore provide a method for a polynomial interpolation of the derived optical constants to any temperature within the $10 - 300$\,K range.  This interpolation is performed over the wavelength region where the MAC varies with temperature, that is for $\lambda \geqslant 80\,\mu$m. The details of the interpolations are given in Appendix~\ref{Appendices_n_k_interpol_T} for all of the samples. To illustrate the results, Fig.~\ref{fig_interpol_T} shows the interpolated optical constants for the X35 sample. We verified that the agreement between the optical constants ---that is, of all of the samples--- calculated from the experimental data and the interpolated values is reasonable. We find that the error on $n$ is small, namely of the order a few percent for all samples. The error on $k$ depends on the sample and, for a given sample, on temperature and the wavelength. This is because of the complex spectral shape of $k$, which differs from sample to sample and from temperature to temperature. For simplicity, we use the same polynomial form for all samples. Overall, the error on $k$ is of the order of $5\%-20$\%, and is of $5\%\ -10$\% in the $5 - 1000\,\mu$m interval where the peak of the interstellar dust emission occurs. It should be emphasised that the interpolation functions are based on mathematical functions (polynomials) that allow the calculation of the refractive index and absorption coefficient at any temperature. They are not based on any physical models that could take into account thermally activated or relaxation processes, and that could lead to a more realistic or physical dependence of the data on temperature. In particular, the interpolated optical constants must be taken with caution for temperatures far from the experimental temperatures, or when temperature variations are studied in detail within a narrow temperature range around the experimental temperatures.

\section{Comparison with the optical constants of cosmic dust models and implications}
\label{sect_implications} % label for section

\begin{figure*}[!t]
\centering
  \includegraphics[scale=0.41, angle =90, trim={0.5cm 0.5cm 0.5cm 1.5cm}, clip]{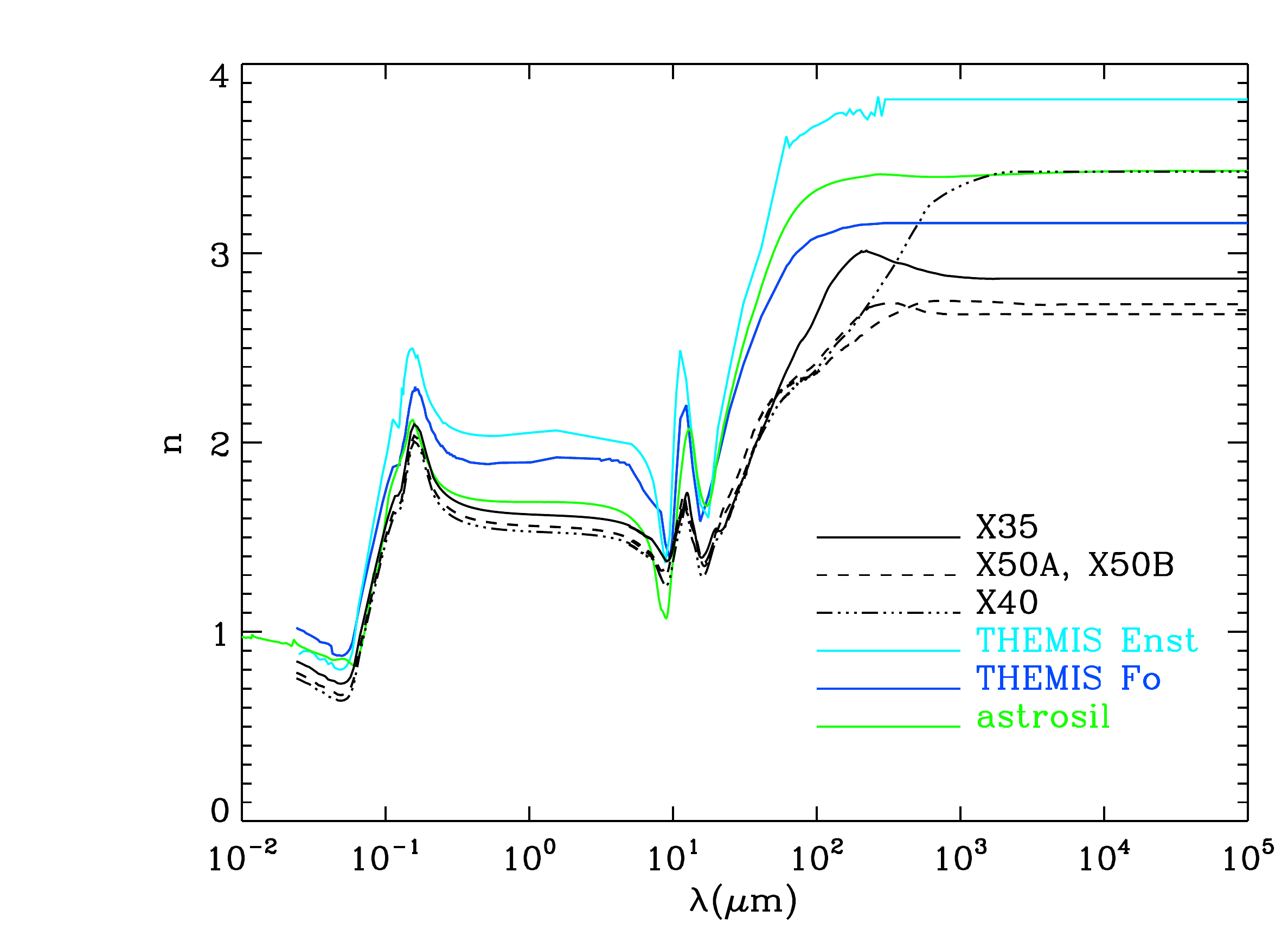}
  \includegraphics[scale=0.41, angle =90, trim={0.5cm 0.5cm 0.5cm 1.5cm}, clip]{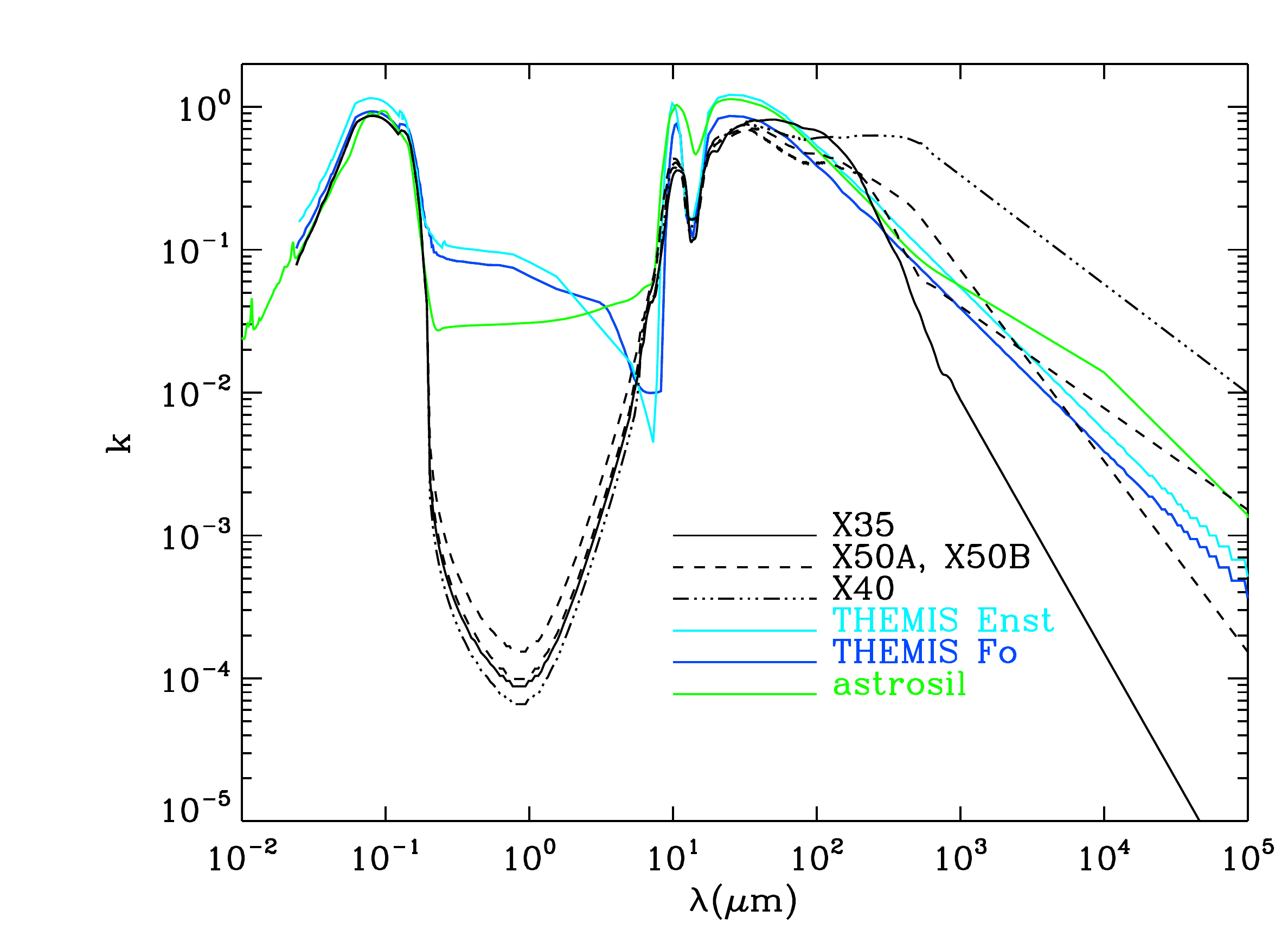}
          \caption{Comparison of the optical constants of magnesium-rich samples at 10\,K with the silicate components of cosmic dust models: X35 (continuous black line), X50A and X50B (dashed lines), and X40 (dotted-dashed line), compared to that for the amorphous forsterite-type (dark blue line) and enstatite-type (light blue line) silicates used in the THEMIS model  and to the {\it astrosil} (green line). }
 \label{fig_comp_nk_dust_model}
\end{figure*}

\begin{figure*}
\centering
  \includegraphics[scale=0.39, angle =90, trim={0.5cm 0.cm 0.5cm 1.5cm}, clip]{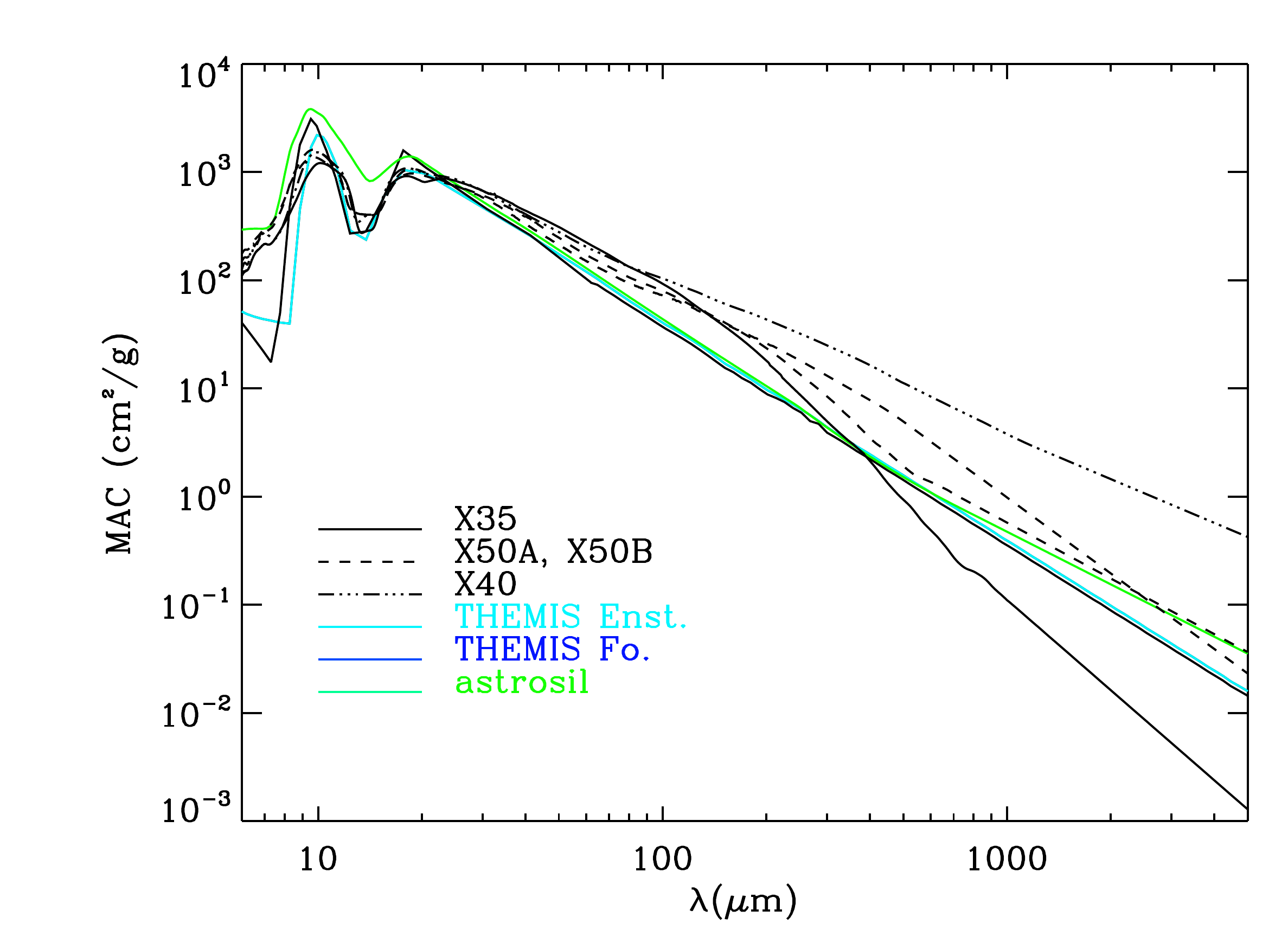}
 \includegraphics[scale=0.39, angle =90, trim={0.5cm 0.cm 0.5cm 0.cm}, clip]{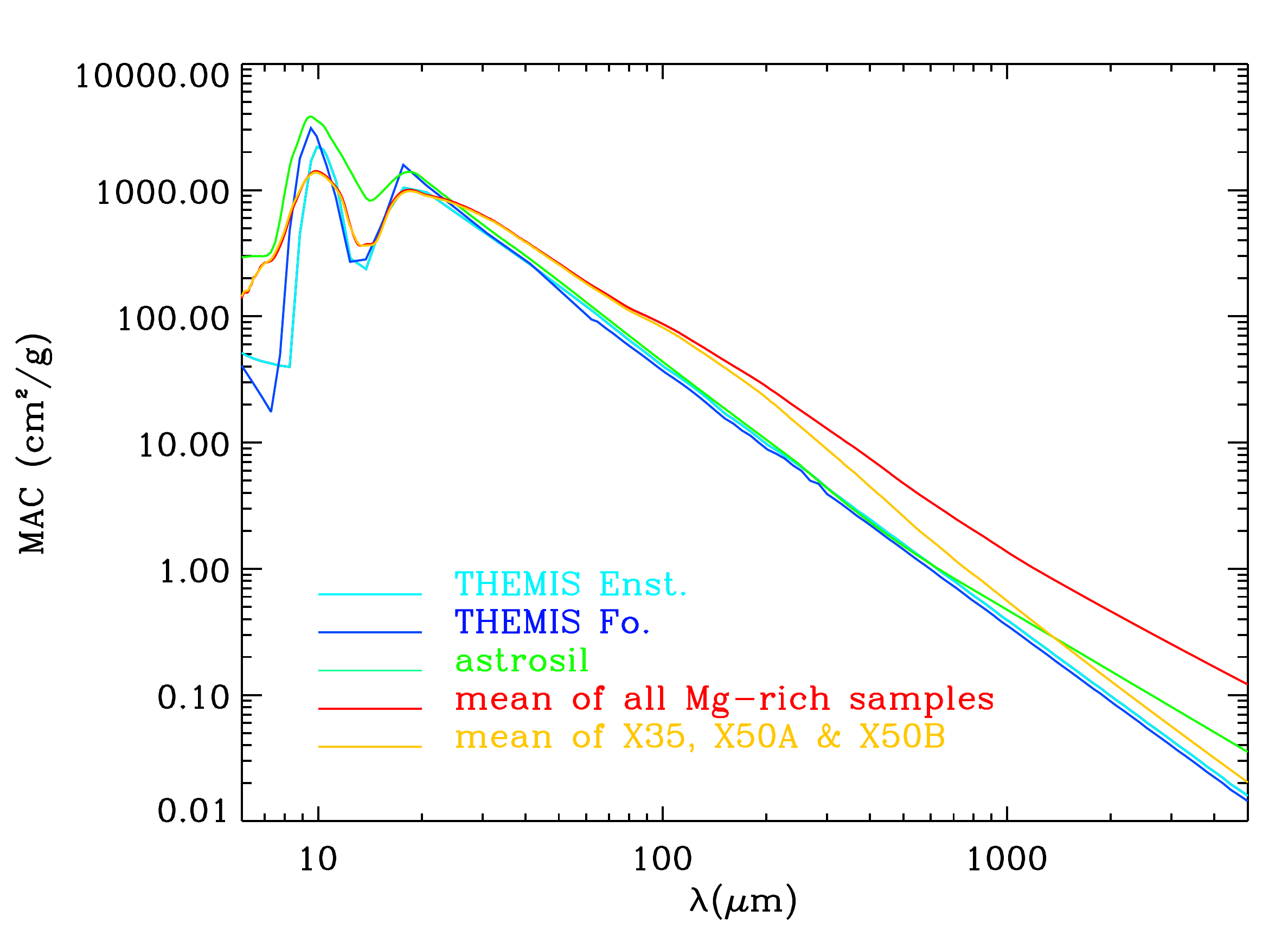}
  \caption{Comparison of the MAC for the silicate components of cosmic dust models with the MAC of the Mg-rich samples. The MACs are calculated for a prolate grain with an axis ratio $a/b = 1.5$ in  the Rayleigh limit. Left panel: Astrosil (green line) and THEMIS (dark and light blue lines) compared with each of the four Mg-rich silicate samples (dark lines). Right panel:  Astrosil (green line) and THEMIS (dark and light blue lines) compared with the average of the four Mg-rich samples (red line) and the average of the X35, X50A, and X50B samples (orange line).}
    \label{fig_comp_kappa_dust_model}% label for figure
\end{figure*}

\begin{table*}[!h]
\caption {Comparison of the averaged mass absorption coefficient of Mg-rich samples with the MAC for the silicate components of the THEMIS model and of the {\it astrosil}.} 
\label{table_kappa}
\begin{center}
\begin{tabular}{c c c c c c }
\hline 
\hline 
                                                                & 100 $\mu$m     &  250 $\mu$m &  500 $\mu$m   & 850 $\mu$m &    1 mm    \\   
\hline 
$\langle$ MAC1 $\rangle$  / MAC$_{\mathrm{{THEMIS}}}$  & 2.1 - 2.3      & 2.8 - 3.0               & 3.0 - 3.3                & 3.3 - 3.7    &   3.5 - 3.8 \\
$\langle$ MAC1 $\rangle$  / MAC$_{astrosil}$                     & 2.0           & 2.7                   & 3.1                      & 3.0                  &  2.9 \\
\hline 
$\langle$ MAC2 $\rangle$   / MAC$_{\mathrm{{THEMIS}}}$ & 2.0 - 2.2      & 2.1 - 2.2       & 1.6 - 1.8       & 1.5 - 1.6   &   1.4 - 1.5 \\
$\langle$ MAC2 $\rangle$  / MAC$_{astrosil}$                     & 1.9           & 2.0                   & 1.7                     &  1.3                 &  1.2  \\
\hline
\hline 
\end{tabular} 
\\
\end{center}
\tablefoot{ $\langle$MAC1$\rangle$ represents the MAC averaged over the four samples X35, X40, X50A, and X50B. $\langle$MAC2$\rangle$ represents the MAC averaged over the sample X35 and the mean of the X50A and X50B samples. The two values for $\langle$MAC1$\rangle$  / MAC$_{\mathrm{{THEMIS}}}$ correspond to the amorphous forsterite-type and enstatite-type silicates of the THEMIS model, respectively.}
\end{table*}

Figure~\ref{fig_comp_nk_dust_model}  compares the optical constants of the THEMIS dust model and of {\it astrosil} with those of the pure magnesium-rich samples. Overall, the refractive indices, $n$, derived in this study are smaller than those used in the dust models, except for the X40 sample which is probably different because it is a solid solution of the amorphous olivine-type and pyroxene-type silicates. The absorption coefficients, $k$, are the same as the ones used in models in the UV and VIS because the same extrapolation as for the dust models is used in this wavelength range. In the NIR range ($\sim 0.08-3\,\mu$m), $k$ is much less absorbent than in some dust models because these samples are Fe-free, whereas the silicates of the THEMIS model incorporates Fe and FeS inclusions into the silicates, and the {\it astrosil} data were modified to account for the observed higher opacity in the NIR in circumstellar shells \citep{draine1984}. The absorption coefficient is smaller than that of the dust models in the wavelength range of the stretching and bending modes of the silicates. However, it is greater in the red wing of the bending mode in the range $\sim 50 - 300\,\mu$m. As discussed in \cite{demyk2017a, demyk2017b}, this enhancement of the MAC is due to the intrinsic physical properties of the grain materials and is probably related to the presence of defects and porosity. Above $\sim 1$\,mm, the absorption coefficient of the enstatite-like samples X50A and X50B is comparable to that of the cosmic dust model silicates, whereas the olivine-like sample X35 exhibits a smaller absorption coefficient characterised by a steeper slope, and the peculiar X40 sample shows a much higher $k$ than the silicates of the cosmic dust models. 

These differences in the optical constants will have an important impact on the MACs calculated for different grain sizes and shape distributions. The left panel of Fig.~\ref{fig_comp_kappa_dust_model} shows the MAC calculated in the Rayleigh limit  for a prolate grain with $a/b = 1.5$ and  for each set of optical constants shown in Fig.~\ref{fig_comp_nk_dust_model}. The differences in the MAC for the Mg-rich silicate analog compared to the MACs for the silicate components of the {\it astrosil} and THEMIS dust models are similar to the difference in the absorption coefficient, $k$, shown in Fig.~\ref{fig_comp_nk_dust_model}. The choice of the model dataset to be used clearly  depends upon the astrophysical environments studied. With regard to the cosmic dust composition, it is almost certainly made of grains of a range of compositions and structures and therefore considering a combination of several materials is reasonable, keeping in mind that the mixture should respect the cosmic elemental abundance constraints. The right panel of Fig.~\ref{fig_comp_kappa_dust_model} compares the MAC of the {\it astrosil} model and that of the silicate components of the THEMIS dust model with that averaged over the four Mg-rich silicate analogues ($\langle MAC1 \rangle$, red line) and over the olivine-type and two enstatite-type stoichiometries ($\langle MAC2 \rangle$, orange line). This figure shows that the MAC averaged over different compositions is higher than the MAC of the silicate components of the cosmic dust models. We note that this is not true for the {\it astrosil} beyond 1.4 mm because {\it astrosil} was modified at long wavelengths to be more absorbent. Table~\ref{table_kappa} gives the value of the ratios between the averaged MAC and the MAC of the cosmic dust models. This ratio varies with the wavelength and the mixture of analogues adopted. At 100 $\mu$m, the Mg-rich silicate analogues are twice as absorbent as the cosmic dust model silicates,  whereas for $\lambda \geqslant 500\,\mu$m they are $\sim 3$ to 3.5 and $\sim 1.2$ to 1.8 times more absorbent for the average of the four Mg-rich samples ($\langle MAC1 \rangle$) and the olivine-type and enstatite-type samples  ($\langle MAC2 \rangle$),  respectively.  

These differences in the MACs, and therefore in the emissivities of the silicates, will have an impact on the modelling of dust emission and on the derived physical parameters, such as the dust mass and temperature estimated from fitting spectral energy distributions. For example, the higher emissivity in the FIR could help to explain the so-called submm  ($\sim 850\,\mu$m) dust excess, which may be, at least partly, due to a deficiency of the current dust models rather than an actual dust emission excess. It is therefore important that cosmic dust models take into account these new laboratory data on silicate dust analogues, which better reflect the physical behaviour of the dust as a function of wavelength and temperature. In addition, the errors on $n$ and $k$ derived in this work need to be fed into optical property calculations for the extinction, absorption, and scattering efficiencies used in cosmic dust models. The impact of these uncertainties on the derived parameters therefore needs to be evaluated. This will be the case for the new version of the THEMIS model which uses some of these optical constants in its amorphous silicate dust component (Ysard et al. 2022).

\section{Conclusions}

We calculated the optical constants of 12 amorphous silicate dust analogues over the wavelength range from $5\,\mu$m to 600 or 1000\,$\mu$m, depending on the sample. For each analog, the optical constants were calculated for grains at 10, 100, 200, and 300\,K and at 30 K for the Mg-Fe samples (the optical constants at 10 and 30\,K are identical for the Mg-rich samples). In order to aid cosmic dust modelling, we provide extrapolations of these optical constants outside the experimental wavelength domain, that is from $0.01\,\mu$m to $10^5\,\mu$m. We also provide a method to interpolate the optical constants at any given temperature within the range $10 - 300\,$K. These 12 sets of optical constants of the amorphous silicate dust analogues, their associated uncertainties, and the extrapolated optical constants are publicly available through the STOPCODA (SpecTroscopy and Optical Properties of Cosmic Dust analogues) database of the SSHADE infrastructure of solid spectroscopy. 

The amorphous silicate analog materials cover a wide range of compositions: from magnesium-rich silicates to silicates containing both iron and magnesium, from olivine-type to pyroxene-type stoichiometry, and from glassy to chaotic  microstructures. These optical constants and their uncertainties can be used to model the dust emission and extinction for any given size and shape distribution and in many astrophysical environments, such as in the ISM of our own Galaxy and in external galaxies, and in the circumstellar envelopes and disks around evolved and young stars. These interstellar silicate analogues are more emissive at long wavelengths than the more commonly used  silicates in current astronomical dust models. Therefore, the dust masses derived from spectral-energy-distribution fitting will be significantly affected and this could perhaps help to explain the so-called emission excess at long wavelengths, which may simply be due to a deficiency of the current dust models.

%_____________________________________________________________

%_____________________________________________________________

\begin{acknowledgements}
 This work was supported by the French \emph{Agence Nationale pour la recherche} project ANR-CIMMES and  by the Programme National “Physique et Chimie du Milieu Interstellaire” (PCMI) of CNRS/INSU with INC/INP co-funded by CEA and CNES.
 \end{acknowledgements}
 
\bibliographystyle{aa}
\bibliography{bib.bib}

\begin{thebibliography}{36}
\expandafter\ifx\csname natexlab\endcsname\relax\def\natexlab#1{#1}\fi

\bibitem[{{Bohren} \& {Huffman}(1998)}]{bohren1998}
{Bohren}, C.~F. \& {Huffman}, D.~R. 1998, {Absorption and Scattering of Light
  by Small Particles}, ed. {Bohren, C.~F.~\& Huffman, D.~R.}

\bibitem[{{Boudet} {et~al.}(2005){Boudet}, {Mutschke}, {Nayral}, {J{\"a}ger},
  {Bernard}, {Henning}, \& {Meny}}]{boudet2005}
{Boudet}, N., {Mutschke}, H., {Nayral}, C., {et~al.} 2005, \apj, 633, 272

\bibitem[{{Brubach} {et~al.}(2010){Brubach}, {Manceron}, {Rouzi{\`e}res},
  {Pirali}, {Balcon}, {Tchana}, {Boudon}, {Tudorie}, {Huet}, {Cuisset}, \&
  {Roy}}]{brubach2010}
{Brubach}, J., {Manceron}, L., {Rouzi{\`e}res}, M., {et~al.} 2010, in American
  Institute of Physics Conference Series, Vol. 1214, American Institute of
  Physics Conference Series, ed. {A.~Predoi-Cross \& B.~E.~Billinghurst},
  81--84

\bibitem[{{Compi{\`e}gne} {et~al.}(2011){Compi{\`e}gne}, {Verstraete}, {Jones},
  {Bernard}, {Boulanger}, {Flagey}, {Le Bourlot}, {Paradis}, \&
  {Ysard}}]{compiegne2011}
{Compi{\`e}gne}, M., {Verstraete}, L., {Jones}, A., {et~al.} 2011, \aap, 525,
  A103+

\bibitem[{{Coupeaud} {et~al.}(2011){Coupeaud}, {Demyk}, {Meny}, {Nayral},
  {Delpech}, {Leroux}, {Depecker}, {Creff}, {Brubach}, \& {Roy}}]{coupeaud2011}
{Coupeaud}, A., {Demyk}, K., {Meny}, C., {et~al.} 2011, \aap, 535, A124

\bibitem[{{Demyk} {et~al.}(2017{\natexlab{a}}){Demyk}, {Meny}, {Leroux},
  {Depecker}, {Brubach}, {Roy}, {Nayral}, {Ojo}, \& {Delpech}}]{demyk2017a}
{Demyk}, K., {Meny}, C., {Leroux}, H., {et~al.} 2017{\natexlab{a}}, \aap, 606,
  A50

\bibitem[{{Demyk} {et~al.}(2017{\natexlab{b}}){Demyk}, {Meny}, {Lu},
  {Papatheodorou}, {Toplis}, {Leroux}, {Depecker}, {Brubach}, {Roy}, {Nayral},
  {Ojo}, {Delpech}, {Paradis}, \& {Gromov}}]{demyk2017b}
{Demyk}, K., {Meny}, C., {Lu}, X.-H., {et~al.} 2017{\natexlab{b}}, \aap, 600,
  A123

\bibitem[{{Desert} {et~al.}(1990){Desert}, {Boulanger}, \&
  {Puget}}]{desert1990}
{Desert}, F., {Boulanger}, F., \& {Puget}, J.~L. 1990, \aap, 237, 215

\bibitem[{{Dorschner} {et~al.}(1995){Dorschner}, {Begemann}, {Henning},
  {Jaeger}, \& {Mutschke}}]{dorschner1995}
{Dorschner}, J., {Begemann}, B., {Henning}, T., {Jaeger}, C., \& {Mutschke}, H.
  1995, \aap, 300, 503

\bibitem[{{Draine} \& {Fraisse}(2009)}]{draine2009}
{Draine}, B.~T. \& {Fraisse}, A.~A. 2009, \apj, 696, 1

\bibitem[{{Draine} \& {Hensley}(2021)}]{draine2021}
{Draine}, B.~T. \& {Hensley}, B.~S. 2021, \apj, 909, 94

\bibitem[{{Draine} \& {Lee}(1984)}]{draine1984}
{Draine}, B.~T. \& {Lee}, H.~M. 1984, \apj, 285, 89

\bibitem[{{Draine} \& {Li}(2007)}]{draine2007}
{Draine}, B.~T. \& {Li}, A. 2007, \apj, 657, 810

\bibitem[{{Flatau} \& {Draine}(2012)}]{flatau2012}
{Flatau}, P.~J. \& {Draine}, B.~T. 2012, Optics Express, 20, 1247

\bibitem[{{Guillet} {et~al.}(2018){Guillet}, {Fanciullo}, {Verstraete},
  {Boulanger}, {Jones}, {Miville-Desch{\^e}nes}, {Ysard}, {Levrier}, \&
  {Alves}}]{guillet2018}
{Guillet}, V., {Fanciullo}, L., {Verstraete}, L., {et~al.} 2018, \aap, 610, A16

\bibitem[{{Hensley} \& {Draine}(2021)}]{hensley2021}
{Hensley}, B.~S. \& {Draine}, B.~T. 2021, \apj, 906, 73

\bibitem[{Hubert {et~al.}(2017)Hubert, Herbin, Visez, Pujol, \&
  Petitprez}]{hubert2017}
Hubert, P., Herbin, H., Visez, N., Pujol, O., \& Petitprez, D. 2017, Journal of
  Quantitative Spectroscopy and Radiative Transfer, 200, 320

\bibitem[{{Huffman} \& {Stapp}(1973)}]{huffman1973}
{Huffman}, D.~R. \& {Stapp}, J.~L. 1973, in Interstellar Dust and Related
  Topics, ed. J.~M. {Greenberg} \& H.~C. {van de Hulst}, Vol.~52, 297

\bibitem[{{J{\"a}ger} {et~al.}(2003){J{\"a}ger}, {Fabian}, {Schrempel},
  {Dorschner}, {Henning}, \& {Wesch}}]{jaeger2003}
{J{\"a}ger}, C., {Fabian}, D., {Schrempel}, F., {et~al.} 2003, \aap, 401, 57

\bibitem[{{J\"ager} {et~al.}(1994){J\"ager}, {Mutschke}, {Begemann},
  {Dorschner}, \& {Henning}}]{jaeger1994}
{J\"ager}, C., {Mutschke}, H., {Begemann}, B., {Dorschner}, J., \& {Henning},
  T. 1994, \aap, 292, 641

\bibitem[{{Jones} {et~al.}(2013){Jones}, {Fanciullo}, {K{\"o}hler},
  {Verstraete}, {Guillet}, {Bocchio}, \& {Ysard}}]{jones2013}
{Jones}, A.~P., {Fanciullo}, L., {K{\"o}hler}, M., {et~al.} 2013, \aap, 558,
  A62

\bibitem[{{Jones} {et~al.}(2017){Jones}, {K{\"o}hler}, {Ysard}, {Bocchio}, \&
  {Verstraete}}]{jones2017}
{Jones}, A.~P., {K{\"o}hler}, M., {Ysard}, N., {Bocchio}, M., \& {Verstraete},
  L. 2017, \aap, 602, A46

\bibitem[{Kitamura {et~al.}(2007)Kitamura, Pilon, \& Jonasz}]{kitamura2007}
Kitamura, R., Pilon, L., \& Jonasz, M. 2007, Appl. Opt., 46, 8118

\bibitem[{{K{\"o}hler} {et~al.}(2014){K{\"o}hler}, {Jones}, \&
  {Ysard}}]{koehler2014}
{K{\"o}hler}, M., {Jones}, A., \& {Ysard}, N. 2014, \aap, 565, L9

\bibitem[{{Li} \& {Draine}(2001)}]{li2001}
{Li}, A. \& {Draine}, B.~T. 2001, \apj, 554, 778

\bibitem[{{Mennella} {et~al.}(1998){Mennella}, {Brucato}, {Colangeli},
  {Palumbo}, {Rotundi}, \& {Bussoletti}}]{mennella1998}
{Mennella}, V., {Brucato}, J.~R., {Colangeli}, L., {et~al.} 1998, \apj, 496,
  1058

\bibitem[{Nitsan \& Shankland(1976)}]{nitsan1976}
Nitsan, U. \& Shankland, T.~J. 1976, Geophysical Journal International, 45, 59

\bibitem[{{Purcell} \& {Pennypacker}(1973)}]{purcell1973}
{Purcell}, E.~M. \& {Pennypacker}, C.~R. 1973, \apj, 186, 705

\bibitem[{{Schmitt} {et~al.}(2018){Schmitt}, {Bollard}, {Garenne}, {Albert},
  {Bonal}, \& {Poch}}]{schmitt2018}
{Schmitt}, B., {Bollard}, P., {Garenne}, A., {et~al.} 2018, in European
  Planetary Science Congress, EPSC2018--529

\bibitem[{{Scott} \& {Duley}(1996)}]{scottduley1996}
{Scott}, A. \& {Duley}, W.~W. 1996, \apjs, 105, 401

\bibitem[{{Siebenmorgen} {et~al.}(2017){Siebenmorgen}, {Voshchinnikov},
  {Bagnulo}, \& {Cox}}]{siebenmorgen2017}
{Siebenmorgen}, R., {Voshchinnikov}, N.~V., {Bagnulo}, S., \& {Cox}, N.~L.~J.
  2017, \planss, 149, 64

\bibitem[{Siegel(1974)}]{siegel1974}
Siegel, G. 1974, Journal of Non-Crystalline Solids, 13, 372

\bibitem[{{Thompson} {et~al.}(2016){Thompson}, {Demyk}, {Day}, {Evans},
  {Leroux}, {Depecker}, {Parker}, {Connor}, {Wilhelm}, \&
  {Cibin}}]{thompson2016}
{Thompson}, S.~P., {Demyk}, K., {Day}, S.~J., {et~al.} 2016, Journal of Non
  Crystalline Solids, 447, 255

\bibitem[{{van de Hulst}(1957)}]{vandehulst1957}
{van de Hulst}, H.~C. 1957, {Light Scattering by Small Particles}

\bibitem[{{Ysard} {et~al.}(2018){Ysard}, {Jones}, {Demyk}, {Bout{\'e}raon}, \&
  {Koehler}}]{ysard2018}
{Ysard}, N., {Jones}, A.~P., {Demyk}, K., {Bout{\'e}raon}, T., \& {Koehler}, M.
  2018, \aap, 617, A124

\bibitem[{{Zubko} {et~al.}(2004){Zubko}, {Dwek}, \& {Arendt}}]{zubko2004}
{Zubko}, V., {Dwek}, E., \& {Arendt}, R.~G. 2004, \apjs, 152, 211

\end{thebibliography}

\FloatBarrier

\appendix

\section{Optical constants and estimated uncertainties}
\label{uncertainty_n_k}

\begin{figure*}[!ht]
\centering
 \includegraphics[scale=.33, trim={0 1cm 0 1.5cm}, clip]{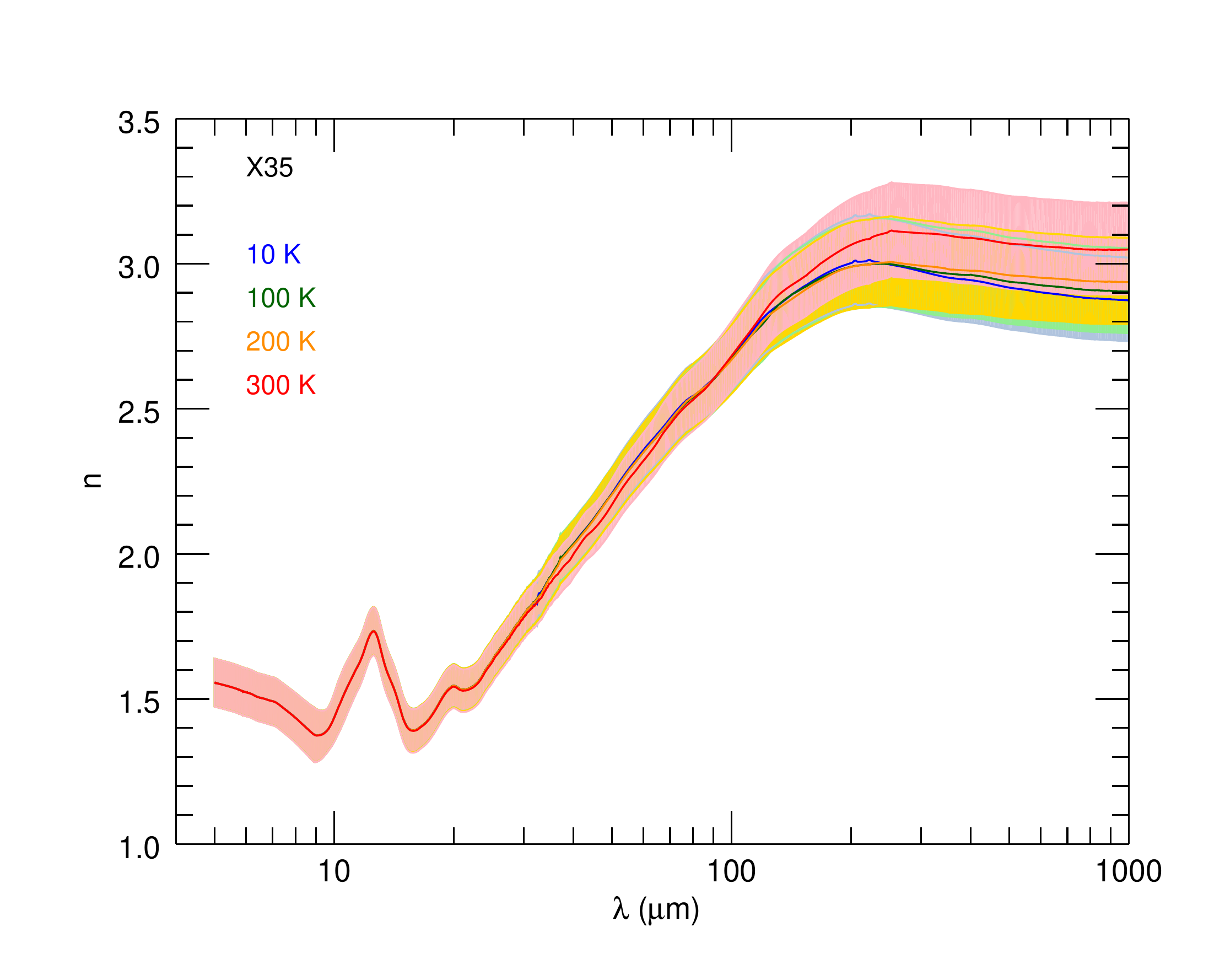}
 \includegraphics[scale=.33, trim={0 1cm 0 1.5cm}, clip]{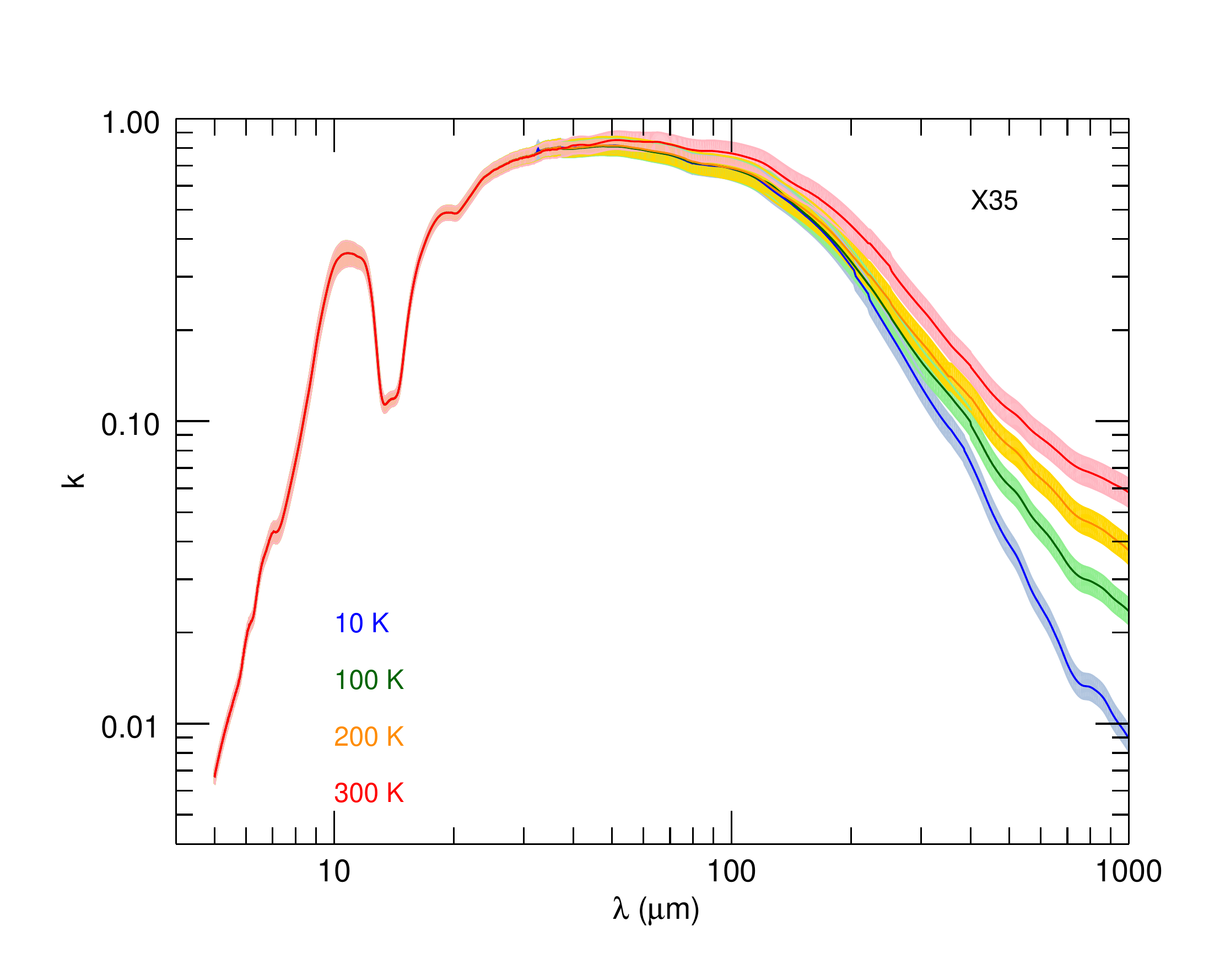}
 \includegraphics[scale=.33, trim={0 1cm 0 1.5cm}, clip]{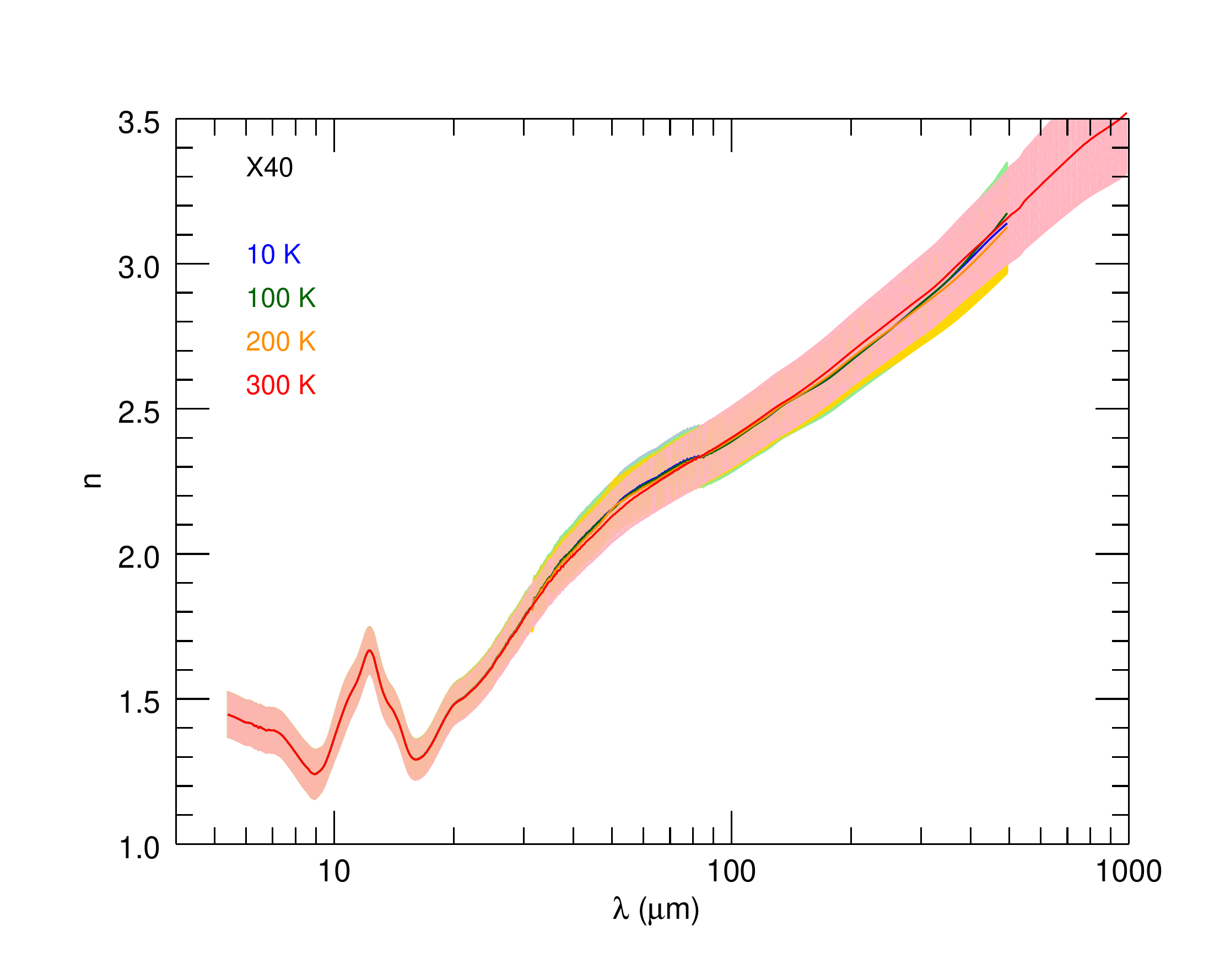}
 \includegraphics[scale=.33, trim={0 1cm 0 1.5cm}, clip]{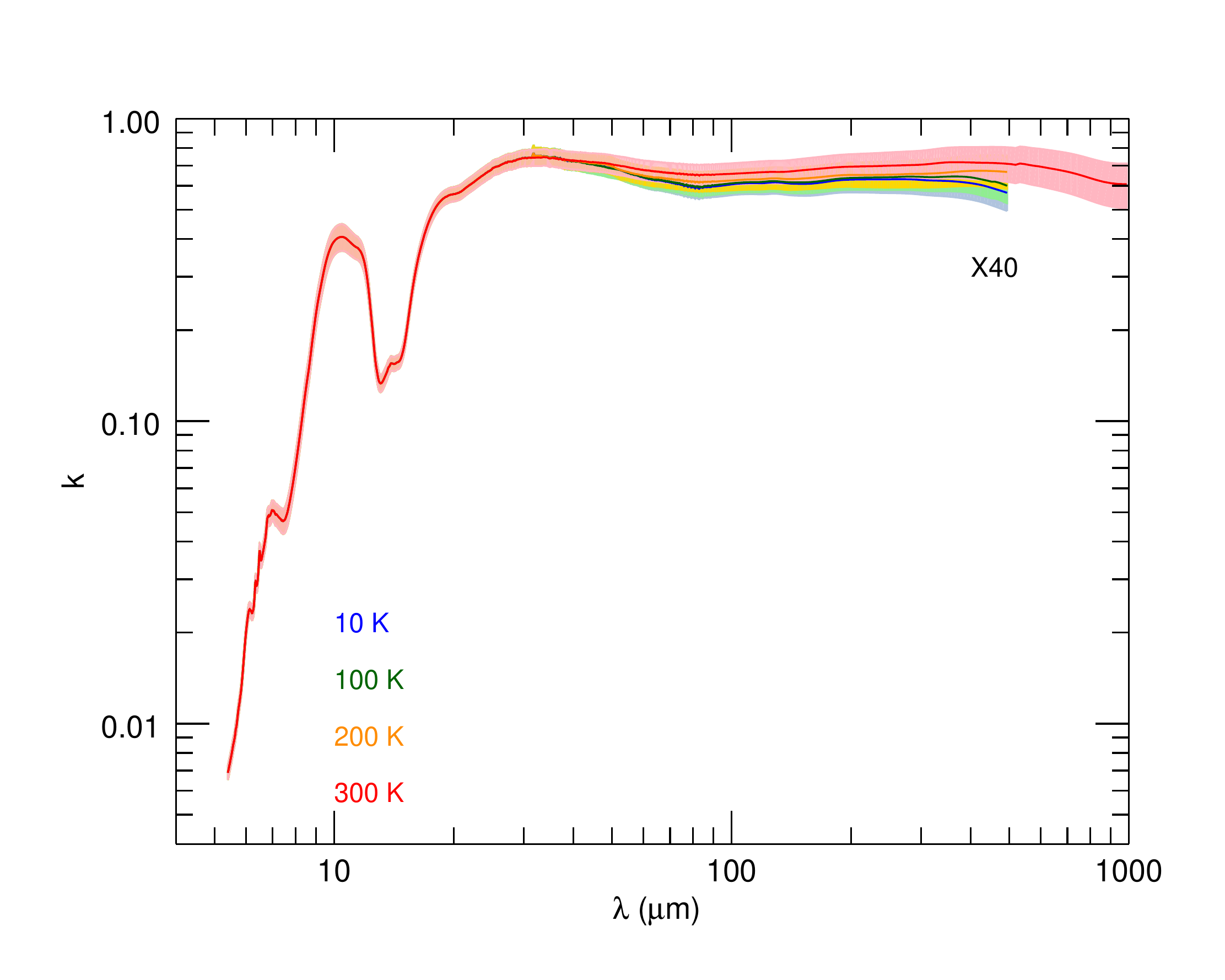}
  \includegraphics[scale=.33, trim={0 1cm 0 1.5cm}, clip]{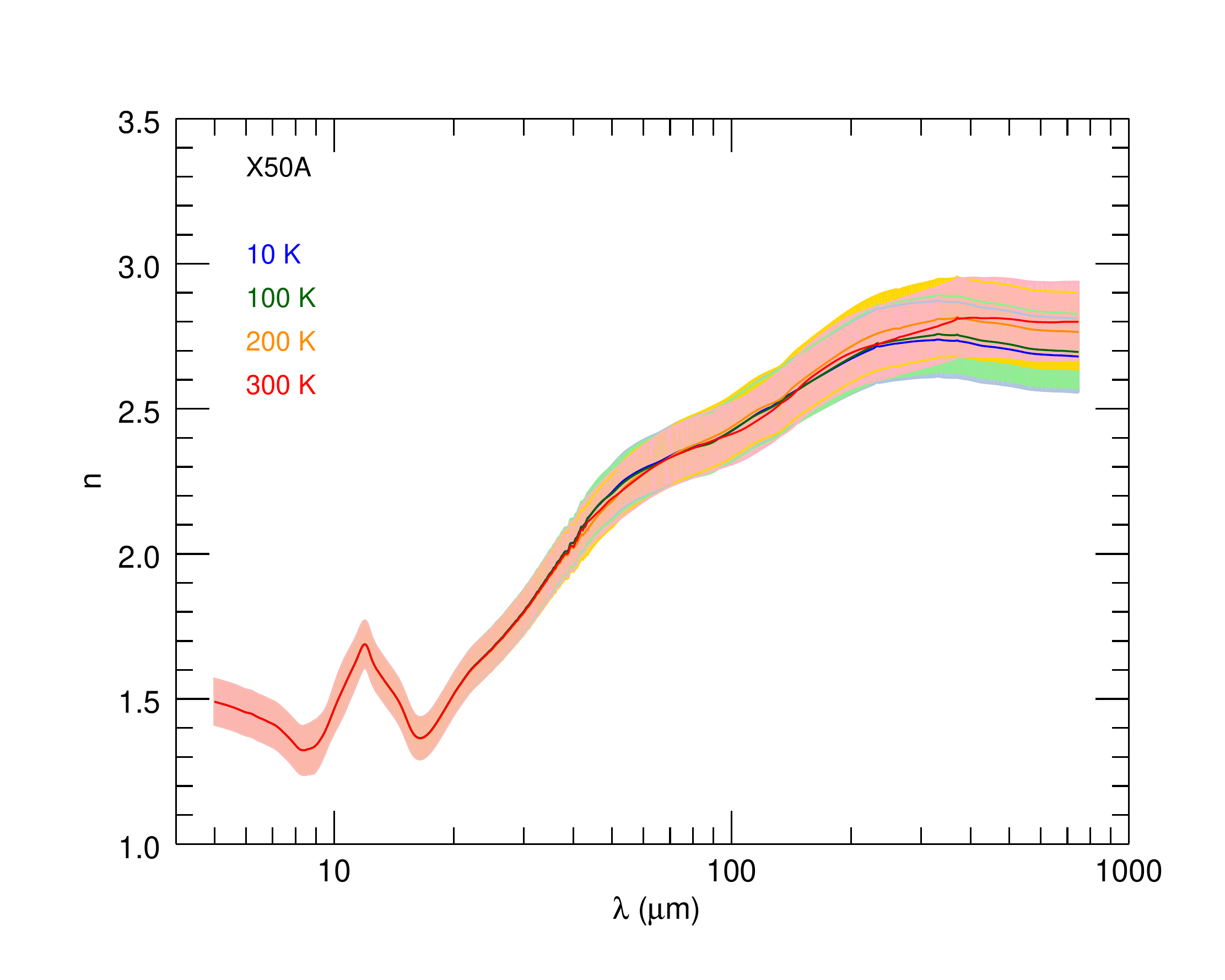}
  \includegraphics[scale=.33, trim={0 1cm 0 1.5cm}, clip]{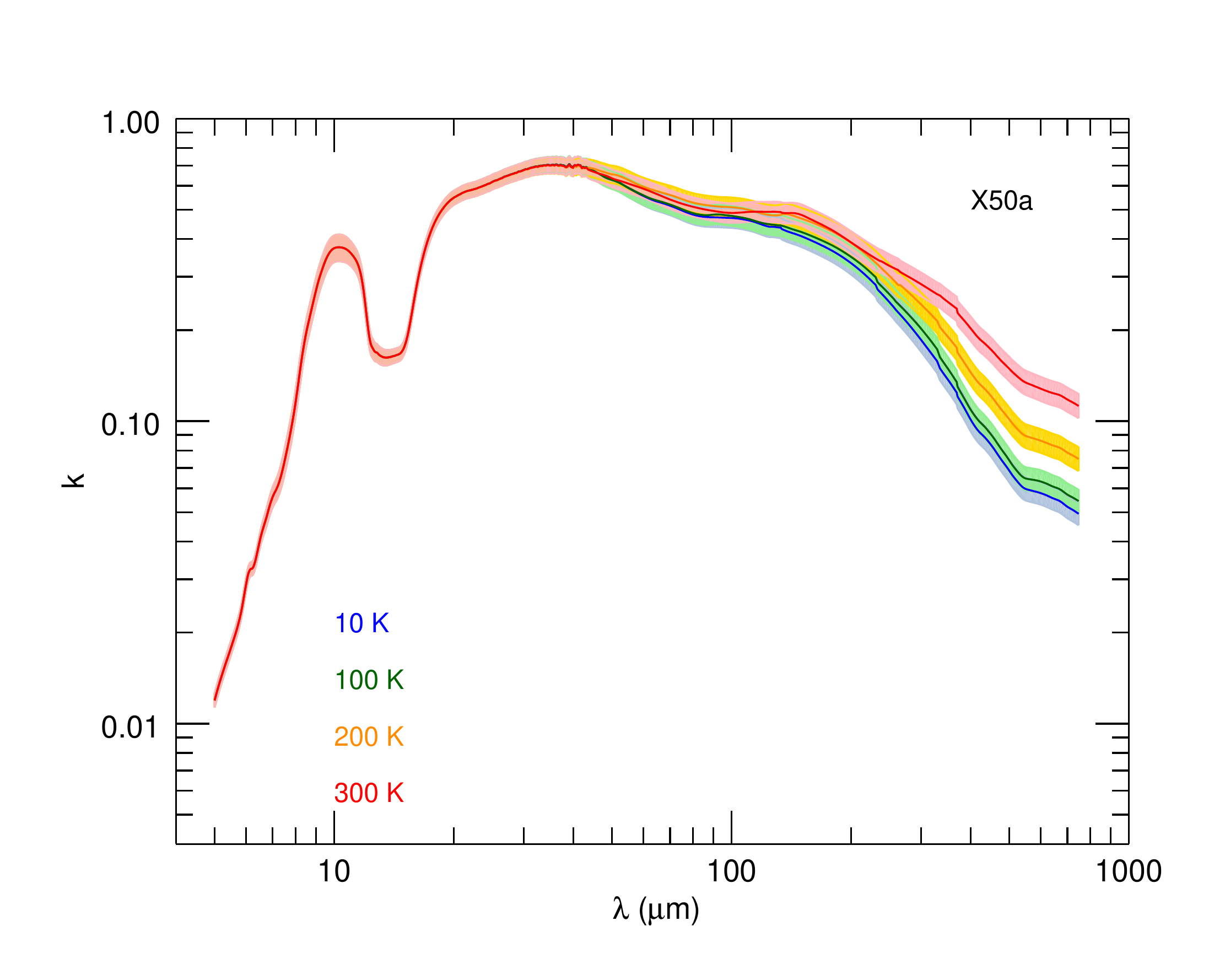}
 \includegraphics[scale=.33, trim={0 1cm 0 1.5cm}, clip]{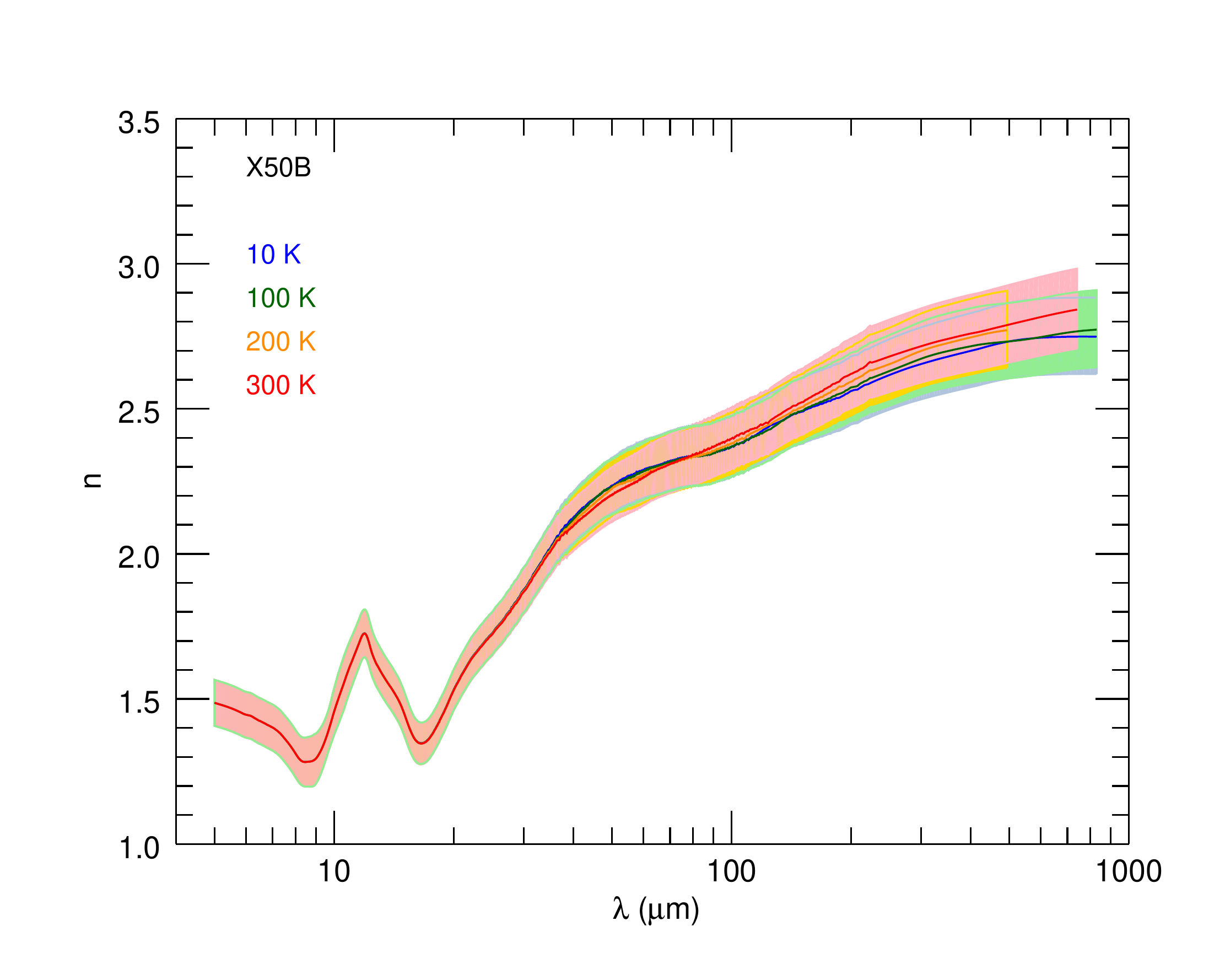}
 \includegraphics[scale=.33, trim={0 1cm 0 1.5cm}, clip]{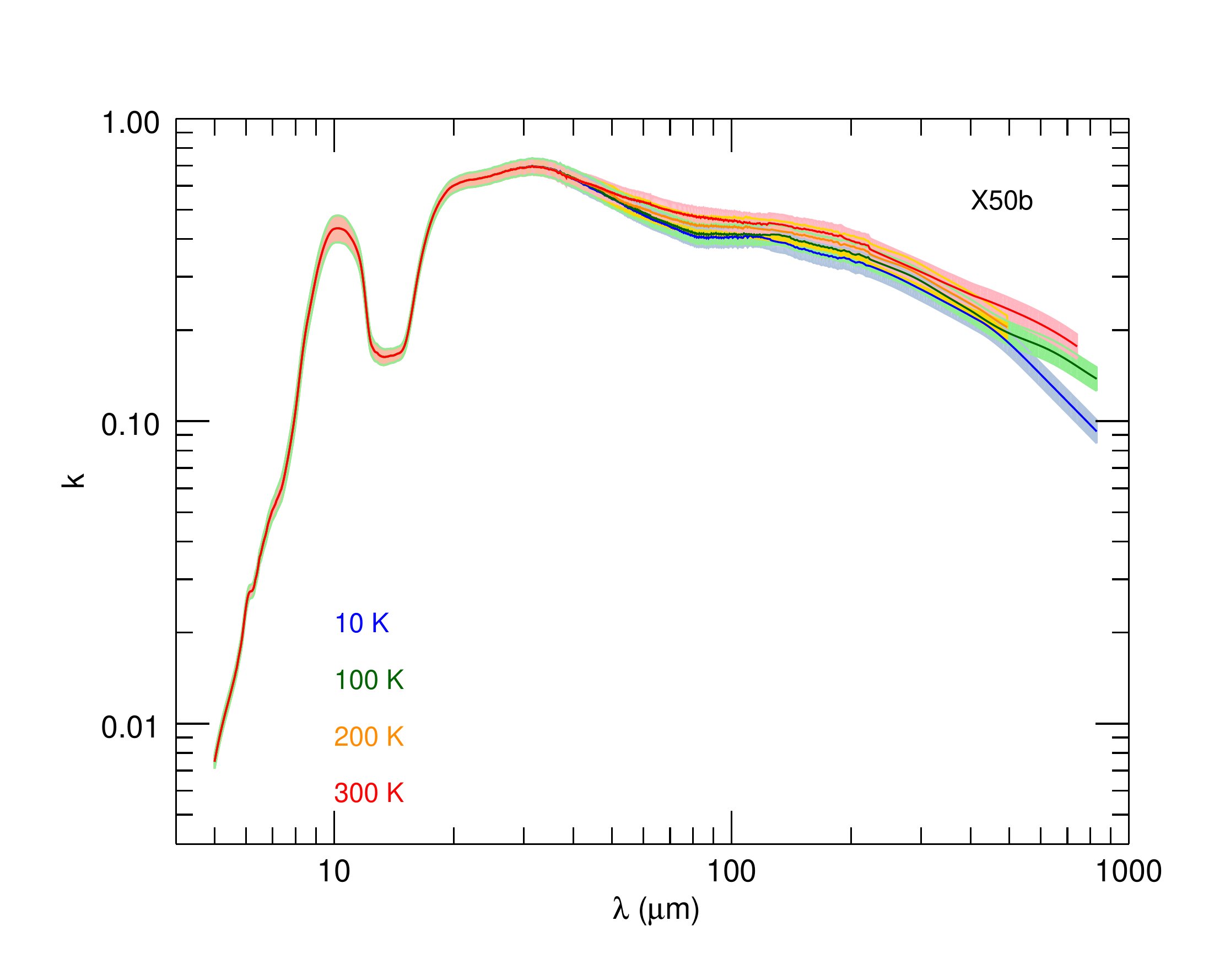}
          \caption{Optical constants for the Mg-rich silicate samples as a function of temperature: 10 K (blue), 100 K (green), 200K (orange), and 300 K (red), with their associated total uncertainties (shaded area).  }
    \label{netk_mgrich_error}% label for figure
\end{figure*}

\begin{figure*}[t]
\centering
  \includegraphics[scale=.33, trim={0 1cm 0 1.5cm}, clip]{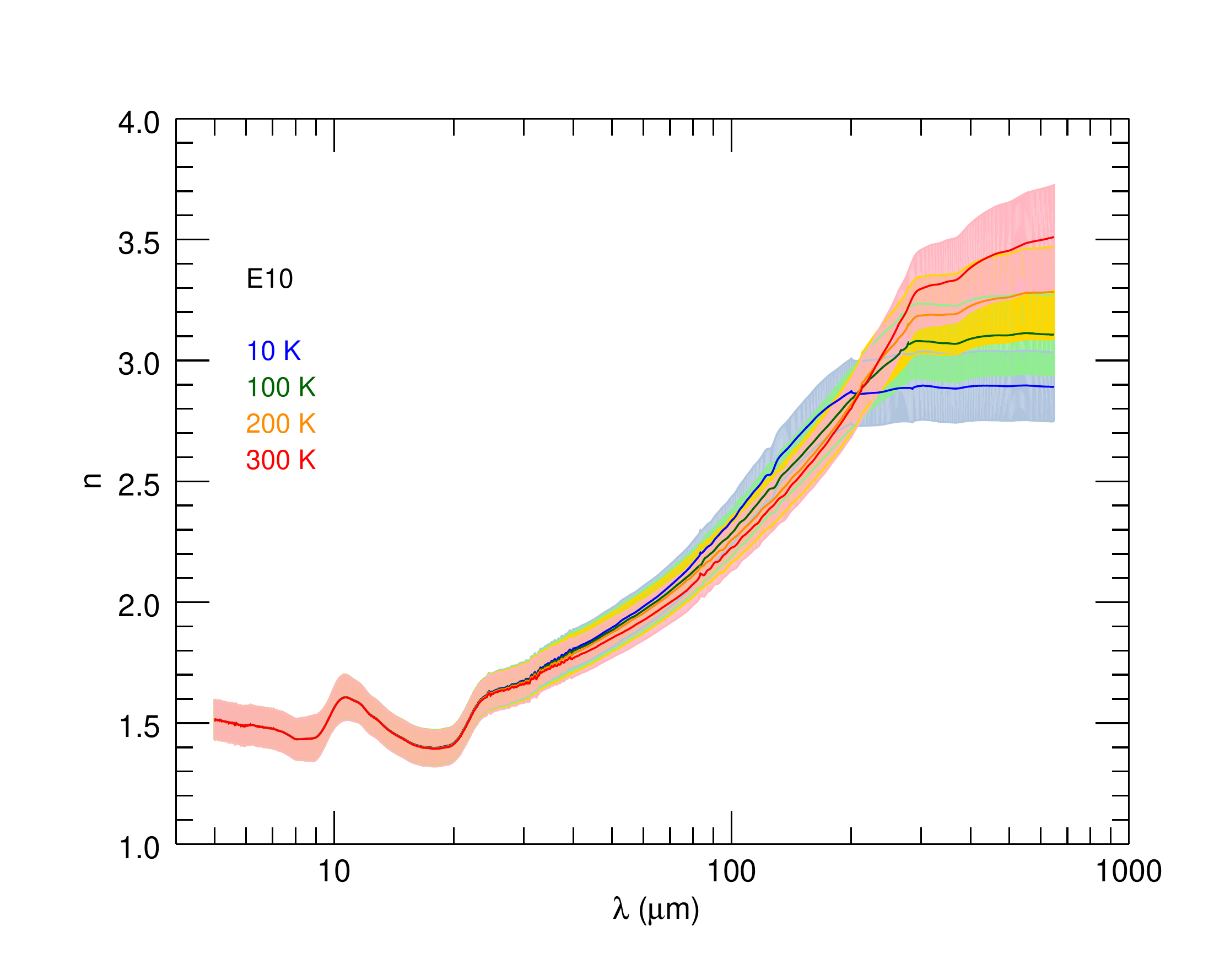}
  \includegraphics[scale=.33, trim={0 1cm 0 1.5cm}, clip]{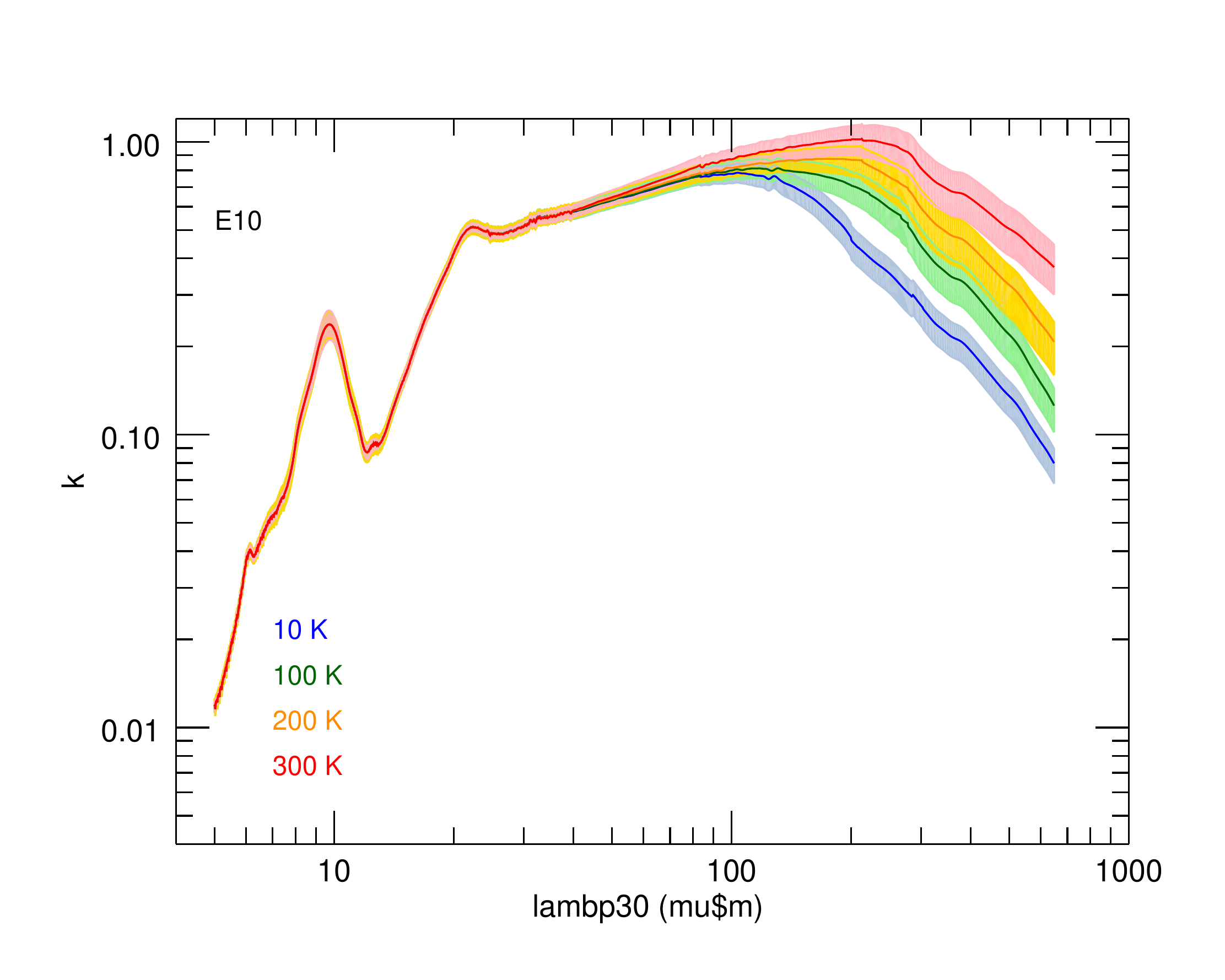}
  \includegraphics[scale=.33, trim={0 1cm 0 1.5cm}, clip]{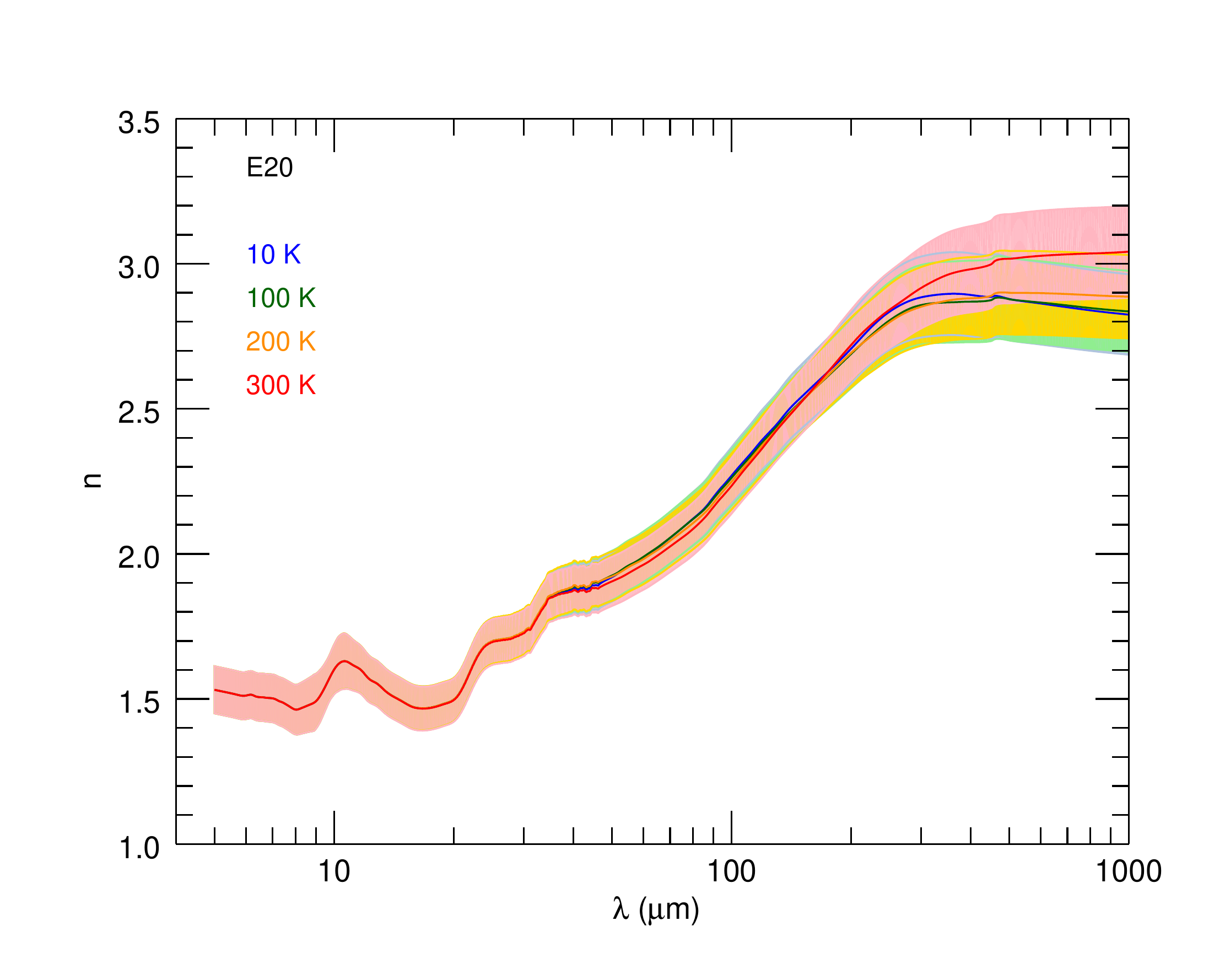}
  \includegraphics[scale=.33, trim={0 1cm 0 1.5cm}, clip]{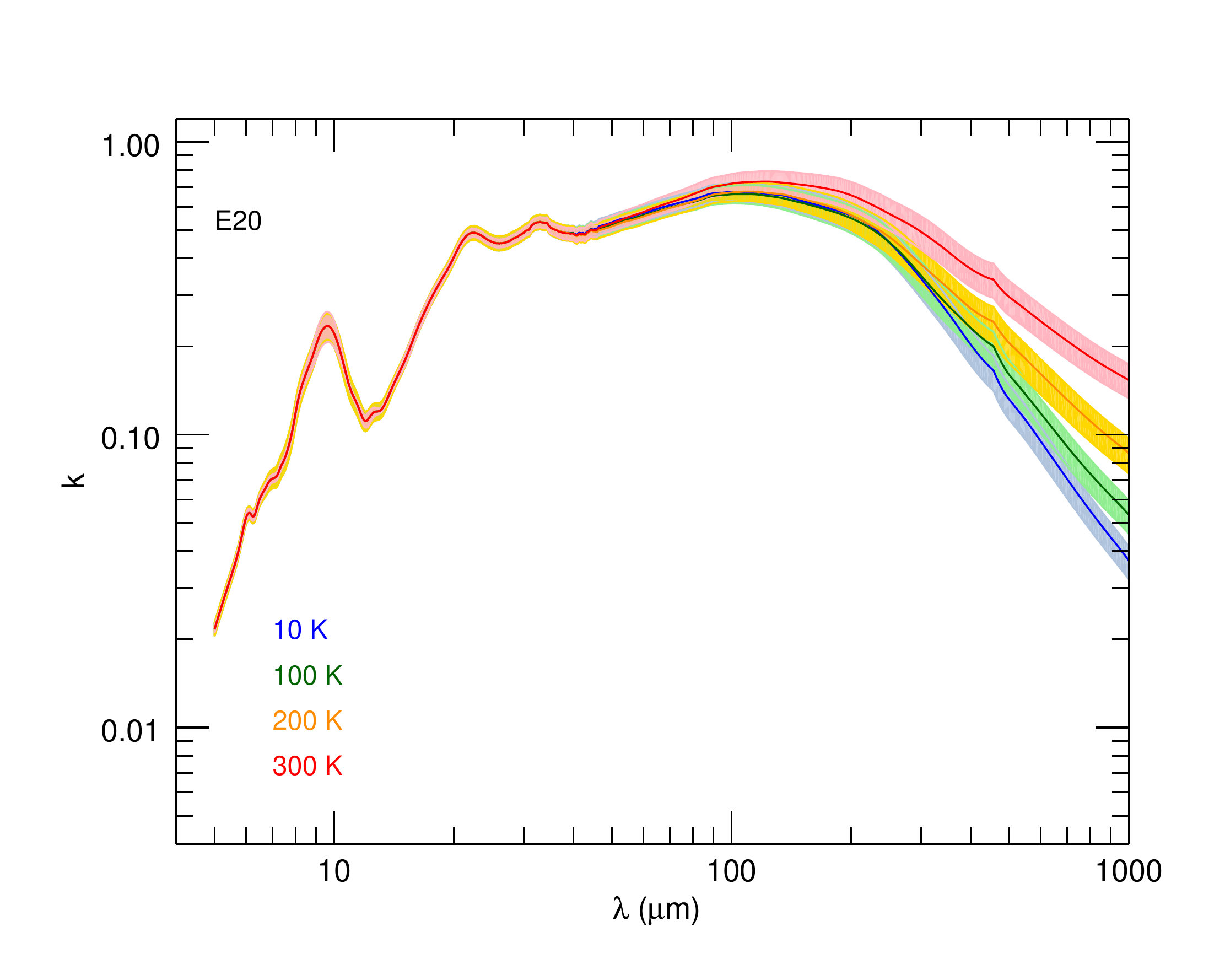}
  \includegraphics[scale=.33, trim={0 1cm 0 1.5cm}, clip]{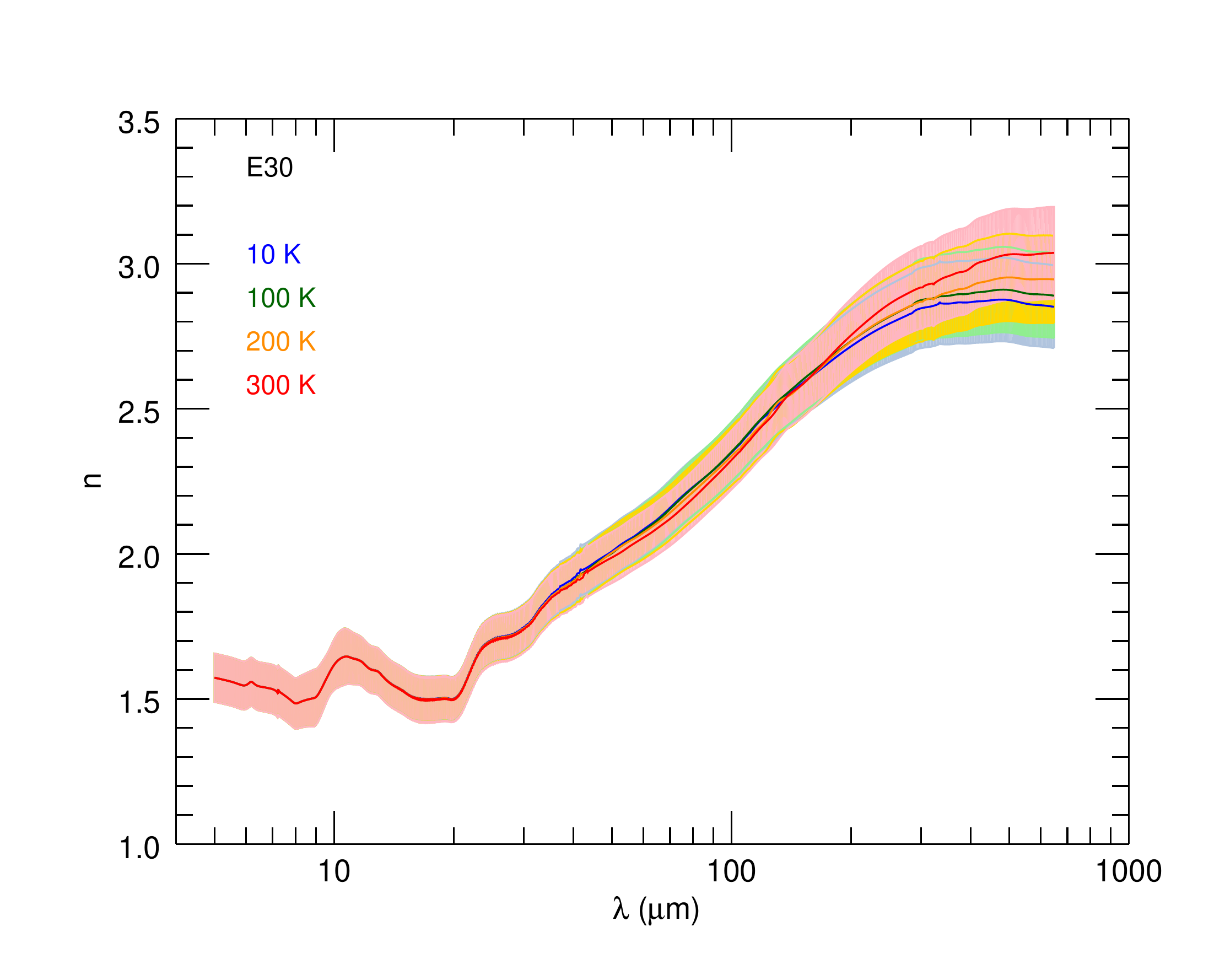}
  \includegraphics[scale=.33, trim={0 1cm 0 1.5cm}, clip]{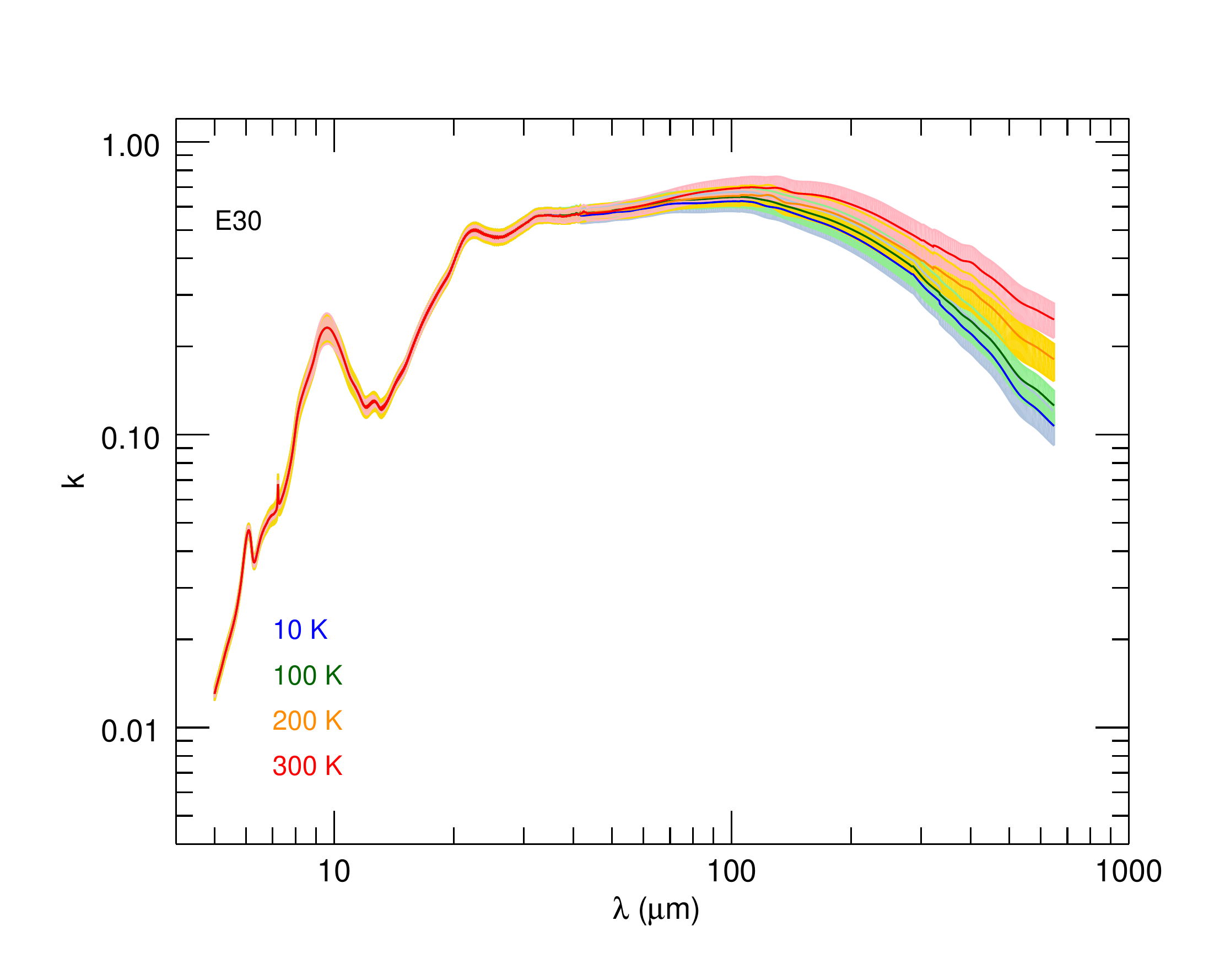}
  \includegraphics[scale=.33, trim={0 1cm 0 1.5cm}, clip]{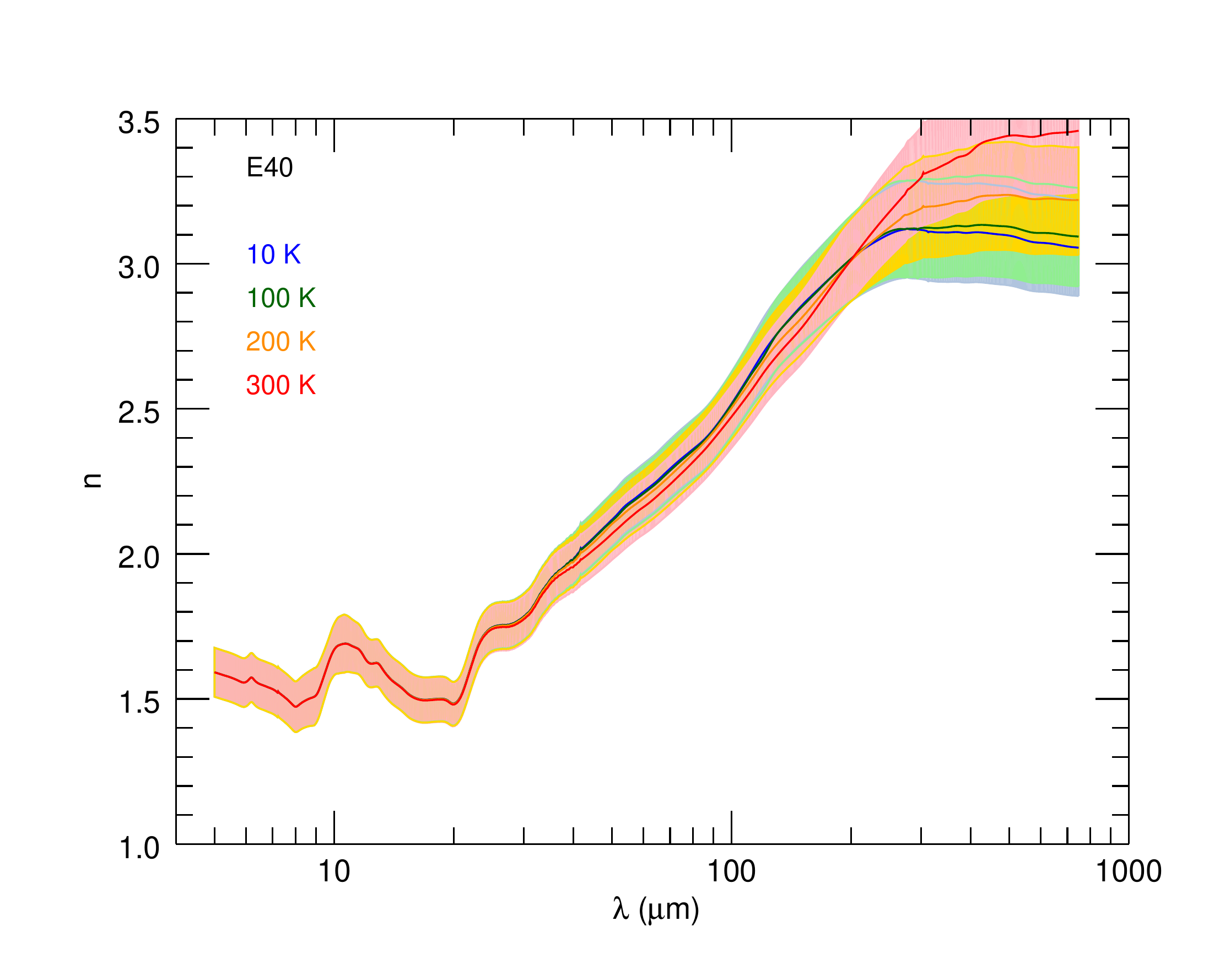}
  \includegraphics[scale=.33, trim={0 1cm 0 1.5cm}, clip]{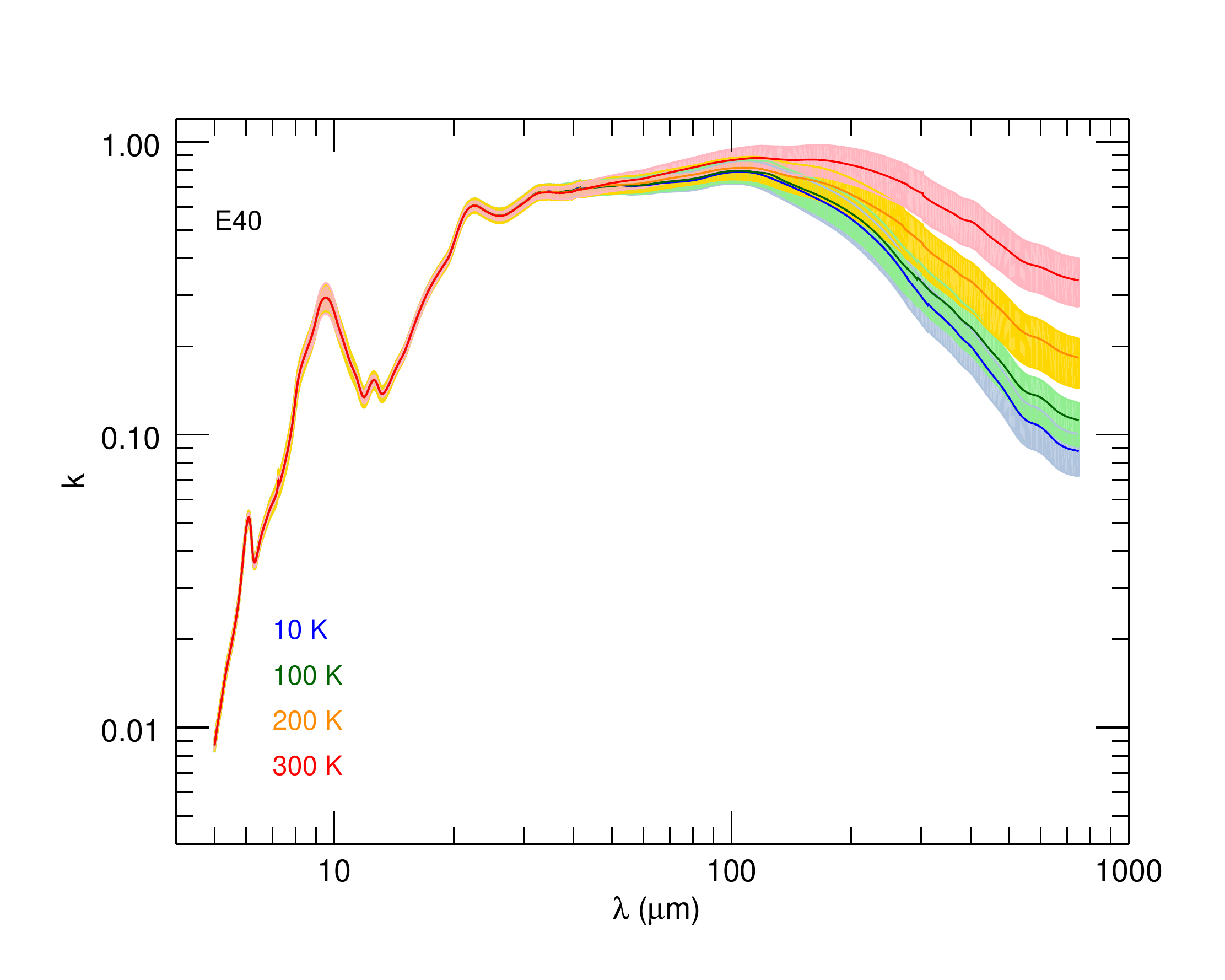}
          \caption{Optical constants for the Fe-rich silicate samples as a function of temperature: 10 K (blue), 100 K (green), 200 K (orange), and 300 K (red), with their associated total uncertainties (shaded area).  }
    \label{netk_ferich_error}% label for figure
\end{figure*}

\begin{figure*}[t]
\centering
  \includegraphics[scale=.33, trim={0 1cm 0 1.5cm}, clip]{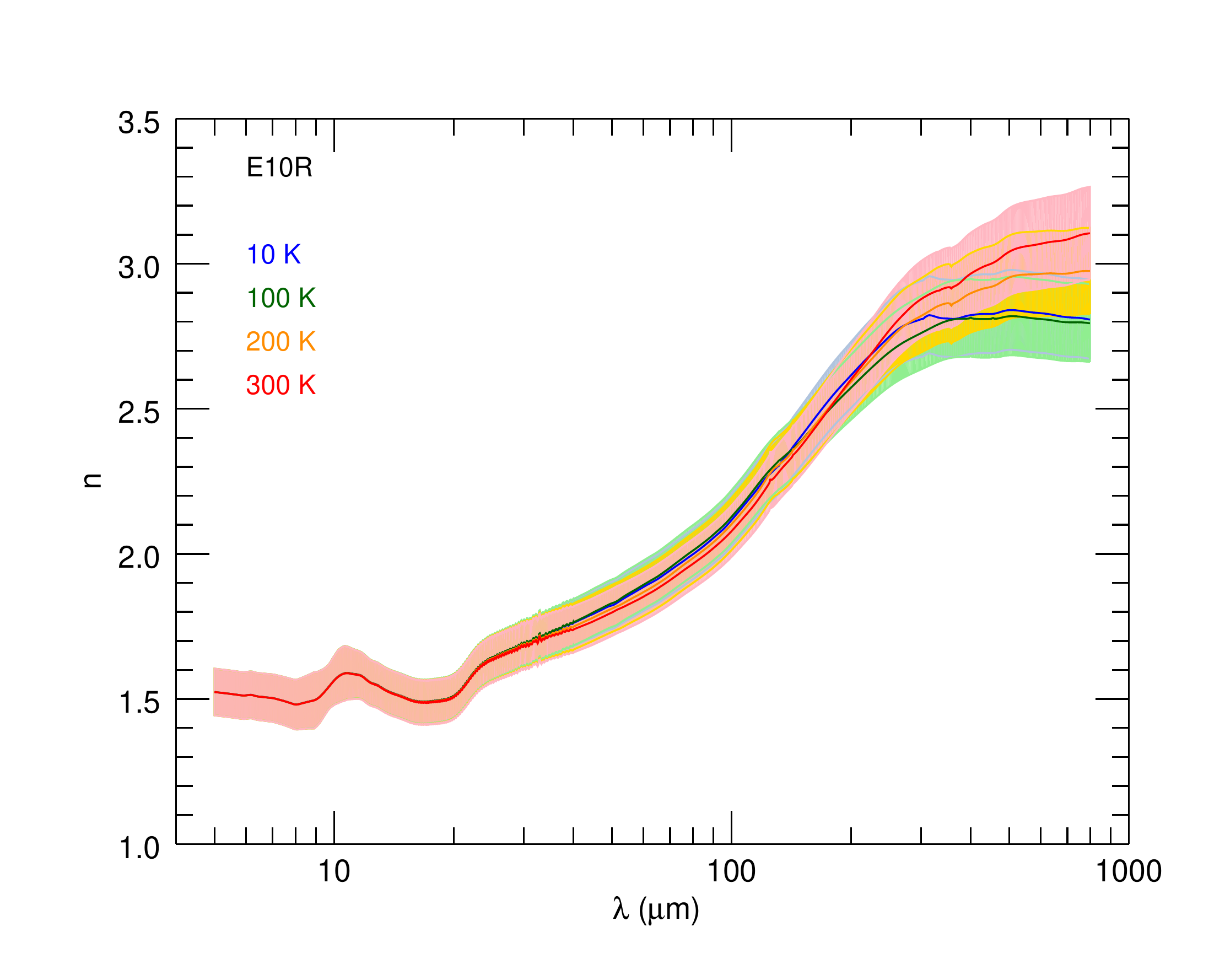}
  \includegraphics[scale=.33, trim={0 1cm 0 1.5cm}, clip]{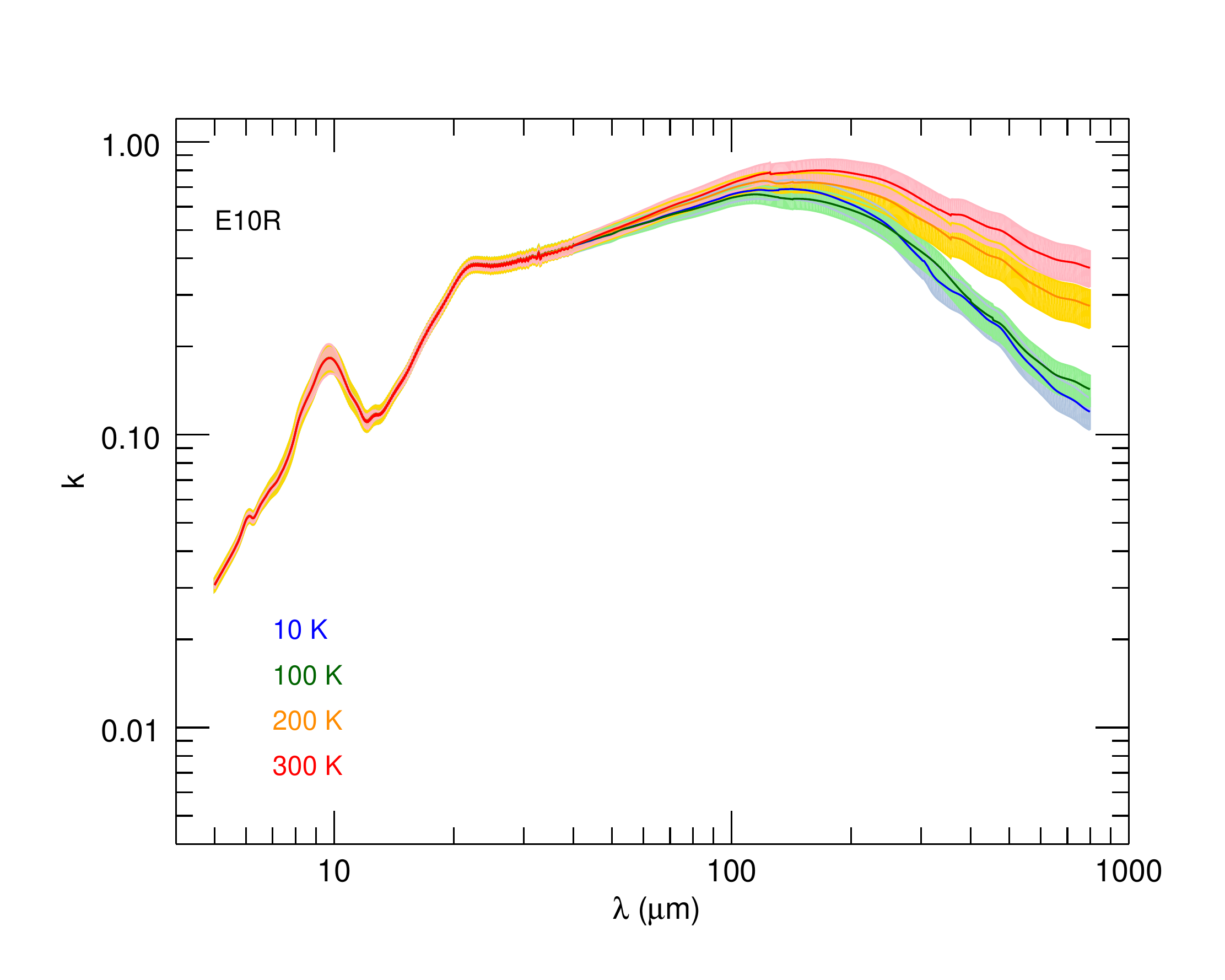}
  \includegraphics[scale=.33, trim={0 1cm 0 1.5cm}, clip]{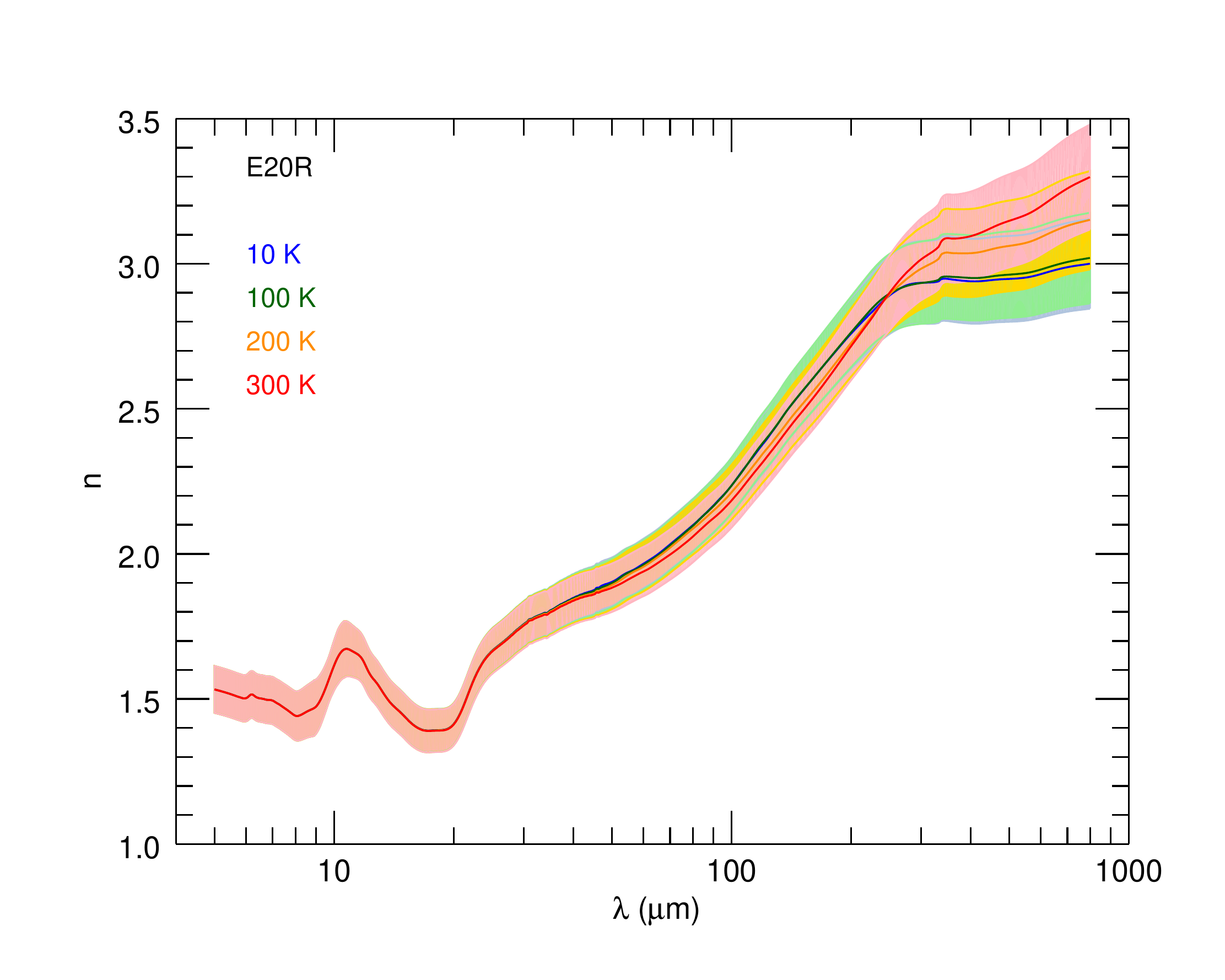}
  \includegraphics[scale=.33, trim={0 1cm 0 1.5cm}, clip]{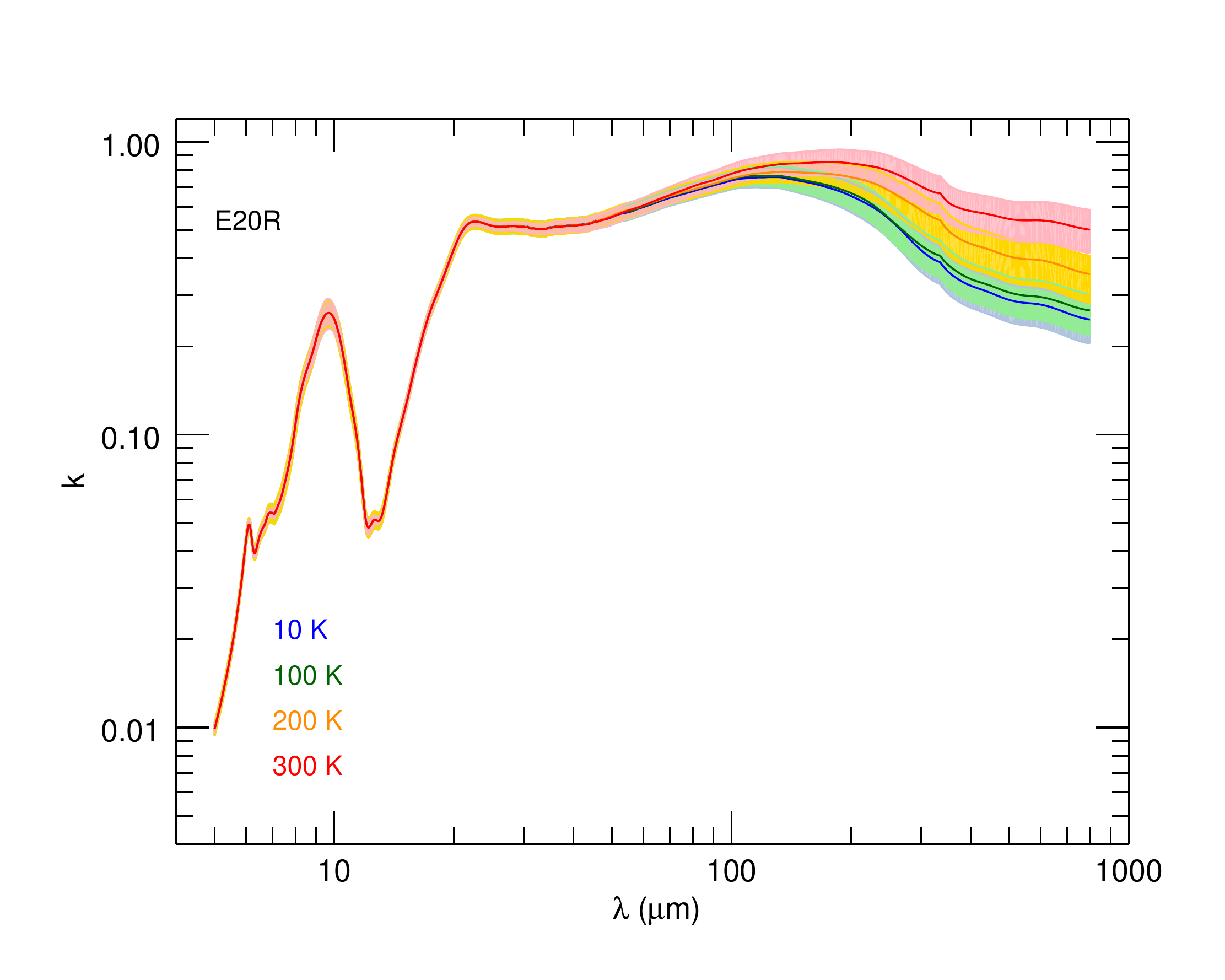}
  \includegraphics[scale=.33, trim={0 1cm 0 1.5cm}, clip]{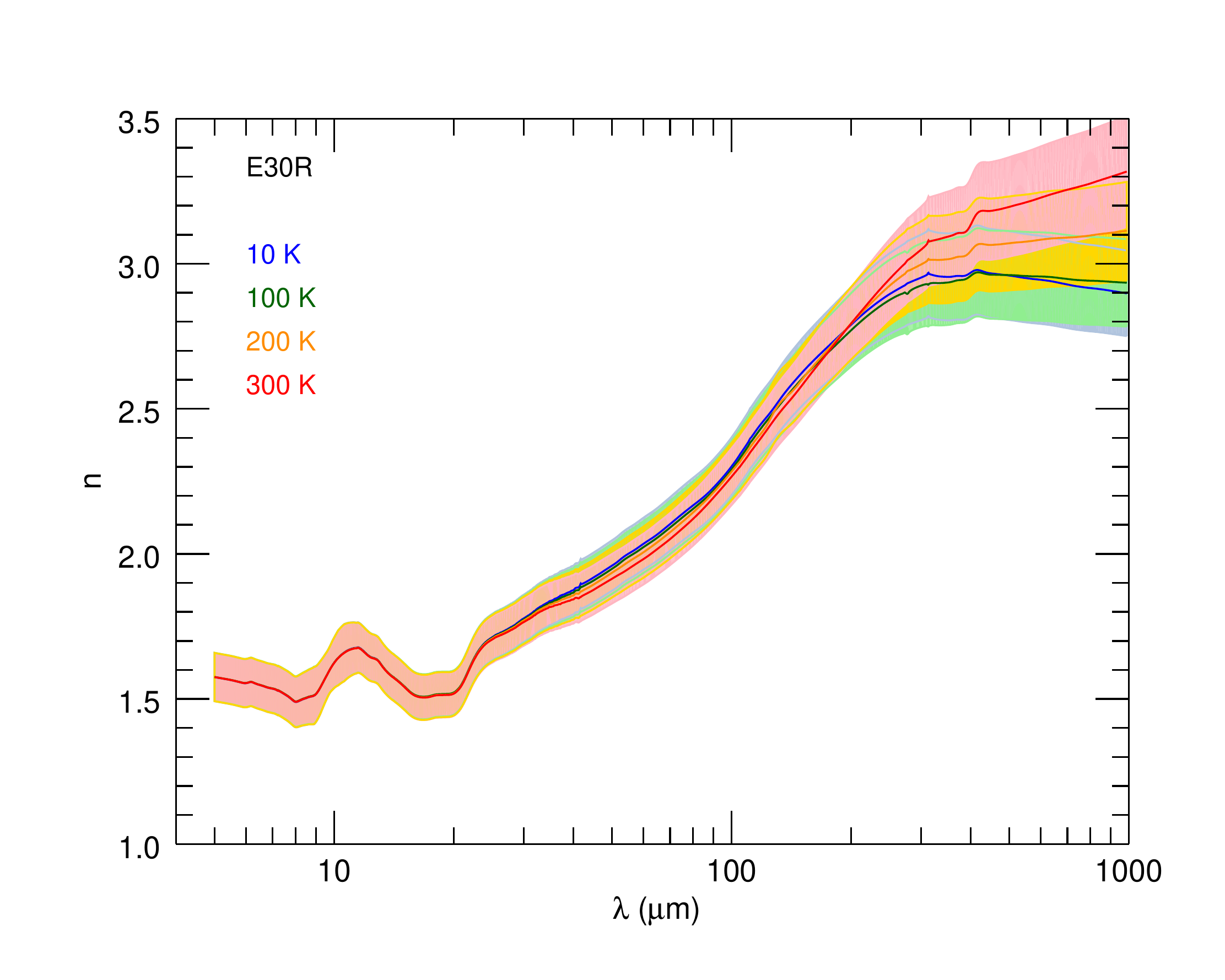}
  \includegraphics[scale=.33, trim={0 1cm 0 1.5cm}, clip]{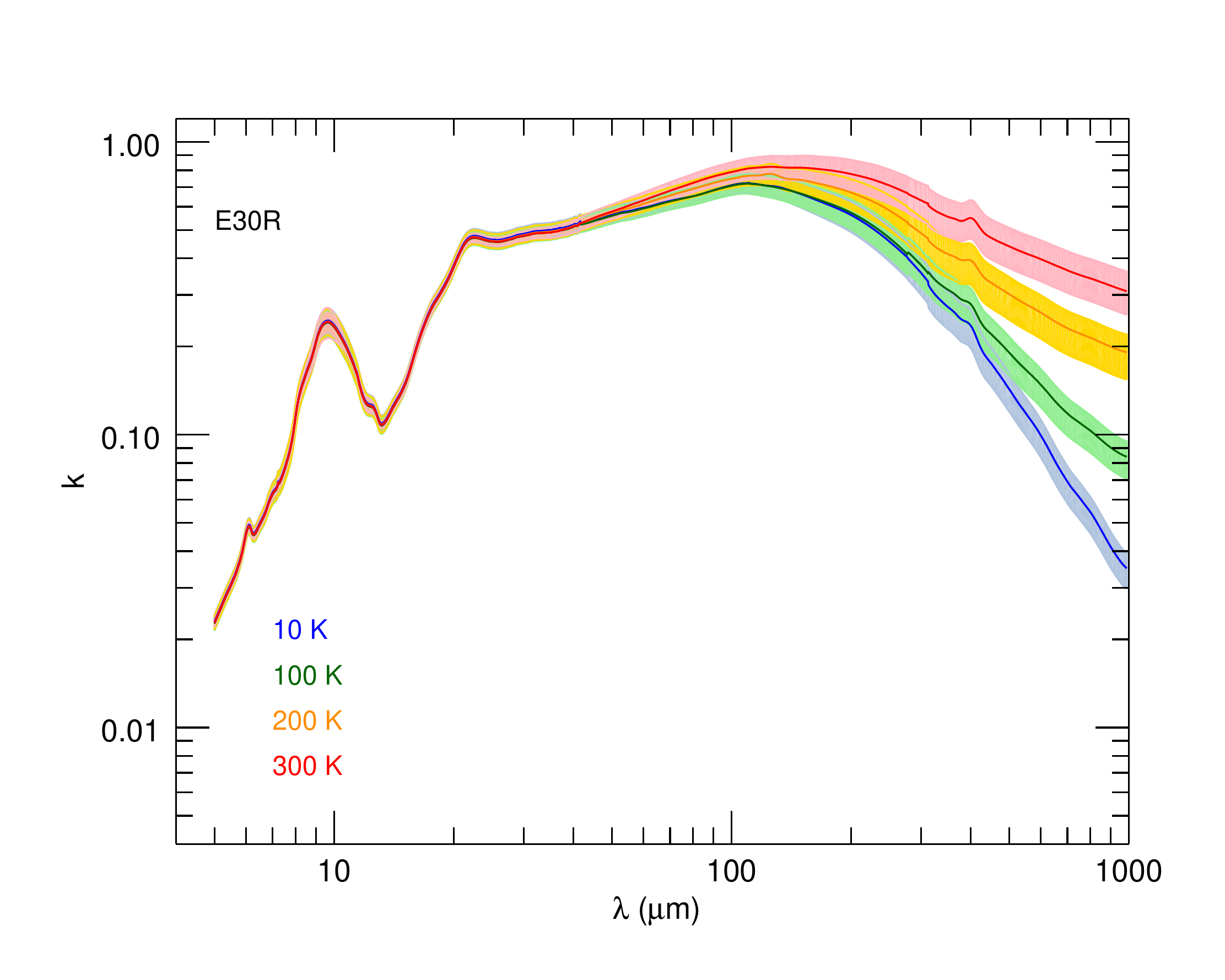}
  \includegraphics[scale=.33, trim={0 1cm 0 1.5cm}, clip]{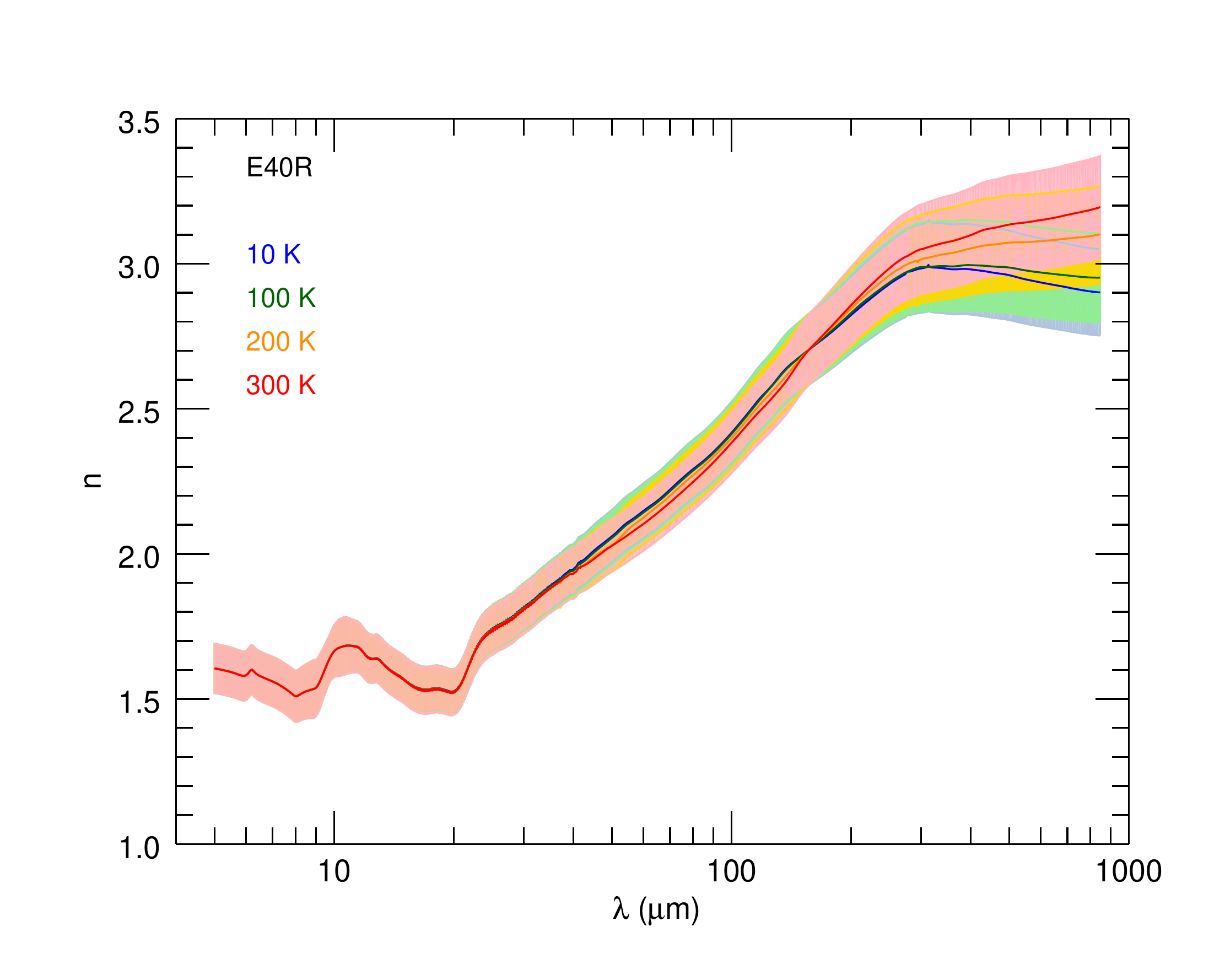}
  \includegraphics[scale=.33, trim={0 1cm 0 1.5cm}, clip]{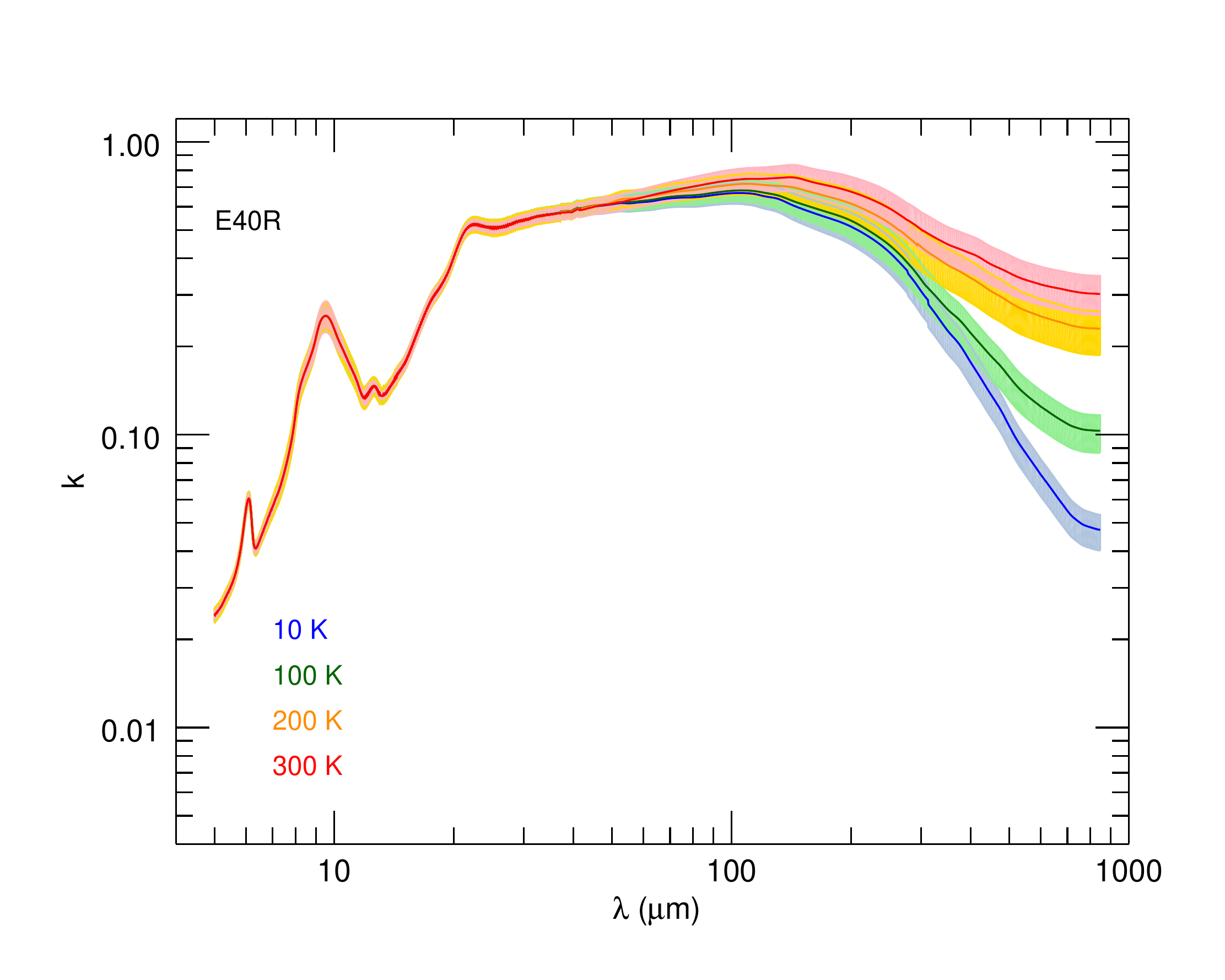}
          \caption{Optical constants for the reduced Fe-rich silicate samples as a function of temperature: 10 K (blue), 100 K (green), 200 K (orange), and 300 K (red), with their associated total uncertainties (shaded area).  }
    \label{netk_ferich_reduced_error}% label for figure
\end{figure*}

\FloatBarrier

\section{Extrapolated optical constants}
\label{extrapol_n_k}

%Figures~\ref{netk_mgrich_extrapolated}, ~\ref{netk_ferich_extrapolated} and~\ref{netk_ferich_reduced_extrapolated} show the optical constant extrapolated outside the experimental spectral range (see Sect~\ref{sect_extrapolated} for details).

\begin{figure*}[!htbp]
\centering
  \includegraphics[scale=0.9, angle =0]{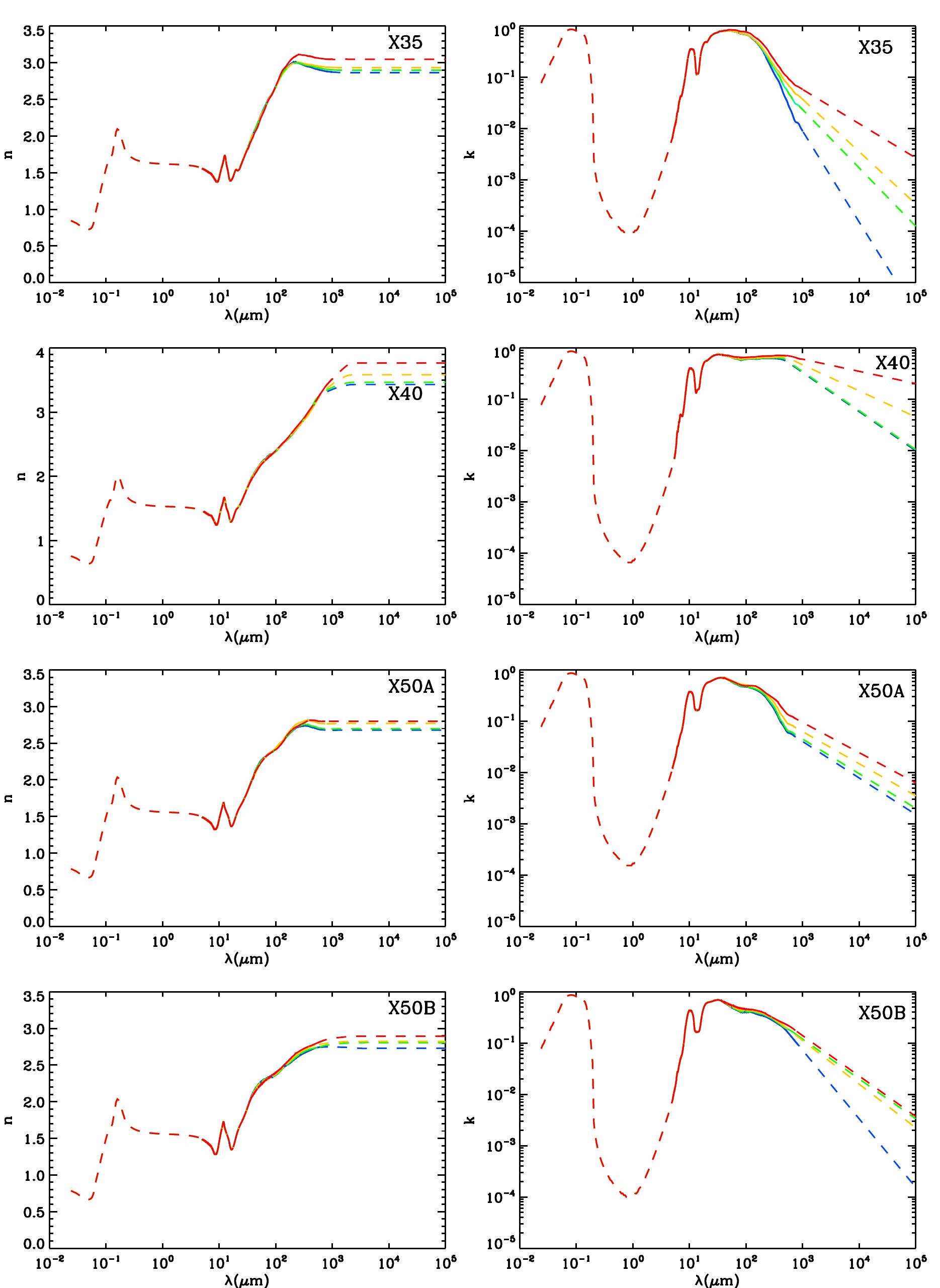}
          \caption{Optical constants for the Mg-rich silicate samples  as a function of temperature: 10 K (blue), 100 K (green), 200 K (orange), and 300 K (red). The dashed lines show the extrapolated $n$ and $k$ outside the wavelength range of the spectroscopic measurements. }
    \label{netk_mgrich_extrapolated}% label for figure
\end{figure*}

\begin{figure*}[htbp]
\centering
 \includegraphics[scale=0.9, angle =0]{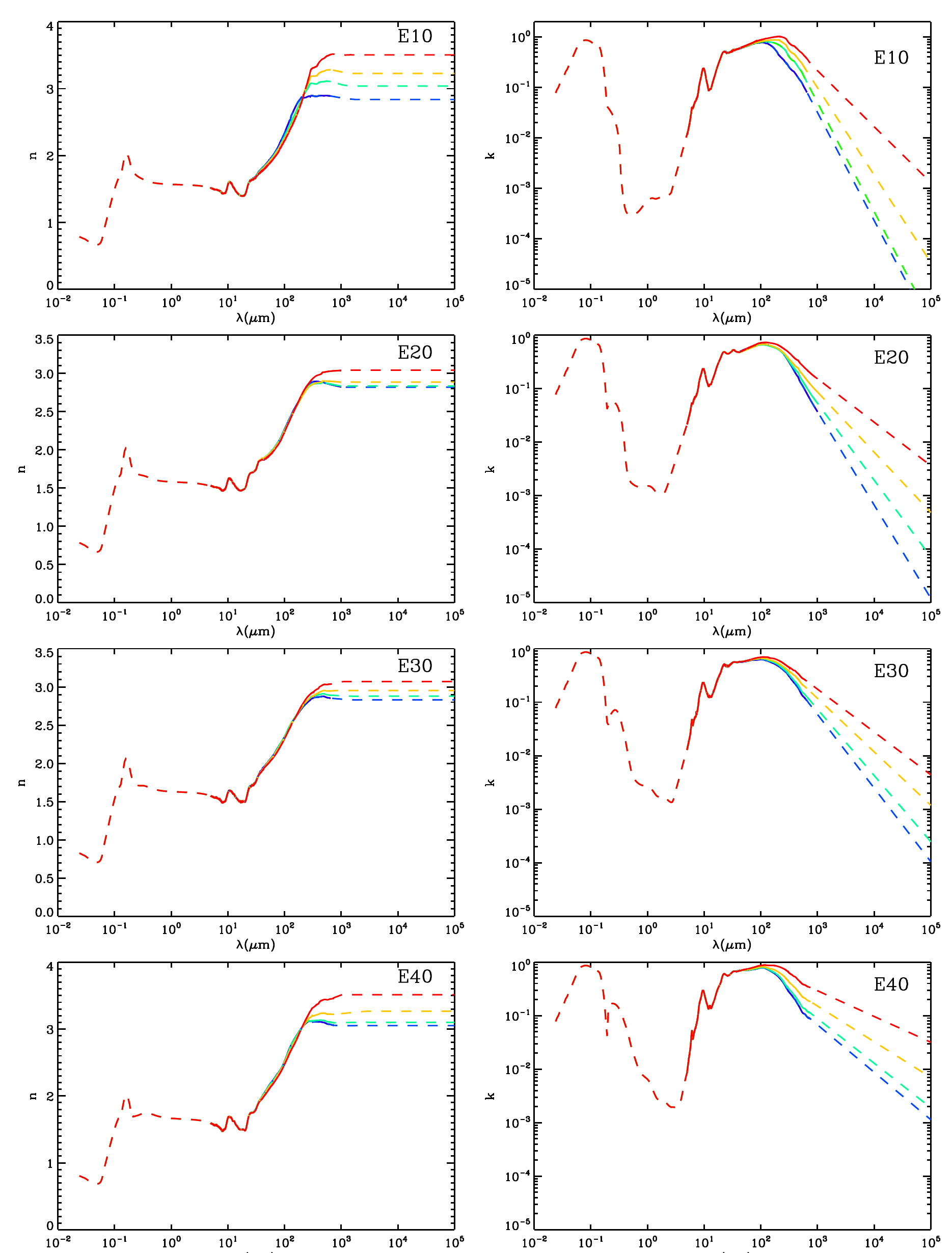}
          \caption{Optical constants for the Fe-rich silicate samples  as a function of temperature: 10 K (blue), 100 K (green), 200 K (orange), and 300 K (red). The dashed lines show the extrapolated $n$ and $k$ outside the wavelength range of the spectroscopic measurements. }
    \label{netk_ferich_extrapolated}% label for figure
\end{figure*}

\begin{figure*}[htbp]
\centering
  \includegraphics[scale=0.9, angle =0]{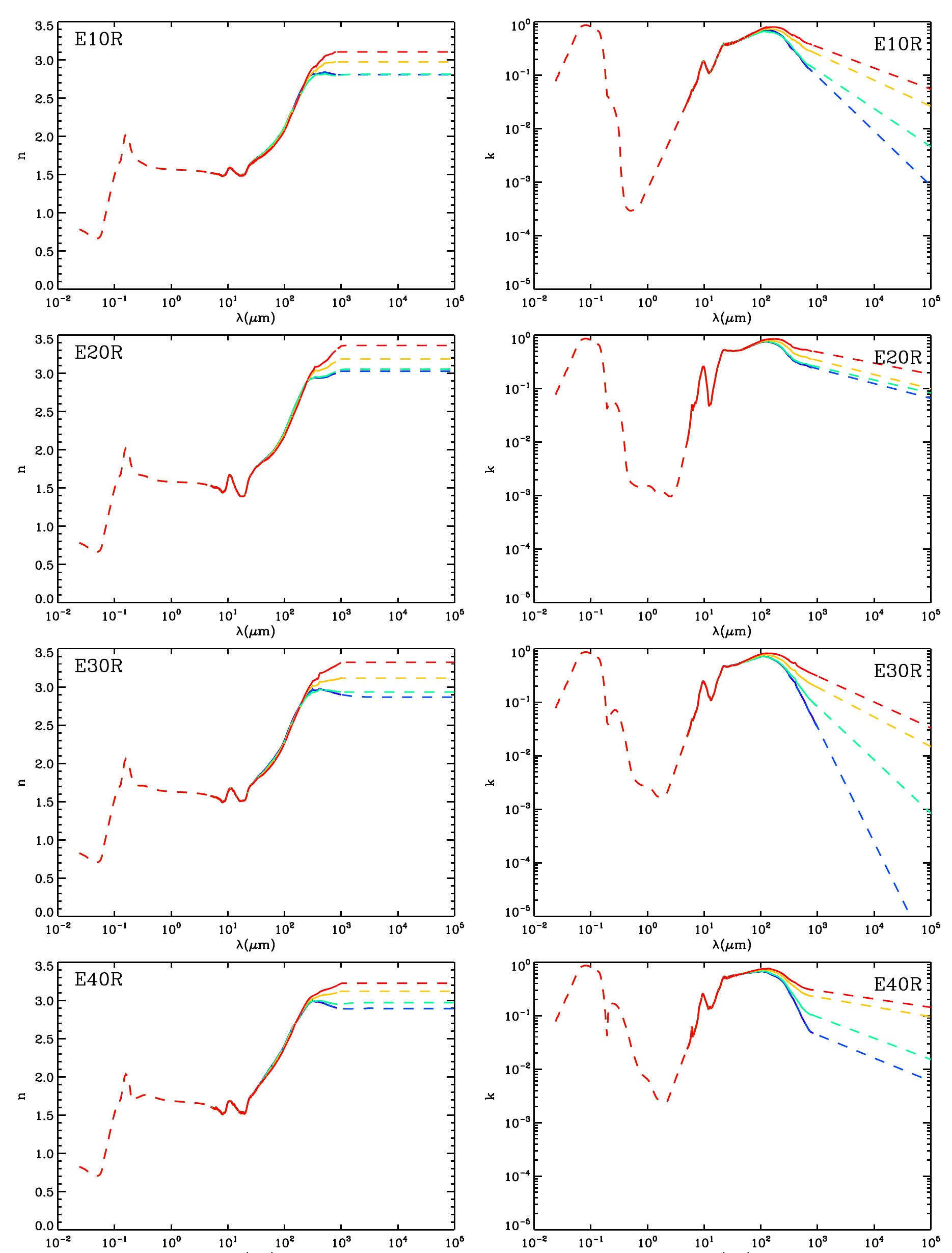}
          \caption{Optical constants for the reduced Fe-rich silicate samples  as a function of temperature: 10 K (blue), 100 K (green), 200 K (orange), and 300 K (red). The dashed lines show the extrapolated $n$ and $k$ outside the wavelength range of the spectroscopic measurements. }
    \label{netk_ferich_reduced_extrapolated}% label for figure
\end{figure*}

\FloatBarrier

\section{Optical constant interpolation as a function of temperature}
\label{Appendices_n_k_interpol_T}

The optical constants of the samples were determined from experiments performed at specific temperatures (10, 30, 100, 200, and 300\,K). Here, we propose a method to perform simple polynomial interpolations of the calculated optical constants to any temperature in the range $10 - 300$\,K. The MACs of the studied samples do not show variations with temperature at short wavelengths and so the interpolations are only made for $\lambda \geqslant 80\,\mu$m. We used the five sets of optical constants corresponding to the five temperatures measured  for the Fe-rich samples. For the Mg-rich samples the MAC at 30\,K is identical to that at 10\,K and the optical constants at 30\,K were therefore set to those at 10\,K, except for the X40 sample for which a better fit was obtained when the data at 30\,K were not included.

We checked that the interpolated optical constants are consistent with the Kramers-Kronig relations by comparing the interpolated refractive indices with the refractive indices calculated from the interpolated absorption coefficient using the Kramers-Kronig relations. The difference between the two does not exceed 2\% for $\lambda \geqslant 80\,\mu$m for all samples and temperatures.

\subsection{Interpolation of the absorption coefficient, $k$}

For all samples, the absorption coefficient, k($\lambda$, T), was interpolated in two steps corresponding to two separate spectral domains. The value of $\lambda_{cut}$ for each sample was chosen to be as close as possible to that reported in Table~\ref{table:extrapol}. For $80\,\mu$m $\leqslant \lambda \leqslant \lambda_{cut}$, a 2D polynomial fit was performed using Eq.~(\ref{Eq_k_T_1}) and the squared matrix $A$ (see Table~\ref{table_param_interpol_k_T_A}) for each sample:
\begin{equation}
\label{Eq_k_T_1}
   \mathrm{log}\,k(\lambda, T)=\sum A_{j,i} \left( \frac{\mathrm{log}\,\lambda- 1.7781602}{1.2518668 \times 10^{-4}} \right)^i \times \left( \frac{T-10}{10} \right) ^j.       
\end{equation}
For $\lambda_{cut} < \lambda \leqslant 10$\,mm, a 1D polynomial fit was used to reproduce the power-law extrapolation at long wavelengths using Eq.~(\ref{Eq_k_T_2}), the column matrix $B$ (see Table~\ref{table_param_interpol_k_T_B}), and a normalisation at $\lambda_{cut}$:
\begin{equation}
\label{Eq_k_T_2}
   \mathrm{log}\,k(\lambda, T)=\sum B_j \times T^j \times log\,\lambda .
 \end{equation}

%Values of $A_{j,i}$, $B_{j}$ and $\lambda_{cut,1}$  for each sample are given in Tables~\ref{table_param_interpol_k_T_A} and ~\ref{table_param_interpol_k_T_B}.

\subsection{Interpolation of the refractive index, $n$}

Different interpolation strategies are needed to interpolate the refractive index, $n(\lambda, T)$, for the different samples. The value of $\lambda_{cut}$ for each sample was chosen to be as close as possible to that reported in Table~\ref{table:extrapol}. 
 
For all samples except E10, E20R, and X50B, for $80\,\mu$m $\leqslant \lambda \leqslant \lambda_{cut}$, a 2D polynomial fit was performed using Eq.~(\ref{Eq_n_T_1}), which is analogous to Eq.~(\ref{Eq_k_T_1}) (see Table~\ref{table_param_interpol_n_T_C} for the value of the parameters $C_{j,i}$ and $\lambda_{cut}$):
\begin{equation}
\label{Eq_n_T_1}
  \mathrm{log}\,n(\lambda, T)=\sum C_{j,i} \left( \frac{\mathrm{log}\,\lambda- 1.7781602}{1.2518668 \times 10^{-4}} \right)^i \times \left( \frac{T-10}{10} \right) ^j  .    
\end{equation}
For  $\lambda_{cut} $ $< \lambda \leq 10$ mm, the spectra can be reproduced by a single constant, that is 
\begin{equation}
\label{Eq_n_T_2}
  \mathrm{log}\,n(\lambda,T) = log\,n(\lambda_{cut},T) .
   \end{equation}
   
For sample X50B for $80\,\mu$m $\leqslant \lambda \leqslant \lambda_{cut}$ the interpolation is made using Eq.~(\ref{Eq_n_T_1}). 
For  $\lambda_{cut} < \lambda \leqslant 10$\,mm the interpolation is made using  Eq.~(\ref{Eq_n_T_3}) (analogous to Eq.~\ref{Eq_k_T_2}):
\begin{equation}
\label{Eq_n_T_3}
   \mathrm{log}\,n(\lambda, T)=\sum E_j \times T^j \times log\,\lambda .
 \end{equation}\\
See Table~\ref{table_param_interpol_n_T_X50B} for the values of the parameters $C_{j,i} $ and $E_j $. 

For samples E10 and E20R, the interpolation of the refractive indices of samples E10 and E20R requires three steps (see Table~\ref{table_param_interpol_n_T_C} for the value of the parameters $C_{j,i}$, $\lambda_{cut,1}$ and $\lambda_{cut,2}$). For $80\,\mu$m $\leqslant \lambda \leqslant \lambda_{cut,1}$: we use Eq~(\ref{Eq_n_T_1}). 
For $\lambda_{cut,1} < \lambda \leqslant  \lambda_{cut,2}$ we use Eq~(\ref{Eq_n_T_E10}) for sample E10 and  Eq~(\ref{Eq_n_T_E20r}) for sample E20R, 
\begin{equation}
\label{Eq_n_T_E10}
   \mathrm{log}\,n(\lambda, T)=\sum C_{j,i} \left( \frac{\mathrm{log})\,\lambda- 2.2431035}{1.2518668 \times 10^{-4}} \right)^i \times \left( \frac{T-10}{10} \right) ^j  ,
\end{equation}
\begin{equation}
\label{Eq_n_T_E20r}
   \mathrm{log}\,n(\lambda, T)=\sum C_{j,i} \left( \frac{\mathrm{log}\,\lambda- 2.3979594}{1.2518668 \times 10^{-4}} \right)^i \times \left( \frac{T-10}{10} \right) ^j   .
\end{equation}
For $\lambda_{cut,2} < \lambda \leqslant  10$\,mm: We use with Eq.~(\ref{Eq_n_T_2}).

\begin{table*}[!t]
\caption {Parameters A$_{i,j}$ and $\lambda_{cut}$ ($\mu$m) for the interpolation of $k$ with the temperature.} 
\label{table_param_interpol_k_T_A}
%\begin{center}
\begin{tabular}{c c c c c c c c }
\hline 
\hline 
 Sample & $\lambda_{cut}$  ($\mu$m) & A$_{0,*}$    & A$_{1,*}$  &   A$_{2,*}$ &  A$_{3,*}$ &   A$_{4,*}$ \\
\hline 
   \multirow{5}{*}{X35}  &   \multirow{5}{*}{700}  & -0.121953    & -1.21069e-05 &  3.58788e-10 & -6.58218e-12 &  4.67800e-16 \\
                                 & &  0.00265459  &  6.39436e-06 & -3.56717e-09 &  6.02274e-13 & -2.90468e-17 \\
                                 & & -0.000529290 & -5.27210e-07 &  2.37585e-10 & -2.35475e-14 &  1.17028e-18 \\
                                 & &  2.65051e-05 &  2.44621e-08 & -1.05439e-11 &  8.77301e-16 & -5.23239e-20 \\
                                 & & -3.51532e-07 & -4.39600e-10 &  2.11014e-13 & -2.21600e-17 &  1.37940e-21 \\
\hline 
   \multirow{5}{*}{X40}  &   \multirow{5}{*}{500}   & -0.209020    &  -2.00847e-05 &   9.15400e-09 &  -9.75224e-13 &   2.42572e-18 \\
                                & &  0.00668272  &  -6.35738e-06 &   1.87448e-09 &  -1.80587e-13 &   3.26331e-18 \\
                                & & -0.000988584 &   9.59802e-07 &  -2.77814e-10 &   2.59587e-14 &  -2.50312e-19 \\
                                & &  5.24272e-05 &  -4.89185e-08 &   1.47407e-11 &  -1.61910e-15 &   4.33710e-20 \\
                                & & -8.58184e-07 &   8.06823e-10 &  -2.51998e-13 &   3.14248e-17 &  -1.22688e-21 \\
\hline 
   \multirow{5}{*}{X50A} &   \multirow{5}{*}{600}     &  -0.242087    &  -0.000137117 &   7.97991e-08 &  -1.96225e-11 &   1.22800e-15 \\
                                 & &   0.00246197  &  -7.39531e-06 &   3.16079e-09 &  -4.97218e-13 &   2.59573e-17 \\
                                & &  -0.000182316 &   7.62682e-07 &  -2.66058e-10 &   4.00088e-14 &  -2.09571e-18 \\
                                & &   1.63704e-05 &  -2.83665e-08 &   6.60904e-12 &  -5.89712e-16 &   1.98529e-20 \\
                                 & &  -4.39536e-07 &   4.35947e-10 &  -8.54503e-14 &   4.95379e-18 &  -4.21175e-23 \\
\hline 
   \multirow{5}{*}{X50B} &   \multirow{5}{*}{650}      &  -0.339182    &  -7.29662e-05 &   2.77285e-08 &  -5.07244e-12 &   2.15659e-16 \\
                                 & &   0.00974328  &   7.15397e-06 &  -6.32309e-09 &   1.19790e-12 &  -6.44846e-17 \\
                                & &  -0.00146659  &  -1.20825e-06 &   1.23257e-09 &  -2.58715e-13 &   1.58166e-17 \\
                                & &   8.29297e-05 &   6.48944e-08 &  -6.96473e-11 &   1.50952e-14 &  -9.56299e-19 \\
                                & &  -1.45688e-06 &  -1.05352e-09 &   1.18445e-12 &  -2.62042e-16  &  1.69527e-20 \\
\hline 
\hline 
   \multirow{5}{*}{E10}  &   \multirow{5}{*}{800}     &  -0.227340    &   0.000195383 &  -8.34077e-08 &   8.87264e-12 &  -3.89490e-16 \\
                                & &   0.0125135   &  -1.56769e-05 &   7.16345e-09 &  -1.06888e-12 &   4.91452e-17 \\
                                & &  -0.00117512  &  -2.91829e-08 &   4.07884e-10 &  -6.51584e-14 &   3.15281e-18 \\
                                & &   5.53934e-05 &   2.84879e-08 &  -3.86775e-11 &   6.07355e-15 &  -2.81824e-19 \\
                                & &  -9.12813e-07 &  -5.49626e-10 &   7.10969e-13 &  -1.10690e-16 &   5.01930e-21 \\
\hline 
   \multirow{5}{*}{E20}  &   \multirow{5}{*}{800}    &  -0.233412    &   1.67341e-05 &   2.11304e-08 &  -8.38879e-12 &   4.89454e-16 \\
                                & &  -0.000519525 &   5.58154e-06 &  -2.51737e-09 &   4.23944e-13 &  -2.17767e-17 \\
                                & &  -0.000187900 &  -5.04594e-07 &   1.12740e-10 &  -8.22954e-15 &   2.97533e-19 \\
                                & &   8.09438e-06 &   3.23416e-08 &  -7.12496e-12 &   4.71199e-16 &  -1.13154e-20 \\
                                & &  -4.59533e-08 &  -6.66470e-10 &   1.74736e-13 &  -1.53888e-17 &   4.95433e-22 \\
\hline 
   \multirow{5}{*}{E30} &   \multirow{5}{*}{600}   &  -0.217081    &   2.53956e-06 &   9.41585e-09 &  -5.10141e-12 &   3.12351e-16 \\
                                 & &   0.00273654  &   7.35505e-07 &   1.20453e-10 &  -1.03003e-13 &   8.16713e-18 \\
                                 & &  -0.000340049 &   3.56602e-08 &  -7.14945e-11 &   2.46715e-14 &  -1.58730e-18 \\
                                 & &   1.12405e-05 &   3.77161e-09 &   3.82978e-13 &  -5.23170e-16 &   3.70018e-20 \\
                                 & &  -9.81321e-08 &  -1.33020e-10 &   4.23137e-14 &  -2.17822e-18 &   3.02127e-23 \\
\hline 
   \multirow{5}{*}{E40} &   \multirow{5}{*}{550}   &  -0.164876    &   6.46273e-05 &  -9.13256e-09 &  -4.66734e-12 &   4.00926e-16 \\
                                 & &   0.000919627 &  -1.15910e-05 &   6.71286e-09 &  -1.22710e-12 &   7.07783e-17 \\
                                 & &   5.96465e-05 &   1.55389e-06 &  -9.68726e-10 &   1.87070e-13 &  -1.08814e-17 \\
                                 & &  -3.59281e-06 &  -8.07070e-08 &   5.10517e-11 &  -9.61230e-15 &   5.47749e-19 \\
                                 & &   8.93506e-08 &   1.34138e-09 &  -8.48850e-13 &   1.57858e-16 &  -8.90460e-21 \\
\hline 
\hline 
   \multirow{5}{*}{E10R} &   \multirow{5}{*}{800}    & -0.284776    &   6.84536e-05 &   6.24114e-09 &  -6.31142e-12 &   4.35156e-16 \\
                                 & & -0.000875946 &  -7.84760e-06 &   3.68212e-09 &  -6.74293e-13 &   3.87900e-17 \\
                                & &  0.000444425 &   3.45193e-07 &  -2.80505e-10 &   6.84448e-14 &  -4.31516e-18 \\
                                &  & -2.78723e-05 &   1.41346e-08 &   1.47475e-12 &  -1.27081e-15 &   1.00830e-19 \\
                                &  &  5.01599e-07 &  -5.70982e-10 &   1.27045e-13 &  -6.71221e-18 &  -1.46959e-22 \\
\hline 
   \multirow{5}{*}{E20R} &   \multirow{5}{*}{570}    &  -0.191012    & -1.32422e-05 &   4.80900e-08 &  -1.52537e-11 &   1.09666e-15 \\
                                &  &  -0.0135292   &  2.25999e-05 &  -9.63083e-09 &   1.46679e-12 &  -7.50636e-17 \\
                                &  &   0.00182778  & -2.87862e-06 &   1.16985e-09 &  -1.68998e-13 &   8.27820e-18 \\
                                &  &  -8.48704e-05 &  1.27210e-07 &  -5.03760e-11 &   7.53475e-15 &  -3.88715e-19 \\
                                &  &   1.29591e-06 & -1.84083e-09 &   7.16051e-13 &  -1.11176e-16 &   6.04518e-21 \\
\hline 
   \multirow{5}{*}{E30R} &   \multirow{5}{*}{700}   &  -0.222912    &  4.12030e-05 &   6.45691e-09 &  -5.56317e-12 &   3.20268e-16 \\
                                &  &  -0.00750121  &  8.61072e-06 &  -3.69774e-09 &   5.23149e-13 &  -2.12664e-17 \\
                                 & &   0.00101619  & -1.03770e-06 &   3.96737e-10 &  -5.21719e-14 &   2.69455e-18 \\
                                 & &  -4.52185e-05 &  5.16242e-08 &  -1.80256e-11 &   2.38635e-15 & -1.33377e-19 \\
                                &  &   6.89414e-07 & -8.59942e-10 &   2.85543e-13 &  -3.72940e-17 &  2.09229e-21 \\
\hline 
   \multirow{5}{*}{E40R} &   \multirow{5}{*}{750}    &  -0.174407    &  -6.22860e-05 &   5.20280e-08 &  -1.36887e-11 &   7.86332e-16 \\
                                &  &  -0.00538704  &   3.26963e-06 &  -9.00717e-10 &   4.91515e-14 &   4.33190e-18 \\
                                &  &   0.000678700 &   2.04638e-07 &  -2.17996e-10 &   4.98814e-14 &  -2.59934e-18 \\
                                &  &  -3.34451e-05 &  -7.11253e-09 &   8.90405e-12 &  -1.77037e-15 &   6.83819e-20 \\
                                &  &   5.44433e-07 &   1.77550e-11 &  -8.23190e-14 &   1.33103e-17 &  -9.70346e-23 \\
\hline 
\hline 
%\end{center}
\end{tabular} 
\end{table*}

\begin{table*}
\caption {Parameters s$_i$ for the interpolation of $k$ with the temperature.} 
\label{table_param_interpol_k_T_B}
%\begin{center}
\begin{tabular}{c c c c c }
\hline 
\hline 
          &  X35                        &  X40          &       X50A    & X50B             \\
\hline 
%$\lambda_{cut}$  ($\mu$m) &        700           &          500           &             600             & 650 \\
B$_0$ &   -1.77941         &    -0.763913    &    -0.711429   &     -1.34416  \\
B$_1$ &    0.00382619   & -0.000584345  & 6.13839e-06 &   0.00360042   \\
B$_2$ &  -0.000444882  & 4.36954e-05   & -1.21576e-05 & -0.000415103  \\
B$_3$ &   1.60200e-05   & -1.35571e-06  &   5.52623e-07 &  1.49386e-05  \\
B$_4$ &  -2.11636e-07   & 1.82103e-08   &  -7.51126e-09 & -1.98052e-07  \\
B$_5$ &   1.39624e-0 9  &-1.18050e-10  &   5.10392e-11  &  1.30725e-09  \\
B$_6$ &  -4.98240e-12   & 4.06062e-13  & -1.88955e-13&  -4.66199e-12  \\
B$_7$ &   9.23759e-15   & -7.19782e-16 &   3.64151e-16&   8.63537e-15   \\
B$_8$ &  -6.99977e-18   &  5.20154e-19 &  -2.86281e-19 & -6.53681e-18  \\
\hline 
  &  &  &    \\
  & E10  & E20 & E30 &  E40  \\
\hline 
%$\lambda_{cut}$  ($\mu$m)&         800           &      800           &        600             & 550 \\
B$_0$ &       -2.10829    &    -1.79874     &      -1.37358    &      -0.865443 \\
B$_1$ &    -0.00282659 &   0.00496601 &  -7.61072e-05 &   -0.00160168 \\
B$_2$ &   -4.95562e-05 &   6.80127e-05 &  -3.47446e-05 &  -8.00290e-05 \\
B$_3$ &    2.23095e-06 &  -2.79818e-06 &   1.71218e-06 &   3.37313e-06 \\
B$_4$ &   -2.82256e-08 &   3.73929e-08 &  -2.32938e-08 &  -4.49327e-08 \\
B$_5$ &    1.89849e-10 &  -2.50987e-10 &   1.59461e-10 &  3.02241e-10 \\
B$_6$ &   -7.10639e-13 &   9.17059e-13 &  -5.96980e-13 &  -1.10943e-12 \\
B$_7$ &    1.39156e-15 &  -1.74437e-15 &  1.16400e-15 &   2.12127e-15 \\
B$_8$ &   -1.11132e-18 &   1.35496e-18 &  -9.24940e-19 &  -1.65593e-18\\
\hline 
  &  &  &    \\
  & E10R & E20R & E30R & E40R \\
\hline 
%$\lambda_{cut}$  ($\mu$m)&         800           &      570           &        700             & 750 \\
B$_0$ &      -1.05712   &   -0.283288   &  -2.25928     &    -0.484317          \\
B$_1$ &     0.00342552  &  0.000722081&    0.0118293    &  0.00313481           \\
B$_2$ &    5.37581e-06  & -3.08444e-05  & -0.000196327&  0.000261087    \\
B$_3$ &   -1.29599e-08  &  8.66218e-07  &  7.52955e-06  & -9.58241e-06          \\
B$_4$ &   -9.91118e-10  & -1.08170e-08  & -1.03664e-07  &  1.26258e-07          \\
B$_5$ &    9.73318e-12  &  6.80201e-11  &  6.93325e-10  & -8.34625e-10          \\
B$_6$ &   -4.27986e-14  & -2.28765e-13  & -2.49065e-12  &  2.99118e-12          \\
Bs$_7$ &    9.29408e-17 &  3.96458e-16  &  4.64027e-15  & -5.57403e-15          \\
B$_8$ &   -8.00080e-20  & -2.79818e-19  & -3.53060e-18  &  4.24534e-18          \\
\hline 
\hline 
%\end{center}
\end{tabular} 
\end{table*}

\begin{table*}[!t]
\caption {Parameters c$_{i,j}$ and $\lambda_{cut}$ ($\mu$m) for the interpolation of $n$ with the temperature.} 
\label{table_param_interpol_n_T_C}
%\begin{center}
\begin{tabular}{c c c c c c c c}
\hline 
\hline 
Sample & $\lambda_{cut}$  ($\mu$m) & C$_{0,*}$    &  C$_{1,*}$  &   C$_{2,*}$ &  C$_{3,*}$ &   C$_{4,*}$ \\
\hline 
\multirow{5}{*}{X35} &   \multirow{5}{*}{ 700}  &    0.369289 &  4.23448e-05 &  -2.97006e-09 &  -4.57994e-13 &   4.31452e-17  \\
 &  &     0.00162165 &  -4.04859e-06 &   1.90709e-09 &  -2.99994e-13 &   1.52380e-17 \\
 & &    -0.000285441 &   6.19443e-07 &  -2.94318e-10 &   4.70445e-14 &  -2.40857e-18 \\
 & &     1.59422e-05 &  -3.19213e-08  &  1.49219e-11 &  -2.37573e-15 &   1.21414e-19 \\
 &  &   -2.83148e-07 &   5.27905e-10 &  -2.39178e-13  &  3.77860e-17 &  -1.92725e-21 \\
\hline 
\multirow{5}{*}{X40} &   \multirow{5}{*}{ 1900}  &  0.355562 &  4.98757e-06 &  4.55709e-09 & -4.41670e-13  & 1.09071e-17 \\
 &  &     0.000312686 &  5.05894e-07 & -2.27231e-10 &  3.10411e-14 & -1.33888e-18 \\
 &  &    -6.91695e-05 & -7.16158e-08 &  3.29694e-11 & -4.20651e-15 &  1.70653e-19 \\
 &  &     2.67311e-06 &  6.04336e-09 & -2.61881e-12 &  3.22174e-16 & -1.23395e-20 \\
 &  &    -3.60552e-08  &-1.26708e-10 &  5.50867e-14 & -6.76692e-18 &  2.56606e-22 \\
\hline 
\multirow{5}{*}{X50A} &   \multirow{5}{*}{600}  &       0.370630   &-1.61102e-05   & 1.74702e-08   &-3.16679e-12    &1.68045e-16 \\
  &  &    -0.00153840  &  7.15684e-06  & -3.98608e-09  &  7.07782e-13  & -3.97369e-17 \\
 &  &     8.44297e-05  & -8.54571e-07  &  5.02182e-10  & -9.00154e-14  &  5.06358e-18 \\
 &  &    -9.38403e-07  &  3.92471e-08  & -2.35561e-11  &  4.25758e-15  & -2.40424e-19 \\
 &  &    -1.80534e-08  & -6.07714e-10  &  3.66824e-13  & -6.65944e-17  &  3.77334e-21 \\
\hline 
\hline 
\multirow{7}{*}{E10} &   \multirow{3}{*}{175}  &         0.291927  &   4.54945e-05  &  -1.96681e-10 & & \\
  &  &   0.000144810 &   -1.01807e-06   &  8.48279e-11 & - & - \\
 &  &   -1.47214e-05  &   2.13280e-08 &   -1.19527e-12 & - & - \\
\cline{3-7}
         &   \multirow{4}{*}{1200}  &        0.454723   &1.43063e-06  &-5.12403e-10   &2.65641e-14 & \\
 &  &    -0.00189216   &4.65902e-06 - &9.87335e-10   &6.27582e-14 & - \\
 &  &     4.61189e-05  &-1.40322e-07  & 3.43893e-11  &-2.35977e-15 & - \\
 &  &    -5.99311e-07   &1.84060e-09  &-4.37328e-13  & 3.04827e-17 & - \\
\hline 
\multirow{5}{*}{E20} &   \multirow{5}{*}{950}  &   0.303521  &  1.58975e-05  &  1.07446e-08  & -2.11759e-12  &  1.03193e-16 \\
&  &    -0.00348286  &  3.87721e-06  & -1.34805e-09  &  1.85428e-13  &-8.54107e-18 \\
&  &    0.000493440  & -5.19204e-07  &  1.62986e-10  & -2.16487e-14  & 9.96719e-19 \\
&  &   -2.52135e-05  &  2.59807e-08  & -7.91240e-12  &  1.04137e-15  & -4.78603e-20 \\
&  &    4.10722e-07  & -4.29314e-10  &  1.31039e-13  & -1.71048e-17  &  7.80037e-22 \\
\hline 
\multirow{5}{*}{E30} &   \multirow{5}{*}{1000}  &  0.289103   & 4.06453e-05  &  1.82651e-09  & -9.75425e-13  &  5.49498e-17 \\
&  &        0.0230209 &  -1.10844e-05 &   1.46127e-09 &  -2.54755e-14 &  -3.17078e-18 \\
&  &      -0.00380266 &   2.06731e-06 &  -3.48739e-10 &   1.99446e-14 &  -1.99619e-19 \\
&  &      0.000209123 &  -1.22496e-07 &   2.30080e-11 &  -1.62835e-15 &   3.46314e-20 \\
 &  &    -3.64811e-06 &   2.23781e-09 &  -4.45059e-13  &  3.45245e-17  & -8.78266e-22 \\
\hline 
\multirow{5}{*}{E40} &   \multirow{5}{*}{650}  &   0.345753 &   2.51081e-05 &   7.58865e-09 &  -1.85812e-12  &  9.88297e-17 \\
 &  &     -0.00331237 &   4.09577e-06 &  -1.41252e-09 &   1.83864e-13 &  -8.02126e-18 \\
 &  &     0.000441963 &  -5.86659e-07 &   2.06129e-10 &  -2.68441e-14 &   1.17311e-18 \\
 &  &    -2.10372e-05 &   2.81631e-08 &  -1.01798e-11 &   1.37669e-15 &  -6.20162e-20 \\
 &  &     3.24317e-07 &  -4.50714e-10  &  1.66822e-13 &  -2.29372e-17 &   1.04671e-21 \\
\hline 
\hline 
\multirow{5}{*}{E10R} &   \multirow{5}{*}{750}  &  0.283509 &    6.21468e-07 &    1.82945e-08 &   -3.22552e-12 & 1.57445e-16 \\
&  &       -0.00160464 &    2.06506e-06 &   -8.75247e-10 &    1.19234e-13 &   -5.41886e-18 \\
&  &       0.000160037 &   -3.99429e-09 &   -2.78930e-11 &    7.66229e-15 &   -4.79439e-19 \\
&  &      -7.69713e-06 &   -6.43001e-09 &    4.86659e-12 &   -9.02578e-16 &    4.93893e-20 \\
&  &       1.24242e-07 &    1.56065e-10 &   -1.06439e-13 &    1.86961e-17 &   -9.99181e-22 \\
\hline 
\multirow{7}{*}{E20R} &   \multirow{3}{*}{250}  &      0.286205 &    3.86873e-05 &   -3.45899e-10 &   &   \\
 &  &    0.000185806  &-1.49199e-07  & 1.48446e-11& -   & -   \\
&  &    -8.79273e-06  &-5.70200e-09 &  1.55098e-12& -   & -   \\
\cline{3-7}
&   \multirow{5}{*}{950}  &   0.465402   &  4.73369e-06  &  -1.60286e-09   &  2.96056e-13 &   \\
&  &    -0.000102724  &-8.85351e-07 &  3.33436e-10 & -3.95540e-14 &  -  \\
&  &    -3.70759e-06 &  1.44039e-07  &-4.91816e-11 &  5.53056e-15 &  -  \\
&  &     2.90672e-07 & -3.23663e-09  & 1.17669e-12 & -1.31112e-16 &  -  \\
\hline 
\multirow{5}{*}{E30R} &   \multirow{5}{*}{1000}  & 0.313227 &  1.66762e-05 &  1.04173e-08 & -2.05911e-12  & 1.00058e-16 \\
&  &         -0.00487897 &  3.96007e-06 & -1.27857e-09 &  1.56551e-13 & -6.47686e-18 \\
&  &         0.000643109 & -4.73774e-07 &  1.37489e-10 & -1.55152e-14 &  6.14124e-19 \\
&  &        -3.16418e-05 &  2.11753e-08 & -5.50482e-12 &  5.87376e-16 & -2.24657e-20 \\
&  &         5.06526e-07 & -3.20817e-10 &  7.55759e-14 & -7.47099e-18 &  2.68804e-22 \\
\hline 
\multirow{5}{*}{E40R} &   \multirow{5}{*}{900}  &  0.328945 &  2.61617e-05 &  4.84095e-09 & -1.22844e-12 &  6.14259e-17 \\
 &  &         0.00123525 & -5.85481e-07 &  2.26149e-11  &  5.49969e-15 & -3.48379e-19 \\
&  &        -0.000189193 &  9.84340e-08  &-6.98533e-12 &  -1.81117e-16 & 3.10827e-20 \\
&  &         8.28390e-06 & -4.06859e-09 & -6.09666e-14  &  9.10942e-17 & -5.94872e-21 \\
&  &        -1.22710e-07 &  4.91876e-11 &  9.11209e-15  & -2.90851e-18 &  1.68421e-22 \\

\hline 
\hline 
%\end{center}
\end{tabular} 
\end{table*}

\begin{table*}[!t]
\caption {Parameters for the interpolation of $n$ with the temperature for the X50B sample.} 
\label{table_param_interpol_n_T_X50B}
%\begin{center}
\begin{tabular}{ c c c c c c c}
\hline 
\hline 
Sample & $\lambda_{cut}$  &  C$_{0,*}$    &  C$_{1,*}$  &   C$_{2,*}$ &  C$_{3,*}$ &   C$_{4,*}$ \\
\hline 
\multirow{10}{*}{X50B} &   \multirow{10}{*}{850}  &    0.364925   & 2.38971e-06  &  3.49962e-09  & -4.42292e-13  &  1.42006e-17 \\
 & &     -0.00172021  & -1.83862e-07  &  4.99537e-10  & -1.07578e-13  &  6.31790e-18 \\
 & &     0.000229433  & -2.22702e-10   & -4.71730e-11  &  1.07137e-14  & -6.35744e-19 \\
 & &    -1.13573e-05  &  6.46914e-10  &  1.97760e-12  & -4.55778e-16  &  2.67421e-20 \\
 & &     1.79178e-07  & -5.08483e-12  & -3.31158e-14  &  7.22734e-18  & -4.08760e-22 \\
& & & & & &\\
& &     E$_0$                   & E$_1$         & E$_2$          & E$_3$         & E$_4$    \\
\cline{3-7}
& &    -0.00342689      & 4.86813e-05  & -5.13022e-06 & 1.79511e-07 &  -2.36441e-09  \\
& &    E$_5$                    & E$_6$                  &E$_7$         & E$_8$           &  -   \\
\cline{3-7}
& &    1.55339e-11   & -5.51124e-14  &1.01526e-16  & -7.64389e-20 & - \\
%$\lambda_{cut}$ ($\mu$m) & 850 &  &   & \\
% s$_0$    &  -0.00342689  &    &  &   &   &   \\
% s$_1$    &  4.86813e-05  &    &   &   &   &  \\
% s$_2$    &  -5.13022e-06  &   &  &    &   & \\
% s$_3$    &  1.79511e-07   &   &   &   &   &   \\
% s$_4$    &  -2.36441e-09  &    &  &   &   &   \\
% s$_5$    &  1.55339e-11  &    &   &   &   &  \\
% s$_6$    &  -5.51124e-14 &   &   &    &   &  \\
 %s$_7$    &   1.01526e-16  &   &   &   &   &   \\
 %s$_8$    & -7.64389e-20  &    &  &   &   &  \\
\hline 
\hline 
%\end{center}
\end{tabular} 
\end{table*}

\end{document}